\documentclass[%
 reprint,
nofootinbib,
 amsmath,amssymb,
 aps,
 prd,
]{revtex4-1}

\usepackage{graphicx}
\usepackage{dcolumn}
\usepackage{bm}
\usepackage{epsfig}
\usepackage{amsmath}
\usepackage{enumerate}
\usepackage{mathrsfs}
\usepackage{natbib}
\usepackage{hyperref}
\usepackage{hhline}
\usepackage{color}

\newcommand{\md}{\mathrm{d}}
\newcommand{\sch}{Schr\"{o}dinger }
\newcommand\lsim{\mathrel{\rlap{\lower4pt\hbox{\hskip1pt$\sim$}}
        \raise1pt\hbox{$<$}}}
\newcommand\gsim{\mathrel{\rlap{\lower4pt\hbox{\hskip1pt$\sim$}}
        \raise1pt\hbox{$>$}}}

\begin{document}
\preprint{APS/123-QED}

\title{Numerical and Perturbative Computations of the Fuzzy Dark Matter Model}

\author{Xinyu Li}
\email{xinyu.li@columbia.edu}
\affiliation{%
Center for Theoretical Physics, Department of Physics, Columbia University, New York, NY 10027
}%

\author{Lam Hui}%
 \email{lhui@astro.columbia.edu}
\affiliation{%
Center for Theoretical Physics, Department of Physics, Columbia University, New York, NY 10027
}%


\author{Greg L. Bryan}
\email{gbryan@astro.columbia.edu}
\affiliation{
Department of Astronomy, Columbia University, New York, NY 10027 \\
Center for Computational Astrophysics, Flatiron Institute, New York, NY 10003
}%


\date{\today}

\begin{abstract}
We investigate nonlinear structure formation in the fuzzy dark matter (FDM)
model using both numerical and perturbative techniques.
On the numerical side, we examine the virtues and limitations of
a Schr\"odinger-Poisson solver (wave formulation) versus a fluid
dynamics solver (Madelung formulation).
On the perturbative side, we carry out a
computation of the one-loop mass power spectrum, i.e. up to third order in
perturbation theory. We find that (1) in many situations, the fluid dynamics solver is
capable of producing the expected interference patterns, 
but it fails in situations where destructive interference 
causes the density to vanish -- a generic occurrence in the nonlinear regime.
(2) The Schr\"odinger-Poisson solver
works well in all test cases, but it is demanding in
resolution: suppose one is interested in the mass power spectrum on 
large scales, it's not sufficient to resolve structure on those same
scales; one must resolve the relevant de Broglie scale which is often
smaller. The fluid formulation does not suffer from
this issue. (3) We compare the one-loop mass power spectrum from perturbation
theory against the mass power spectrum from the Schr\"odinger-Poisson
solver, and find good agreement in the mildly nonlinear regime.
We contrast fluid perturbation theory with wave perturbation theory;
the latter has a more limited range of validity.
(4) As an application, we compare the Lyman-alpha forest flux power
spectrum obtained from the Schr\"odinger-Poisson solver versus one from
an N-body simulation (the latter is often used as an approximate
method to make predictions for FDM). At redshift $5$, the two, starting from the same
initial condition, agree to better than $10 \%$ on observationally relevant scales
as long as the FDM mass exceeds $2 \times 10^{-23}$ eV.
We emphasize that the so called quantum pressure is capable of both
enhancing and suppressing fluctuations in the nonlinear regime
- which dominates depends on the scale and quantity of interest.
\end{abstract}

\maketitle


\section{\label{sec:level1}Introduction}

The fuzzy dark matter (FDM) model posits that dark matter consists of
an ultra-light boson with a macroscopic de Broglie wavelength \cite{Baldeschi:1983mq,Hu:2000ke,Lesgourgues:2002hk,Amendola:2005ad,Chavanis:2011zi}.
A pseudo Nambu-Goldstone boson is a fairly natural candidate for such
a particle, for instance an axion-like particle.\footnote{For the purpose of this article, it is not crucial that FDM is an axion or axion-like particle. The main ingredient we assume is that the particle is bosonic, non-relativistic and can be approximated as free except for gravitational interaction. The advantage of thinking of the FDM as an axion is that the relic abundance naturally takes the desired value, under mild assumptions.}
The axion $\phi$ is
an angular field with a periodicity of $2\pi F$, where $F$ is often
called the axion decay constant. Non-perturbative effects give rise to
a potential, breaking the shift symmetry expected for a Goldstone
boson to a discrete symmetry $\phi \rightarrow \phi + 2\pi F$. Assuming a
primordial value of order $F$, the axion goes from being frozen in the
early universe to oscillating in the late universe, leading to a
relic abundance of \cite{Arvanitaki:2009fg,Marsh:2015xka,Hui:2016ltb}:
\begin{eqnarray}
\Omega_{\rm axion} \sim 0.1 \left( \frac{F}{ 10^{17} {\,\rm GeV}}
  \right)^2 \left(\frac{m}{10^{-22} {\,\rm eV}}\right)^{1/2} \, ,
\end{eqnarray}
where $m$ is the axion mass. The abundance is more sensitive to the
value of $F$ than to $m$. Axion candidates in string theory typically
span $F \sim 10^{16} - 10^{18} {\,\rm GeV}$ \cite{Svrcek:2006yi}. There is thus quite a
wide range in $m$ that gives the desired abundance for dark matter.
Interestingly, the rather low $m$ suggested by this argument predicts
a number of astrophysical effects that are potentially observable,
spanning the linear regime (see the seminal paper by \citet{Hu:2000ke} who coined
the term fuzzy dark matter) and the nonlinear one \cite{Schive:2014dra,Schive:2014hza,Mocz:2015sda,Hui:2016ltb,Schwabe:2016rze,Veltmaat:2016rxo,Mocz:2017wlg,Lin:2018whl,Veltmaat:2018dfz,Nori:2018hud}.

For most applications, the non-relativistic approximation is adequate
-- a complex scalar $\psi$ is introduced that relates to the real
axion field $\phi$ as follows:
\begin{eqnarray}
\phi = \sqrt{\frac{\hbar^3}{2m}} \left( \psi e^{-imt /\hbar} + \psi^*
  e^{imt/\hbar} \right) \, ,
\end{eqnarray}
where we set the speed of light $c=1$, but keep $\hbar$
explicit. Assuming $|\dot\psi| \ll m |\psi|$, the Klein-Gordon
equation\footnote{For most applications, the self-interaction of 
the axion can be ignored. Thus, the predictions for structure
formation are identical to that of a light, free scalar.
}
for the axion $\phi$ reduces to the Schr\"odinger equation for
$\psi$:
\begin{eqnarray}
\label{eqn_sch}
i\hbar\left(\partial_t
  \psi+\frac{3}{2}H\psi\right)=\left(-\frac{\hbar^2}{2m
  a^2}\nabla^2+m\Phi\right)\psi \, ,
\end{eqnarray}
where $t$ is proper time, the spatial derivative is with respect to
comoving coordinates, $a$ is the scale factor, $H \equiv \dot a / a$ is the Hubble
parameter and $\Phi$ is the gravitational
potential, obeying:
\begin{eqnarray}
\label{eqn_Poisson}
\nabla^2 \Phi = 4\pi G a^2 \bar\rho \delta
\end{eqnarray}
where $\bar\rho$ is the cosmic mean mass density and $\delta$ is the
overdensity.
We use $\dot{}$ to denote a derivative with respect to proper time $t$.
Note that:
\begin{eqnarray}
\bar\rho = m\bar\psi^2 \quad, \quad \rho = m|\psi|^2 \quad , \quad
  \delta\equiv(\rho-\bar\rho)/\bar\rho \, ,
\end{eqnarray}
where $\bar\psi$ is chosen to be real without loss of generality.

The Schr\"odinger-Poisson system of wave dynamics 
(\ref{eqn_sch}) and (\ref{eqn_Poisson})
can be recast as fluid
dynamics (known as the Madelung formulation \citep{Madelung1927}, 
see the Feynman lectures for a
discussion \citep{1965flp..book.....F}). 
The field $\psi$ is related to the fluid mass density $\rho$ as above,
and the fluid velocity $\boldsymbol{v}$ as follows:
\begin{equation}
	\psi\equiv\sqrt{\frac{\rho}{m}}e^{i\theta} \quad, \quad
\boldsymbol{v}\equiv \frac{\hbar}{m a}\nabla \theta \, .
\end{equation}
With this mapping, the conservation associated with the $U(1)$
symmetry of the Schr\"odinger equation (which originates from particle
number conservation in the non-relativistic limit) becomes mass
conservation:
\begin{eqnarray}
	\dot{\rho}+3H\rho+\frac{1}{a}\nabla\cdot(\rho\boldsymbol{v})=0 \, ,
\label{fluid_den}
\end{eqnarray}
and the conjugate part of the Schr\"odinger equation gives
the analog of the Euler equation:
\begin{eqnarray}
	\dot{\boldsymbol{v}}+H\boldsymbol{v}+\frac{1}{a}(\boldsymbol{v}\cdot\nabla)\boldsymbol{v}
  = -\frac{1}{a}\nabla\Phi-\frac{\hbar^2}{2m^2 a^3}\nabla p \, ,
\label{fluid_vel}
\end{eqnarray}
where 
\begin{equation}\label{qpres}
	p \equiv -\frac{\nabla^2 \sqrt{\rho}}{\sqrt{\rho}} =
        -\frac{1}{2}\nabla^2\log\rho-\frac{1}{4}\left(\nabla\log\rho\right)^2
        \, .
\end{equation}
The quantity $p$ is often referred to as ``quantum pressure''. 
This is a bit of a misnomer (which we adopt nonetheless, following convention)
--- it is in fact not a pressure, but
arises from some particular combination of stress i.e. the stress tensor in
general has non-vanishing
off-diagonal terms.\footnote{The stress tensor takes a special form such
  that its divergence divided by density takes the form of the
  spatial gradient of $p$.}
  
In the fluid formulation, $\hbar$ can be grouped
together with $m$ to define a length scale (the Compton
scale $\hbar/m$), after which $\hbar$ does not appear in the rest of
the equations (\ref{fluid_den}) and (\ref{fluid_vel}). 
In the applications we are interested in, the relevant particle number
occupancy is large, making quantum fluctuations very small. 
The Schr\"odinger equation, despite its appearance, should be
interpreted as an equation for a classical complex scalar $\psi$
(though we will adhere to the common terminology of 
$\psi$ as the wave function).
Wave mechanics effects such as interference are still present,
since $\rho = m |\psi|^2$, but they are classical in nature,
much like the interference of waves in classical electromagnetism.

In the literature, there are investigations of structure formation in
the FDM model using both the wave formulation
\citep{Schive:2014dra,Schwabe:2016rze,Mocz:2017wlg,Mocz:2018ium} and the fluid
formulation \citep{Mocz:2015sda,Veltmaat:2016rxo,Nori:2018hud,Zhang:2017chj}. Our goal is to build on and extend these
investigations in a number of ways. (1) We investigate the strengths
and weaknesses of solvers based on the wave and fluid formulations by
studying test cases. (2) We carry out perturbative computations (in
the fluid formulation, for reasons that will become clear) up to third
order in perturbation theory, and compare the results against
numerical solutions. (3) As an application, we numerically 
compute the Lyman-alpha forest flux power spectrum in the FDM model,
and compare the prediction from solving the Schr\"odinger-Poisson
system versus the prediction from a pure gravity solver (starting from the
same initial conditions). Current constraints on the FDM mass from
the Lyman-alpha forest use N-body simulations (i.e. a gravity solver)
to approximate the dynamics \cite{Irsic:2017yje,Armengaud:2017nkf}\footnote{An exception is \cite{Zhang:2017chj} who incorporated quantum pressure in their simulations. As we will see below, incorporating quantum pressure in a fluid formulation has issues that need to be addressed.}, and we address the question of how the
predictions are sensitive to the presence/absence of quantum pressure.

The paper is organized as follows.
In Sec. \ref{PTcomputation}, we summarize perturbation theory results
that are presented in more details in the Appendix. The perturbative
computation provides a number of insights that are useful for
interpreting the numerical results.
In Sec. \ref{Numericalcomputation}, we describe our code \texttt{SPoS}, a second order
Schr\"odinger-Poisson solver and compare it with a 
grid-based fluid solver that incorporates quantum pressure.
We go over the pros and cons of both, and show the
fluid formulation fails in cases where the
density vanishes due to chance destructive interference (which seem to
generically occur in the highly nonlinear regime). 
Certain details of the fluid solver are relegated to the Appendix.
In Sec. \ref{ForestApplication}, we apply the Schr\"odinger-Poisson solver to
compute the Lyman-alpha forest flux power spectrum, and quantify how
the predictions differ if one were to approximate the dynamics as
purely gravitational (i.e. FDM versus CDM dynamics, from the same
initial conditions). 

In sample cosmological calculations, the parameters used are:
density parameters for dark matter $\Omega_{m,0}=0.268$, dark energy
$\Omega_{\Lambda,0}=0.732$, the Hubble constant in 
unit $100$~/km/s/Mpc $H_0=0.704$, the fluctuation 
amplitude $\sigma_8=0.811$, the spectral index $n_s = 0.961$ and 
the CMB temperature today $T_{CMB} = 2.726$~K. 
Our cosmological numerical simulations start at $z=100$. 

As this article was under preparation, a paper by Nori et al. 
\cite{Nori:2018pka}
appeared that has some overlap with this one, in particular regarding
the Lyman-alpha forest.

\section{Perturbative Computations of the FDM Model}
\label{PTcomputation}

There are two possible perturbative computations,
one in the fluid formulation and one in the wave formulation.
We summarize the main results in this section, and relegate
the details to Appendices \ref{FPT} and \ref{WPT}. 

In the wave formulation, one expands in small perturbation
to the wave function: $\delta\psi / \bar\psi$. 
In the fluid formulation, one expands
in small overdensity
$\delta \equiv (\rho - \bar\rho)/\bar\rho$ and peculiar velocity
$v$. To linear order, they are related by:
\begin{equation}
\delta = (\delta\psi + \delta\psi^*)/ \bar\psi \, ,
\end{equation}
and
\begin{equation}
{\boldsymbol{v}} = {\frac{\hbar} {2iam \bar\psi} }{\mbox{\boldmath$\nabla$}}
(\delta\psi - \delta\psi^*) \, .
\end{equation}
The latter in momentum space reads:
\begin{equation}
\frac{\delta\psi - \delta\psi^*}{\bar\psi}
\sim \frac{am v}{\hbar k} \, .
\end{equation}
where $k$ is the (comoving) momentum. 
The regime of validity in wave perturbation theory is thus different
from that in fluid perturbation theory.
For fluid perturbation theory, $\delta$ and $v$ ought to be small.
For wave perturbation theory, the smallness of the real part of $\delta\psi/\bar\psi$
is equivalent to the smallness of $\delta$, but the
smallness of the imaginary part requires $v \ll \hbar k/(am)$. 
Since the physical length scale of interest $a/k$ is generally longer
than the Compton scale $\hbar/m$, the latter requirement 
is more stringent than simply requiring a small $v$. 
To put it in another way, wave perturbation theory requires
$a/k \ll \hbar /(mv)$ i.e
the proper length scale of interest must be shorter than the de
Broglie scale:
\begin{equation}\label{db}
\frac{\lambda_{\rm deB.}}{2\pi} \equiv \frac{\hbar}{mv} = 1.92\;\mathrm{kpc}\left(\frac{10^{-22}\;\mathrm{eV}}{m}\right)\left(\frac{10 \;\mathrm{km/s}}{v}\right).
\end{equation}
This is a rather demanding requirement: for a given comoving $k$,
wave perturbation theory generally breaks down at an earlier time
than fluid perturbation theory (see Appendix \ref{WPT}).

We therefore focus on fluid perturbation theory in the rest of the
paper.
We compute the predictions up to third order in perturbation, and derive the one-loop corrections to the mass power spectrum
(Appendix \ref{FPT}). Let's consider the right hand side of the Euler
equation (\ref{fluid_vel}) - taking its divergence and using the Poisson
equation (\ref{eqn_Poisson}):
\begin{eqnarray}
\label{EulerRHS}
&& {\rm R. \, H. \, S.} = - 4\pi G\bar\rho a\delta \nonumber \\
&& \quad + \frac{\hbar^2}{ 4 a^3 m^2} \nabla^2 \Big( \nabla^2 \delta - \frac{1}{4}
  \nabla^2\delta^2 - \frac{1}{2} \delta\nabla^2\delta + \nonumber \\
&& \quad \quad 
\frac{1}{8} \nabla^2 \delta^3 + \frac{3}{8} \delta^2 \nabla^2\delta 
+ \frac{1}{8} \delta \nabla^2 \delta^2 +
... \Big)
\end{eqnarray}

To linear order in $\delta$, as noted by \cite{Hu:2000ke}, the quantum
pressure leads to a suppression of growth on small length scales
(large momenta). The characteristic momentum where the quantum
pressure term balances gravity is 
$k_{\rm Jeans}$, the (comoving) Jeans scale
(Appendix \ref{FPT}):
\begin{eqnarray}
\label{kJeansdef}
k_{\rm Jeans} \equiv \frac{44.7}{{\rm Mpc}} 
\left( 6a
\frac{\Omega_{m0}}{0.3}\right)^{1/4}
\left( \frac{H_0}{70 {\,\rm km/s}} 
\frac{m}{10^{-22} {\,\rm eV}} \right)^{1/2} \, .
\end{eqnarray}

Eqn. (\ref{EulerRHS}) shows that the second order terms from
the quantum pressure could act in the {\it same} direction as
gravity, enhancing rather than suppressing fluctuations.
Consider the example of a density peak with 
$\delta > 0$ a Gaussian profile, it is straightforward to
check that, at the peak, the second order contributions to quantum 
pressure act in an opposite way to the first order contribution.
Of course, once the second order terms become important, one has to
consider the higher order terms as well which could yet reverse
the sign of the effect -- in such a case,
the perturbative expansion cannot be counted
on to give the correct predictions and one has to resort to numerical
computations. Nonetheless, this exercise shows that in the nonlinear regime,
it is by no means guaranteed that quantum pressure leads to a
suppression of fluctuations (see further discussions in Appendix \ref{FPT}).

Interestingly, the wave formulation offers
useful insights in the nonlinear regime. 
In a (non-solitonic) halo which consists of a
superposition of waves of different momenta, the fact that
$\rho \propto |\psi|^2$ sets up the opportunity for interference
effects. A chance superposition could lead to local constructive or
destructive interference, causing sizable density fluctuations on
the de Broglie scale. This is borne out by numerical simulations
of FDM halos \cite{Schive:2014dra,Schive:2014hza}.

Let us close this section with a sample computation of the mass power
spectrum up to one-loop, that is, including corrections to the power
spectrum that are quadratic in the linear power spectrum (Appendix \ref{FPT}).

\begin{figure}[htpb]
\vspace*{-1.5cm}
\begin{tabular}{c}
\includegraphics[width=0.5\textwidth]{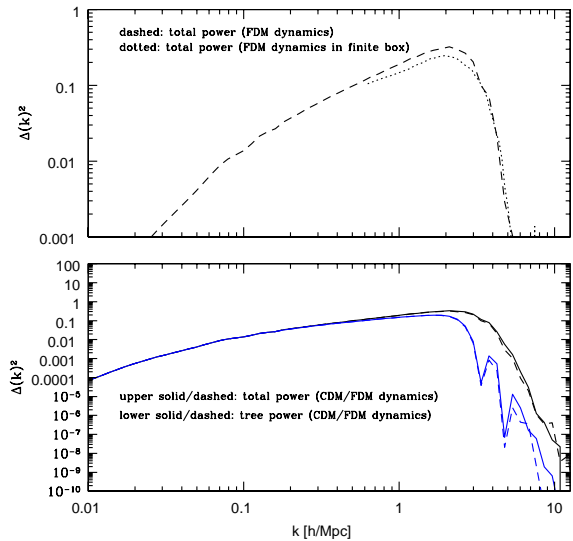} 
\end{tabular}
\vspace*{-1.0in}
\caption{{\it Lower panel:} the upper solid/dashed (black) lines give the
$z=5$ total power (tree + 1-loop) under CDM/FDM dynamics respectively.
The initial condition is the same in both cases, corresponding to
an FDM mass of $10^{-23} {\rm eV}$. The lower solid/dashed (blue)
lines gives the $z=5$ tree power under CDM/FDM dynamics.
{\it Upper panel:} the dashed line is the $z=5$ total power under FDM
dynamics (identical to the upper dashed line in the lower panel); the dotted
line is the total power if the tree power spectrum is cut-off at $k <
2\pi/10$ h/Mpc, which limits the transfer of power from low momenta
to high momenta through the loop integrals
(accounting for the effect of a finite box size of $10$ Mpc/h).
Here, $\Delta (k)^2 \equiv 4\pi k^3 P(k)/(2 \pi)^3$ represents the
dimensionless power.}
\label{pcomparefig}
\end{figure}

The tree (linear) power spectrum at redshift $z=5$ for
an FDM particle mass of $10^{-23}$ eV is shown in the
lower panel of Fig. \ref{pcomparefig} (the lower dashed blue  line).
The tree power spectrum at the same redshift
and with the same initial condition but CDM dynamics is shown
as the lower solid blue line in the same panel.
The fact that they are very similar is due to the fact that
the linear growth factor $D_k$ does not differ a whole lot between
FDM and CDM dynamics for $k < 10$ h/Mpc. 
(Recall from Eqn. (\ref{kJeansdef}) that 
$k_{\rm Jeans} \sim 14$ Mpc$^{-1}$ for a mass of $10^{-23}$ eV and
$a=1/6$.) 
In other words, the suppression of power seen in
Fig. \ref{pcomparefig} at high momenta has nothing
to do with the low redshift FDM dynamics; it's all in the initial
condition. Indeed, the scale at which the initial or primordial power
spectrum is suppressed by a factor of two is \citep{Hu:2000ke}:
\begin{equation}
k_{1/2} = \frac{1.62}{\rm Mpc} \left( \frac{m}{10^{-23} {\,\rm eV}} \right)^{4/9} \, .
\end{equation}
This is substantially smaller (i.e. longer length scale) than
$k_{\rm Jeans}$ at $z=5$. Thus, based on linear perturbation theory
alone, we do not expect CDM and FDM dynamics to give vastly
different power spectra starting from the same initial condition.
The question is whether this persists at higher order
in perturbation theory.

Fig. \ref{pcomparefig} lower panel shows the total (tree + one-loop)
power:
the upper solid (black) line uses CDM dynamics while the upper dashed
(black) line uses FDM dynamics (same initial condition). 
The power is enhanced at high momenta compared
with the tree power spectrum (the lower solid/dashed blue lines).
The power at $k \gsim k_{1/2}$ comes almost entirely
from transferred power from large scales (low momenta).
Just as in the case of the tree power, CDM vs FDM dynamics
does not appear to make a big difference in the total power.
The total power with FDM dynamics is slightly lower than that
with CDM dynamics, except possibly at $k \sim 10$ h/Mpc, though
at that point, the power spectrum is rather low in any case (due to
the suppression of high $k$ power in the initial condition).
One main goal of the following section is to investigate using
numerical simulations to what
extent these conclusions remain valid in the fully nonlinear regime.

As we will see below, numerical simulations place twin demands on
resolution and box size that can be challenging to
satisfy. Perturbative calculation is useful for gauging the effect of
a finite box size. 
Fig. \ref{pcomparefig} upper panel compares the total power
in two cases (both with FDM dynamics): the dashed line
is the same as the black dashed line (power spectrum up to one-loop) 
in the lower panel; the dotted line shows also the
power up to one-loop, except the loop integral is computed with an
infrared cut-off at momentum $2\pi/10$ h/Mpc.
The latter is chosen to mimic the effect of having a finite simulation
box of size $10$ Mpc/h. One can see the transferred power from large
scales is diminished, because the relevant large scale modes are
absent in a finite box. We will return to this issue of a finite box
size in our numerical computations.

\section{Numerical Computations of the FDM Model}
\label{Numericalcomputation}

In this section, we turn to numerical computations.
Just as in the case of perturbative calculations, one
can integrate the Schr\"odinger-Poisson system Eqn. (\ref{eqn_sch}) 
and (\ref{eqn_Poisson}) or the fluid-Poisson system
Eqn. (\ref{eqn_Poisson}), (\ref{fluid_den}) and (\ref{fluid_vel}). 

The advantage of the fluid-Poisson system is that working
hydrodynamics codes already exist. The quantum pressure term $p$ 
encodes the effects of wave dynamics, such as interference.
This is borne out by tests we have performed, detailed in Appendix
\ref{FDMfluidSolver} (see e.g. Fig. \ref{fresnel_rho}). 
Problems arise, however, in test cases where the the density vanishes
at isolated points, which naturally occur due to chance destructive interference
in the wave language. The quantum pressure
is ill-defined in these regions. One might hope that these isolated
problematic points do not affect the overall evolution, but they can,
as illustrated in the example of Fig. \ref{collision}. It is
conceivable that a fluid code can be suitably modified to integrate across
these regions of vanishing density. We defer this to a future
discussion. We will instead largely focus on a numerical code that solves the
Schr\"odinger-Poisson system, though in a few examples, we will
contrast the performance of the fluid and the wave codes.
Let us also mention the possibility of a hybrid approach that
combines the best features of both formulations, for instance a fluid
solver on large scales and a wave solver on small scales (see further
discussions below). An example is given by \cite{Veltmaat:2018dfz} who used the combination of an N-body code and a Schr\"odinger-Poisson solver.

\subsection{Description of \texttt{SPoS} (Schr\"odinger-Poisson Solver)}
By defining a normalized wave function $\tilde{\psi}\equiv a^{3/2}\psi$, the Schr\"odinger-Poisson equation for FDM dynamics takes the following form
\begin{equation}\label{eqn:sch_norm}
\frac{\partial \tilde{\psi}}{\partial t} = (K+V)\tilde\psi\equiv \left(i\frac{\hbar}{ma^2}\nabla^2 -i\frac{m}{\hbar} \Phi\right)\tilde{\psi} \, ,
\end{equation}
where $K=i\frac{\hbar}{ma^2}\nabla^2$ is the kinetic operator and $V= -i m \Phi/\hbar$ is the potential operator.
Our \texttt{SPoS} code integrates Eqn. (\ref{eqn:sch_norm}) using an operator split method similar to \citep{Mocz:2017wlg,Mocz:2018ium}.
The operator split method follows the observation that the unitary
time evolution operator can be split in the following way to give
second order accuracy in time (i.e. corrections are third order):
\begin{eqnarray}\label{eqn:split}
	\tilde\psi(t+\Delta t) &=& e^{(K+V)\Delta t}\tilde\psi(t)\nonumber\\
	&=&e^{V\Delta t/2}e^{K\Delta t}e^{V\Delta t/2}\tilde\psi(t)	+\mathcal{O}(\Delta t^3)\,.
\end{eqnarray}

Therefore, for each time step $\Delta t$, our \texttt{SPoS} code evolves $\tilde\psi$ in the following three sub-steps.
In the first sub-step, $\tilde\psi$ is evolved with the gravitational potential operator
for $\Delta t/2$:
\begin{equation}
\tilde{\psi} \rightarrow \exp\left( -i\frac{m}{\hbar}\Phi\frac{\Delta t}{2}\right)\tilde{\psi}\,.
\end{equation}
In the second sub-step, the kinetic operator is applied to $\tilde\psi$ using the 4th order Runge-Kutta method for time integration with a 4th order finite difference method for spatial derivatives
\begin{equation}
\tilde{\psi} \rightarrow \tilde{\psi} + i\frac{\hbar}{ma^2}\nabla^2\tilde{\psi}\Delta t \, .
\end{equation}
Finally, $\psi$ is evolved with the gravitational potential for another $\Delta t/2$ to finish a full time step of evolution.

The time step $\Delta t$ is set by respecting the
Courant-Friedrichs-Lewy (CFL) condition for a parabolic system with cell
size $\Delta x$ and requiring the phase change in the  gravitational
potential step is smaller than unity i.e.
\begin{equation}
\Delta t<\min\left[\frac{ma^2}{6\hbar}\Delta x^2,  \frac{\hbar}{m\Phi}\right]\, .
\end{equation}

The \texttt{SPoS} code is built as a module into the ENZO code \citep{Bryan:2013hfa} for cosmological simulations with existing modules to calculate the cosmological expansion and gravitational potential; some analysis below is carried out with the {\sc yt} package \citep{yt}.

\subsection{Tests of \texttt{SPoS}
(and Comparisons with a Fluid Solver)}\label{TestOfSolver}

\begin{figure*}[htpb]
\vspace*{-0.3cm}
\begin{tabular}{c}
\includegraphics[width=0.5\textwidth]{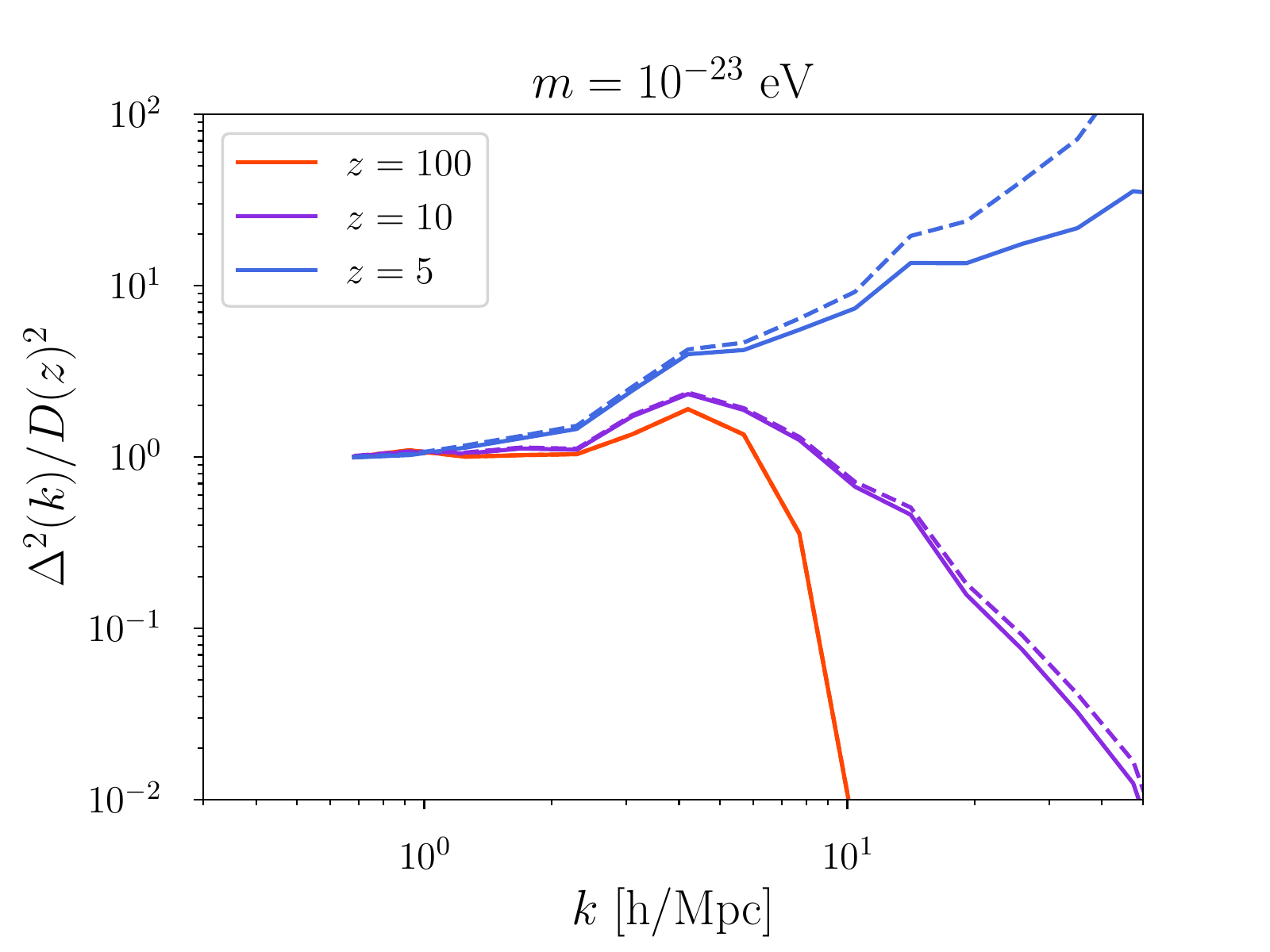} 
\includegraphics[width=0.5\textwidth]{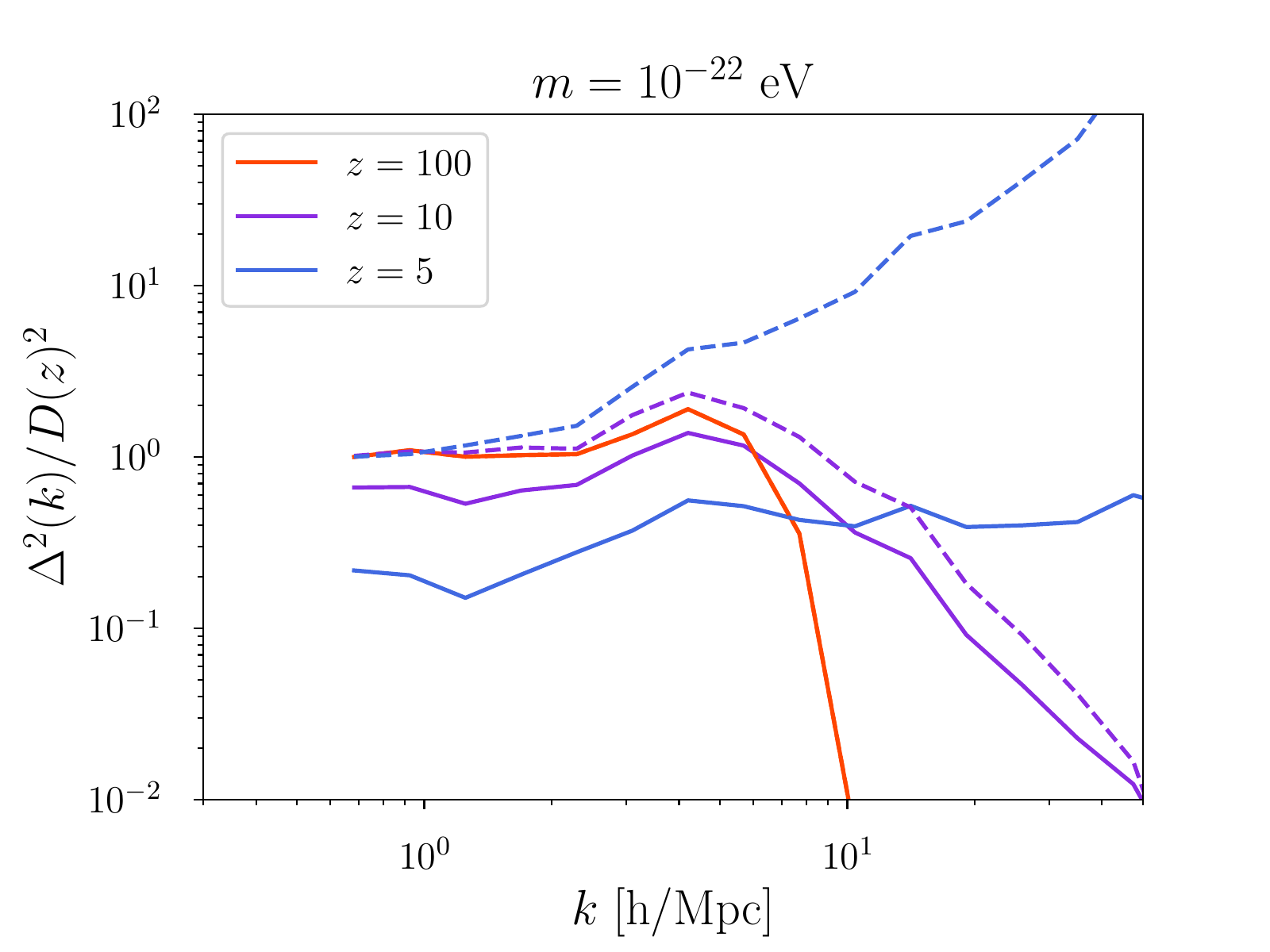} 
\end{tabular}
\caption{Power spectrum evolution in a $10$~Mpc/h box. Here, 
$\Delta^2 (k,z) \equiv
4\pi P(k,z) / (2\pi)^3$ represents the dimensionless power. 
It is normalized by the linear growth factor $D(z)^2$ (according to
CDM, which coincides with that for FDM at low $k$'s). 
The normalization of $D(z)$ is chosen such that the ratio 
$\Delta^2(k,z)/D(z)^2$ should be unity at low $k$'s if the power
spectrum is evolving correctly. The solid lines represent the outputs
from \texttt{SPoS}. The dashed lines represent the
outputs from a fluid-Poisson solver that incorporates quantum
pressure. \textit{Left} -- FDM (dynamics) mass
  $m=10^{-23}$~eV. \textit{Right} -- FDM (dynamics) mass
  $m=10^{-22}$~eV. The failure of the solid lines to converge to unity
at low k's for $z=10$ and $z=5$ is a sign that \texttt{SPoS}
does not give the correct power spectrum evolution, because it is not
resolving the de Broglie scale. On the other hand, the fluid-Poisson solver with quantum pressure gives the correct behavior at low k.
}
\label{ps23}
\end{figure*}

The Schr\"odinger-Poisson solver is very demanding in terms of
resolution. Recall that the velocity is related to the gradient of the
phase of the wave function, i.e. a given velocity translates into a phase that varies on the scale
of the de Broglie wavelength.
It is thus important the de Broglie scale is resolved;
otherwise the velocity field is not represented correctly.
This is true even if the scale of interest is much larger than the de
Broglie scale. This is illustrated in Fig. \ref{ps23}. 

The matter (mass) power spectrum is shown for FDM dynamics
corresponding
to a mass of $10^{-23}$ eV and $10^{-22}$ eV on the left and right panel
of Fig. \ref{ps23} respectively.\footnote{Both simulations use the {\it same} initial power spectrum
\cite{Hu:2000ke} corresponding to an FDM mass of $10^{-22}$ eV.}
In both cases, the simulation box size is $10$ Mpc/h
comoving, with a $1024^3$ grid (the Nyquist frequency is
about $600$ h/Mpc comoving). 
The resolution is adequate
to resolve the de Broglie scale for the case of $10^{-23}$ eV but
not for $10^{-22}$ eV. One can estimate the de Broglie scale
by using the velocity dispersion on the scale of the box (comoving)
$10$ Mpc/h. At $z = 5$, this is about $100$ km/s, giving a (physical
or proper) de Broglie
scale of about $2 {\,\rm kpc}$ for $m=10^{-23}$~eV
and $0.2$ {\,\rm kpc} for $m=10^{-22}$~eV (Eqn. \ref{db}).
Contrast this with the simulation resolution, which is about
$2$ kpc (proper) at $z=5$. Thus the de Broglie scale is marginally
resolved for $m=10^{-23}$ eV but certainly not for
$m=10^{-22}$ eV.

In Fig. \ref{ps23}, we divide the power spectrum
by the square of the linear growth factor $D(z)$ (where $D(z)$ is the
CDM growth factor, which matches the FDM linear
growth factor at low $k$). The normalization of $D(z)$ is chosen such that
at $z=100$, the ratio $\Delta^2(k,z)/D(z)^2$ is unity at the lowest comoving
$k \sim 0.7$ h/Mpc. 
With this convention, the fact that on the left panel of Fig. \ref{ps23}, the power
spectrum at all redshifts (solid lines) converges to unity at low $k$'s means
the large length scale power is evolving correctly i.e. in accordance
with linear perturbation theory expectation. 
The same is not true on the right panel, where the de Brogile scale is
not resolved. The large length scale
growth of the power spectrum is slower than it should be i.e.
at low redshifts, the ratio $\Delta^2(k,z)/D(z)^2$ falls short of
unity at low $k$'s (solid lines). This occurs despite the fact that
those scales (say comoving $k \lsim 1$ h/Mpc) are well resolved.

The Schr\"odinger-Poisson code thus behaves very differently from
an N-body code or a fluid code. {\it For an N-body or fluid-formulation code,
the correct evolution of
power on some large scale requires only that large scale to be 
resolved. For a wave-formulation code, there is the additional
requirement that the de Broglie scale be resolved, which is often more
demanding.}

Also plotted in Fig. \ref{ps23} are the results of
a fluid-formulation solver that incorporates quantum pressure (dashed
lines). One can see that the low $k$ (large length scale) power
spectrum evolves correctly in both cases i.e. a fluid solver is less
demanding in terms of spatial resolution if one is only interested in
large length scales. Note that the wave-formulation solver and the
fluid-formulation solver differs in their predictions at high $k$'s
(small length scales); the fluid solver predicts generally more power
at high $k$'s. We interpret this as a by-product of the fluid code's
failure to evolve correctly past regions of destructive interference
(see Fig. \ref{collision} in Appendix
\ref{FDMfluidSolver}). 

Fig. \ref{slice_pic} shows a density slice of the wave-formulation
simulation on the left and fluid-formulation simulation on the right (color coded for density)
at $z=5$, for the case of $m = 10^{-23}$ eV. One can see that
the structures on large scales are broadly consistent.
But the fluid simulation suffers from too much collapse and less wave-like structure on small
scales, due to its inaccuracy in evolving past regions of destructive
interference (see discussion in Appendix \ref{FDMfluidSolver}, around Fig. \ref{collision})-- the density structure becomes inaccurate
on small scales, which compromises also the computation of quantum pressure.

\begin{figure*}[htpb]
\vspace*{-0.3cm}
\begin{tabular}{c}
\includegraphics[width=0.5\textwidth]{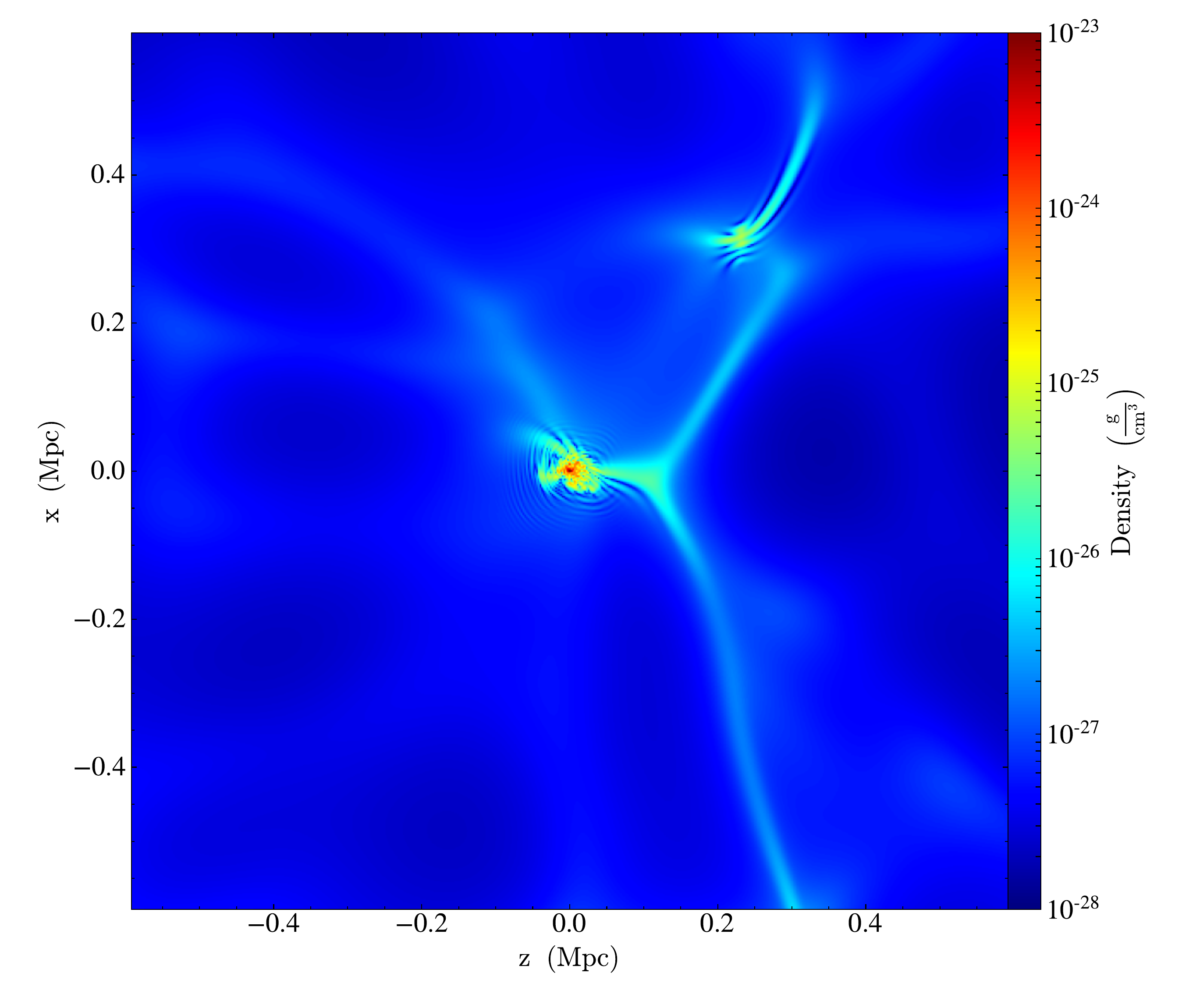} 
\includegraphics[width=0.5\textwidth]{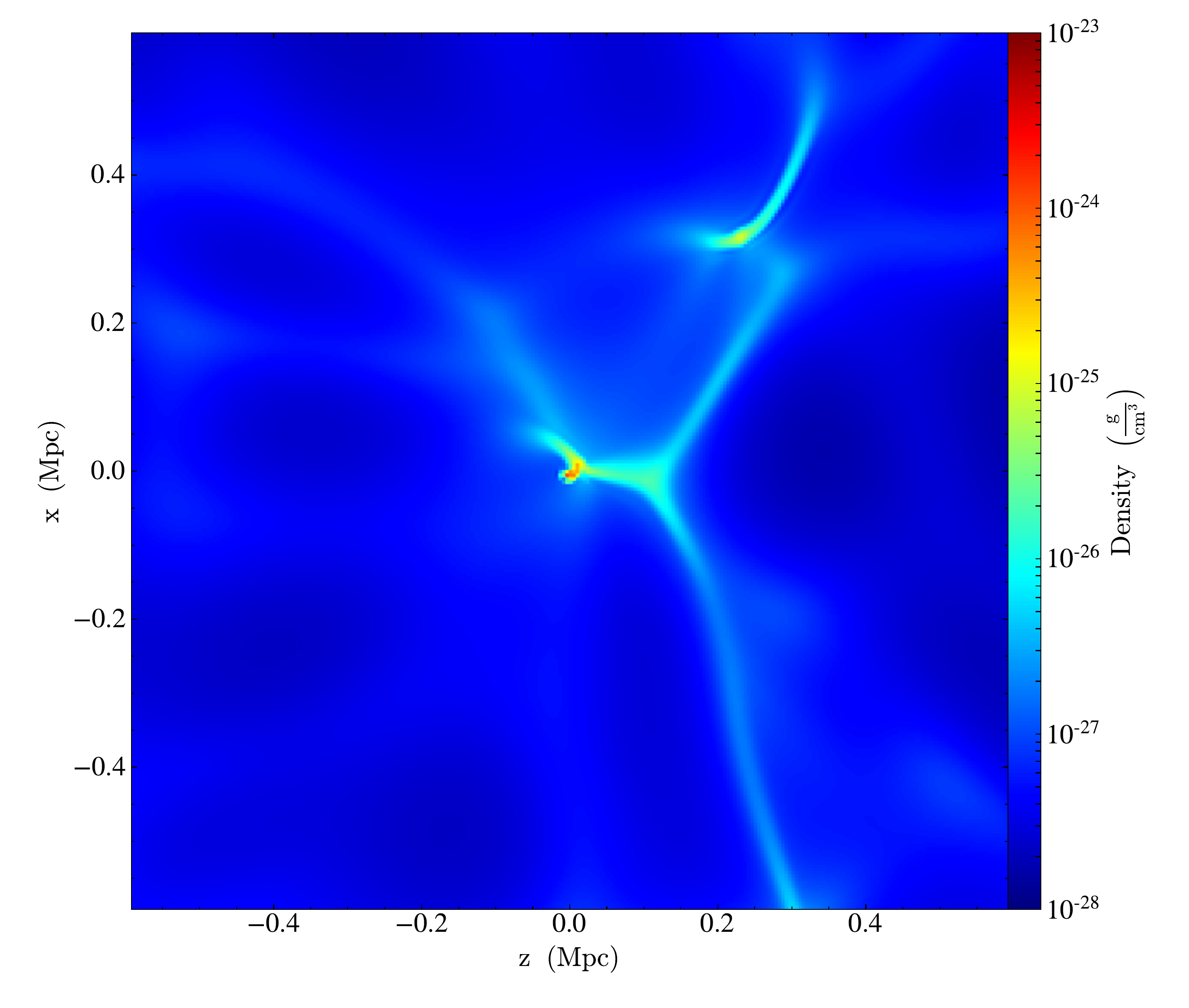} 

\end{tabular}
\caption{A slice at $z=5$ depicting density from
simulations with $m=10^{-23}$ eV (corresponding to Fig. \ref{ps23}). \textit{Left} -- wave-formulation simulation. \textit{Right} -- fluid-formulation simulation.
}
\label{slice_pic}
\end{figure*}

\begin{figure}[htpb]
\vspace*{-0.3cm}
\begin{tabular}{c}
\includegraphics[width=0.48\textwidth]{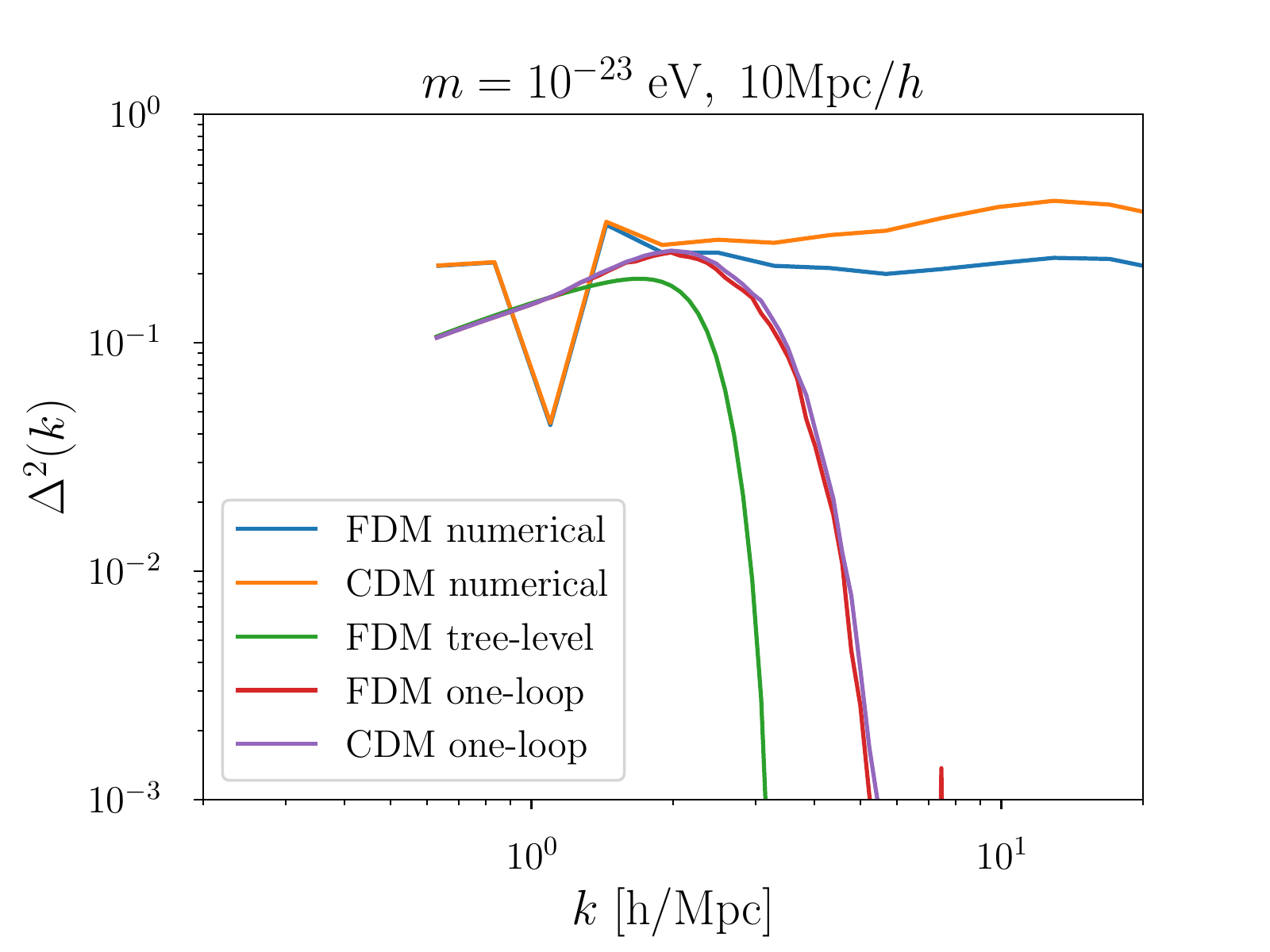} \\
\includegraphics[width=0.48\textwidth]{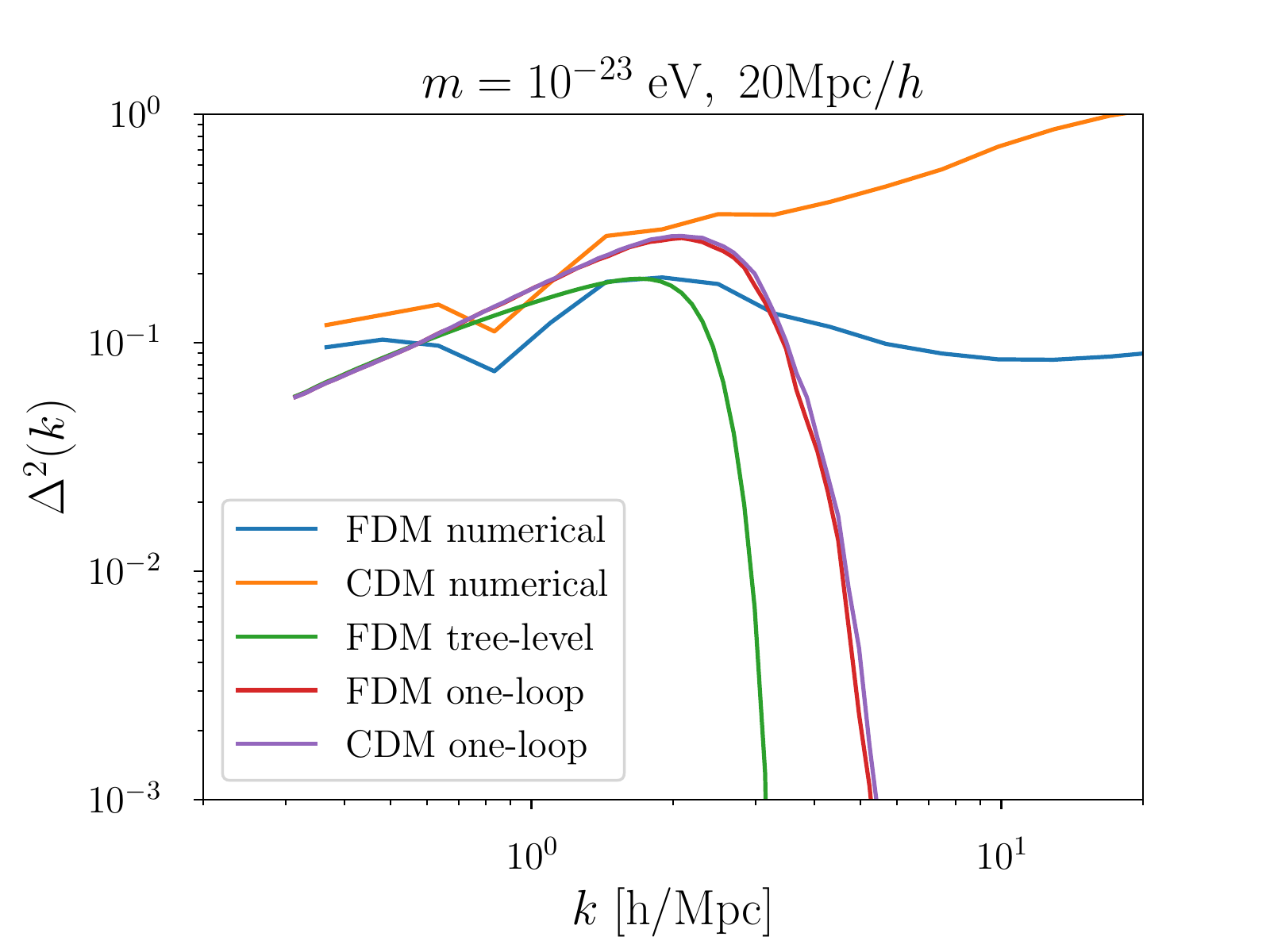} \\
\includegraphics[width=0.48\textwidth]{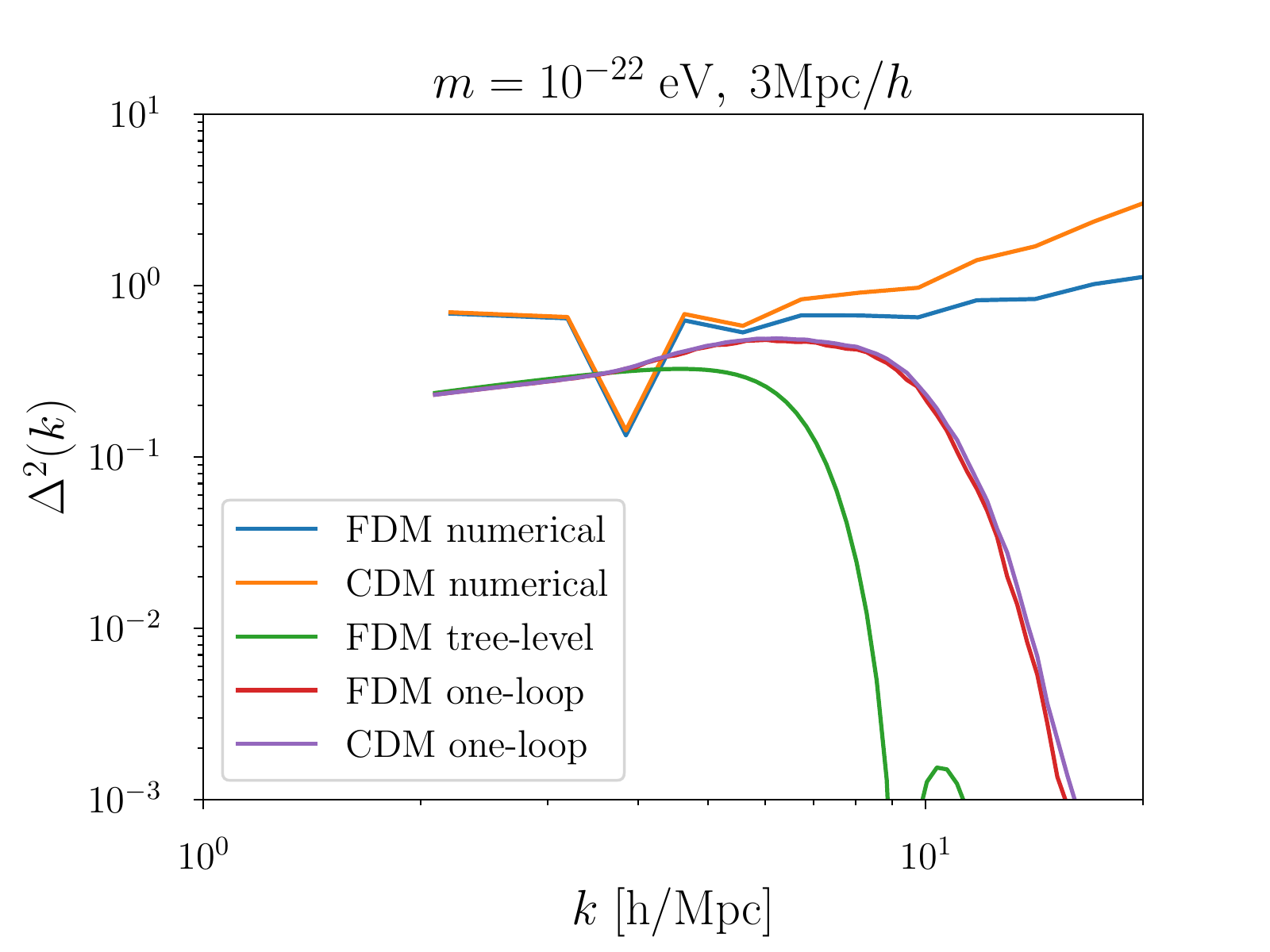} 
\end{tabular}
\caption{Comparison of mass power spectrum between simulations and
  perturbative calculations.
  The top (orange) line is from an N-body simulation i.e. CDM
  dynamics but from the same initial condition. The next lower (blue)
  line is from \texttt{SPoS}. The next two lines (red
  and purple) are the one-loop predictions form FDM and CDM dynamics
  respectively. The lowest (green) line is the linear (tree) FDM power spectrum. 
  \textit{Top} -- FDM mass $10^{-23}$~eV in a $10$ Mpc/h
  box.
  \textit{Middle} -- FDM mass $10^{-23}$~eV in a $20$ Mpc/h
  box.
  \textit{Bottom} -- FDM mass $10^{-22}$~eV in a $3$ Mpc/h
  box.
}
\label{pert10}
\end{figure}

As a further test, we compare the predictions of third-order
perturbation theory (specifically,  the mass power spectrum including the
one-loop corrections) with outputs from \texttt{SPoS}. The top panel of Fig. \ref{pert10} shows the
mass power spectrum for the case of an FDM mass of $10^{-23}$ eV,
with a $10$ Mpc/h comoving box and a $1024^3$ grid.\footnote{For Fig. \ref{pert10}, the
  initial condition uses an FDM mass that is consistent with the FDM
  dynamics i.e. $10^{-23}$ eV for the top and middle panel,
and $10^{-22}$ eV for the bottom panel. In all cases, FDM predictions
are computed with \texttt{SPoS}. The one-loop power
spectrum is computed with an infra-red cut-off in the loop integral
that corresponds to the box size in each case.}
One can see that the one-loop power spectrum matches the
numerical power spectrum up to a comoving $k \sim 3$ h/Mpc.
We also show the numerical power spectrum with the same initial
condition, box size and resolution, 
but with CDM dynamics (i.e. a pure N-body simulation
with an initial power spectrum that corresponds to that of an FDM
model). CDM dynamics appears to predict more power at high $k$'s
compared to FDM dynamics.

The middle panel of Fig. \ref{pert10} shows the same but with a $20$ Mpc/h comoving box.
One can see that the FDM numerical predictions fall short at low $k$ because
of the failure to resolve the de Broglie scale, while the CDM
numerical predictions seem fine (at least at low $k$).
The bottom panel of Fig. \ref{pert10} shows the analog for an FDM
mass of $10^{-22}$ eV. In this case, in order to resolve the de
Broglie scale, the box size has to be lowered to $3$ Mpc/h comoving
(the grid remains $1024^3$). The lack of large scale power in the
initial condition due to the small box size means the predicted
matter power spectrum is lower than it should be.
This should be kept in mind in applications of the code;
we will return to this point below in the application to the
Lyman-alpha forest.

\section{Application to the Lyman-alpha Forest}
\label{ForestApplication}

The Lyman-alpha forest consists of hydrogen absorption lines in the spectra of distant quasars. They provide useful constraints on the matter power spectrum at redshift $z=2-6$ on scales down to about $0.1$~Mpc comoving \citep{Croft:1997jf,Hui:1998hq,Croft:2000hs,McDonald:2004xn,Viel:2006yh}.
The small scale linear matter power spectrum is sensitive to assumptions about the nature of dark matter. 
For instance, warm dark matter causes a suppression of power on scales below its free-streaming scale, and can be constrained by the Lyman-alpha forest observations \cite{Viel:2013apy}.
The warm dark matter constraint \cite{Viel:2013apy} can be roughly translated into a constraint on FDM which also suppresses power on small scales e.g. \cite{Hui:2016ltb}.
More recently, Lyman-alpha forest constraints on FDM were obtained by
\cite{Irsic:2017yje,Armengaud:2017nkf}, who simulated the FDM model using N-body simulations; the effect of ``fuzziness" is accounted for purely through the initial conditions.
In other words, the assumption was that CDM dynamics (with FDM initial condition) is sufficient to capture accurately the predictions of the FDM model.
We are in a position to check this assumption, by comparing the outputs of
an N-body code (CDM dynamics) and \texttt{SPoS} (FDM dynamics), starting from the same initial condition. Let us stress that physical effects that cause additional sources of fluctuations, such as fluctuations in the ionizing background, are not addressed in this paper, even though they are probably important in assessing the reliability of the forest constraints on FDM or warm dark matter (see e.g. \cite{Hui:2016ltb}).

\subsection{The 1D Flux Power Spectrum - Methodology and Convergence
  Tests}
\label{convergencetests}
The main observationally relevant quantity in the context of the Lyman-alpha forest is the one-dimensional (1D) flux power spectrum. The transmitted flux $f$ is related to the optical depth $\tau$ by $f = e^{-\tau}$ (ignoring the continuum). The optical $\tau$ is strictly speaking an integral along the line of sight of the neutral hydrogen density field. Here, we employ a simple approximation (sometimes known as the local Gunn-Peterson approximation):
$\tau = A (1+\tilde{\delta})^2$ where $A$ is a constant, and $\tilde{\delta}$ is a smoothed version of the overdensity $\delta$ i.e. in Fourier space:
\begin{equation}
\tilde{\delta}(k) = \exp\left[-\left( \frac{k}{k_f} \right)^2\right]\delta(k)
\end{equation}
where the smoothing is motivated by considerations of the effects of baryon pressure
and the filtering scale is chosen to be $k_f = 40$ h/Mpc comoving \cite{Gnedin:1997td}.
\footnote{A number of quantities (such as the intergalactic medium temperature, which affects $k_f$, and the equation of state \cite{Hui:1997dp}) are implicitly kept fixed in our simple model for $\tau$. Varying them within observational bounds is not expected to significantly change our conclusions below.}
The constant $A$ is determined by requiring that the mean flux $\langle f \rangle$ agree with the observed value at the redshifts of interest \cite{Viel:2013apy}.
The three-dimensional (3D) flux power spectrum $P_{\rm 3D}^f$ is computed in the usual way: by Fourier transforming the field $f$ in the simulation box.
The 1D flux power spectrum $P^f$ is obtained by integrating $P_{\rm 3D}^f$:
\begin{equation}
\label{P1d3d}
P^f (k) = \int\limits_{k}^{\infty} \frac{k'\;\md k'}{2\pi} P_{\rm 3D}^f (k')\,.
\end{equation}

\begin{figure*}[htpb]
\vspace*{-0.3cm}
\begin{tabular}{c}
\includegraphics[width=0.5\textwidth]{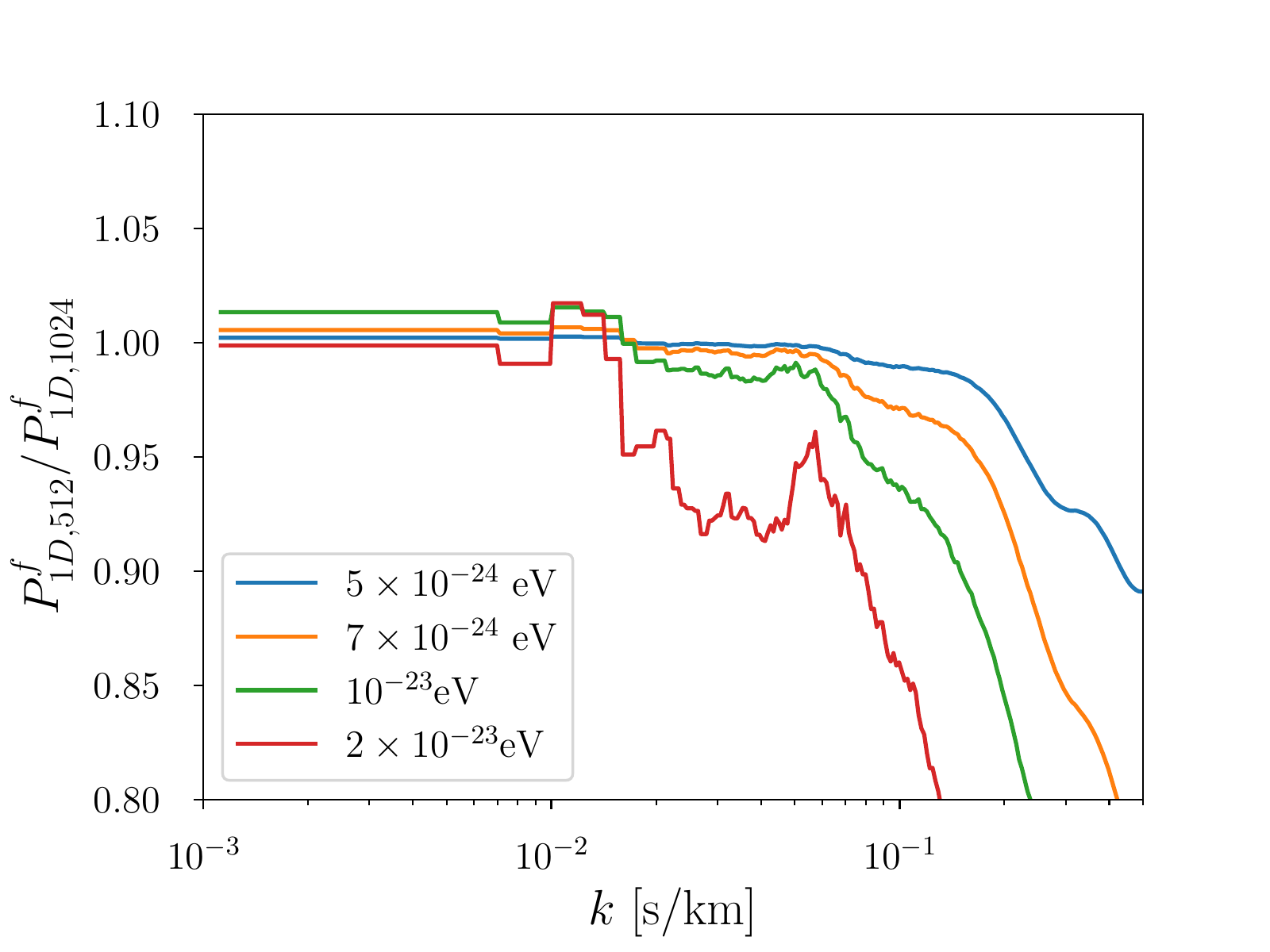} 
\includegraphics[width=0.5\textwidth]{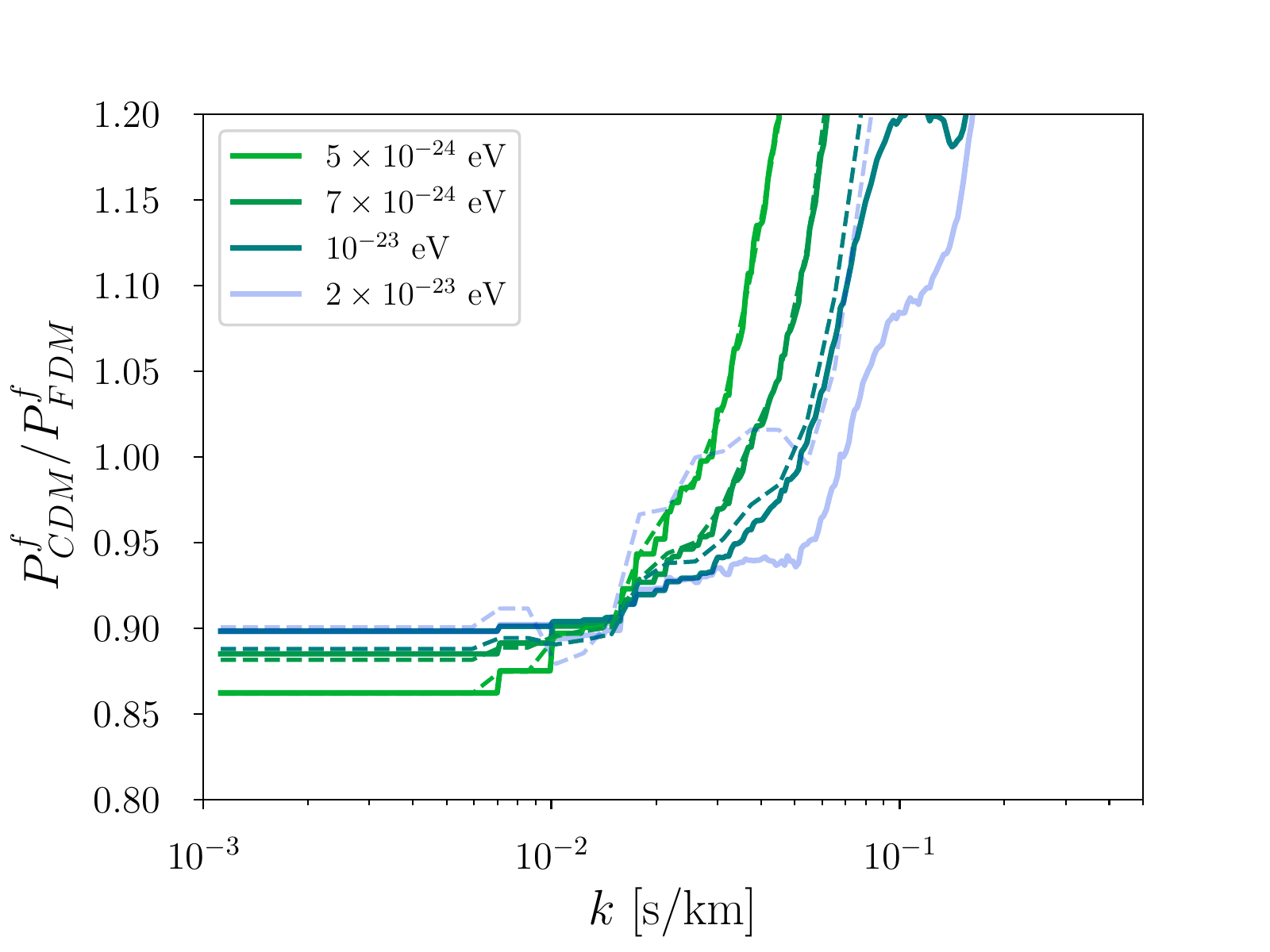} 
\end{tabular}
\caption{Convergence test of 1D flux power spectrum at $z=5$ in a comoving box of $10$ Mpc/h. \textit{Left} -- Ratio of the 1D flux spectrum $P^f$ at $z=5$ from Schr\"odinger-Poisson
  simulations with a $512^3$ grid to $P^f$ from simulations with a $1024^3$ grid,
  for various FDM dynamics masses.
  \textit{Right} -- The 1D flux spectrum $P^f$ with CDM
  dynamics divided by $P^f$ with FDM dynamics, for different
  resolutions and various FDM
  masses. Solid and dashed lines represent the resolution of a
  $1024^3$ grid and a $512^3$ grid respectively.
}
\label{ConvergenceTest}
\end{figure*}

We carry out a number of convergence tests to gauge the accuracy
of our computation of $P^f(k)$ from the (Schr\"odinger-Poisson)
simulations. We focus on results at $z=5$, where most of the FDM
constraints come from \cite{Irsic:2017yje}. 
The left panel of Fig. \ref{ConvergenceTest} shows the ratio of $P^f(k)$ from two simulations
with different resolution: one with a $512^3$ grid and the other with
a $1024^3$ grid, both in a box of size $10$ Mpc/h comoving.
Each line represents the ratio $P^f_{512} / P^f_{1024}$ with a
different FDM mass.\footnote{In the test cases in this sub-section
  (\ref{convergencetests}), the FDM mass refers to
  the FDM mass used in evolving the Schr\"odinger-Poisson system
  i.e. the dynamics mass. The initial conditions were chosen to be
  identical i.e. the initial power spectrum corresponds to that of an
  FDM mass of $10^{-22}$ eV.  
  By doing this, we perform tests more stringently with more initial small scale fluctuations in the computational box. 
  Otherwise if initial conditions consistent with small dynamical FDM mass are used, the initial small scale power will be more suppressed.
 We expect the small scale overdensity to stay longer in the stage of linear growth, and the results to be better than our test simulations.
  }
At low $k$'s, the two different resolutions give fairly close results,
suggesting convergence. Even at $k \sim 0.1$ s/km (the highest $k$
accessible in the relevant data \cite{Irsic:2017yje}), we see that
as long as the FDM mass is below $\sim 2 \times 10^{-23}$ eV,
the $1024^3$ grid simulation (in a $10$ Mpc/h box i.e. a resolution of about 
$0.01$ Mpc/h comoving) should give results accurate at the
$5 \%$ level or better.

Since our interest is mainly in exploring the difference in
implications between FDM and CDM dynamics, we show in
the right panel of Fig. \ref{ConvergenceTest} another version of the resolution convergence
test. Plotted in the figure is the ratio 
$P^f_{CDM}$ (from N-body simulations i.e. CDM dynamics) to
$P^f_{FDM}$ (from Schr\"odinger-Poisson simulations i.e. FDM
dynamics), starting from the same initial condition in a $10$ Mpc/h
box. The solid and dashed lines represent
the results from simulations with a
$1024^3$ grid and a $512^3$ grid respectively.
It appears $P^f_{CDM}/P^f_{FDM}$ is accurate at the few percent
level, for $k \lsim 0.1$ s/km, as long as the FDM mass is less than $\sim 2 \times 10^{-23}$ eV
and the resolution is about $0.01$ Mpc/h comoving (roughly that of 
a $1024^3$ grid in a $10$ Mpc/h box). 

It is worth noting that even though the CDM and FDM dynamics
are identical on large scales, and thus the 
predicted mass power spectrum agrees between the two
(see Sec. \ref{Numericalcomputation}), 
the flux power spectrum $P^f$ does not agree between the two
on large scales (low $k$'s). There are two reasons for this:
(1) the very nonlinear transformation from density to flux mixes scales,
and (2) the 1D flux power spectrum $P^f$ is itself an integral of
the 3D flux power spectrum that receives contributions from
high $k$'s (see Eqn. \ref{P1d3d}).
If the ratio $P^f_{CDM}/P^f_{FDM}$ were scale independent,
then the fact that the ratio differs from unity is not so important, since
there are free parameters, such as $\sigma_8$ and $A$ or the level of
the ionizing background, that can be adjusted in the Lyman-alpha
forest model to produce roughly the same effect (i.e. moving $P^f$ up
and down). From the right panel of Fig. \ref{ConvergenceTest}, we see that this ratio is
dependent on $k$---indeed it rises with increasing $k$---and it is
important to ensure it does not become too large if one were to use
N-body dynamics to approximate FDM dynamics, as is commonly done.

\begin{figure*}[htpb]
\vspace*{-0.3cm}
\begin{tabular}{c}
\includegraphics[width=0.5\textwidth]{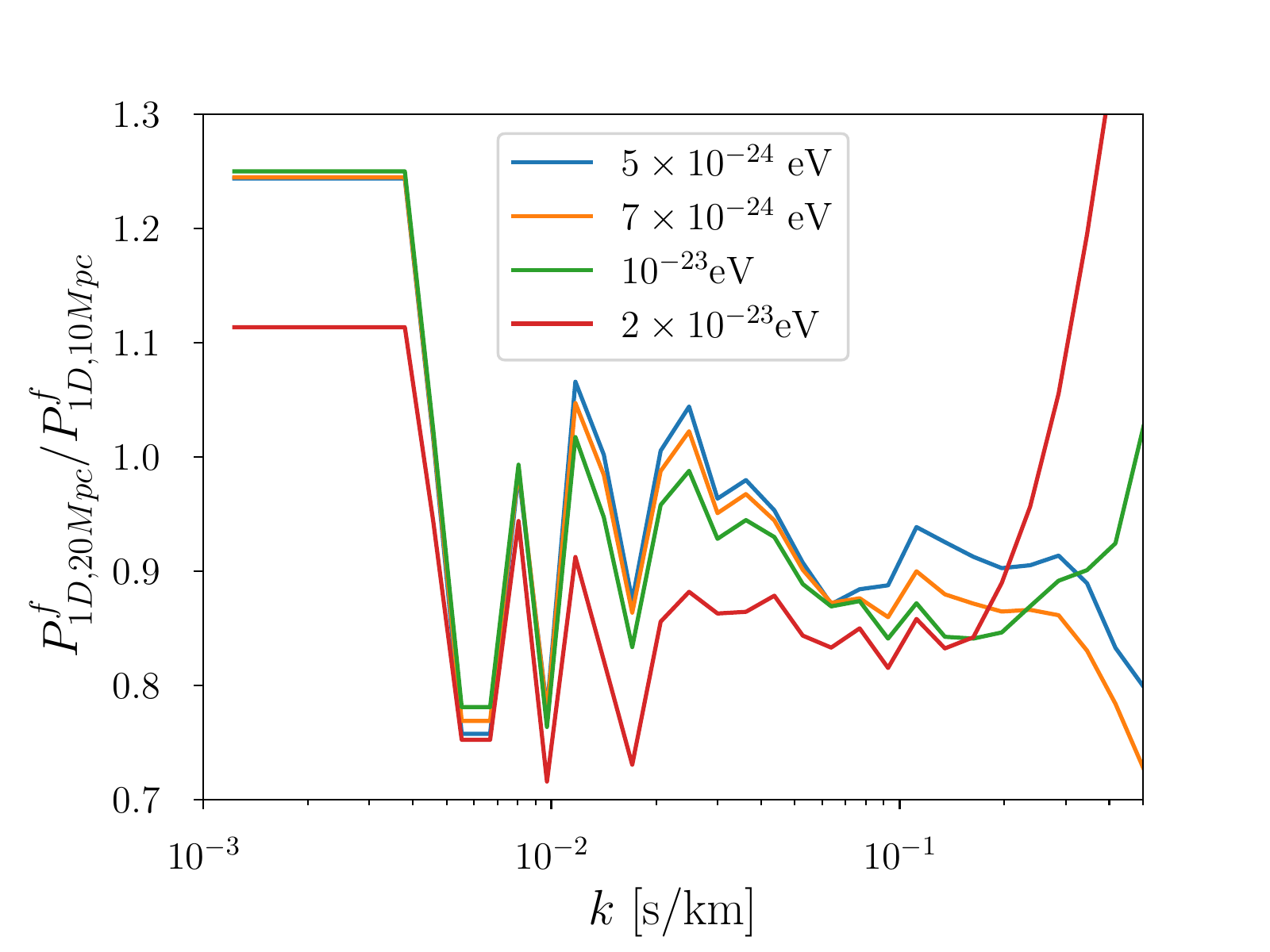} 
\includegraphics[width=0.5\textwidth]{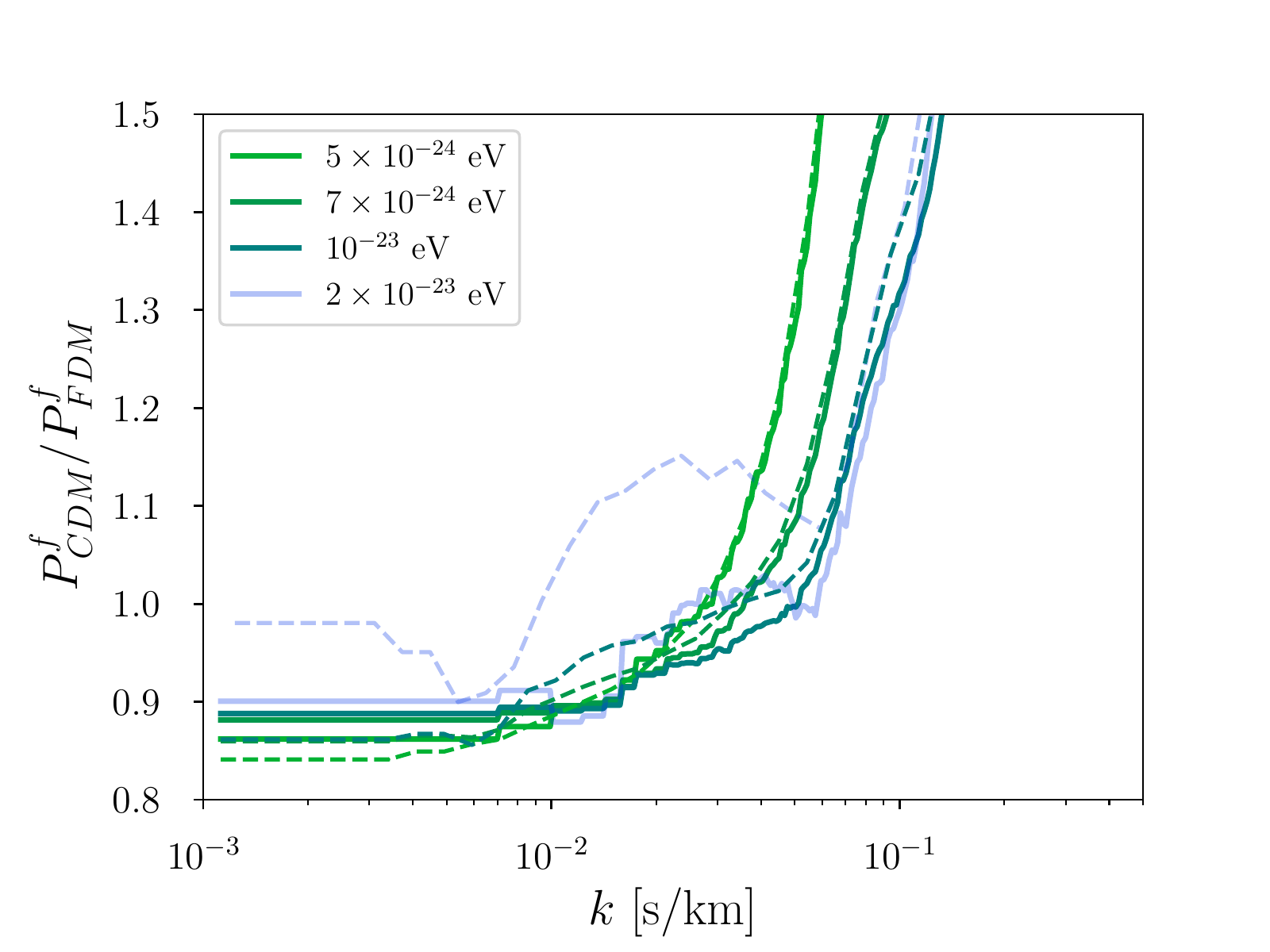} 

\end{tabular}
\caption{Test of finite box size correction of the 1D flux power spectrum at $z=5$. \textit{Left} -- Ratio of $P^f$ from
simulations with a box size of $10$~Mpc/h to $P^f$ from
simulations with a size of $20$~Mpc/h, for various FDM masses, keeping
the resolution fixed.
\textit{Right} -- The 1D flux spectrum $P^f$ with CDM
  dynamics divided by $P^f$ with FDM dynamics, for different box sizes
  and various FDM
  masses. Solid and dashed lines represent
a box size of $10$~Mpc/h and $20$ Mpc/h respectively.
}
\label{BoxSizeTest}
\end{figure*}

The effect of a finite box size is shown in Fig. \ref{BoxSizeTest}.
On the left panel, we compare the flux power spectrum from simulations with different
box sizes: $P^f_{10Mpc}$ from a $10$~Mpc/h box and $P^f_{20Mpc}$ from
a $20$~Mpc/h box. The resolution is kept fixed: a $512^3$ grid is
employed for the $10$ Mpc/h
simulations while a $1024^3$ grid is used for the $20$ Mpc/h
simulations. We see that $P^f_{20Mpc}$ and $P^f_{10Mpc}$ differ by
more than $10 \%$ across a wide range of scales, suggesting
non-negligible box size effects; in other words, a $10$ Mpc/h box is
too small to reliably predict the flux power spectrum.
It might thus seem impossible to satisfy the twin demands of
high resolution and large box size, given that the largest grid we can
practically simulate is about $1024^3$. 

Fortunately, for the present purpose, what we care about is the
difference between CDM and FDM dynamics, exemplified in the ratio
$P^f_{CDM}/P^f_{FDM}$. This is shown on the right panel of Fig. \ref{BoxSizeTest}.
We see that the different box sizes (solid lines for a $10$ Mpc/h box,
dashed lines for a $20$ Mpc/h box) give fairly similar results for this
ratio, as long as the FDM mass is small enough i.e. the de Broglie
scale is resolved -- the case of $2 \times 10^{-23}$ eV is one where
the de Broglie scale is {\it not} resolved, as we already know from
the resolution tests discussed earlier (i.e. a grid of $1024^3$ is
inadequate for a box as big as $20$ Mpc/h).

Henceforth, for a given FDM mass, we choose a box size small
enough so that a $1024^3$-grid simulation can resolve the de Broglie
wavelength. This in general means the flux power spectrum $P^f$ has
large finite-box corrections. However, since such corrections affect the
CDM-dynamics and FDM-dynamics simulations in similar ways, the ratio
$P^f_{CDM} / P^f_{FDM}$ is actually reliable and sufficiently accurate for our
purpose.

\subsection{The 1D Flux Power Spectrum -- CDM versus FDM Dynamics}
\label{CDMvsFDM}

Adopting the methodology laid out above, we carry out a series of
simulations of different FDM masses where the de Broglie scale is resolved (by adjusting the box
size and always using a $1024^3$ grid). Here, the initial power spectrum
has the same FDM mass as that used in the dynamics.
Our aim is to determine the scales at $z=5$ where $P^f_{CDM}$ from
N-body simulations can be used to approximate $P^f_{FDM}$ from FDM
dynamics, as a function of FDM mass.
We run simulations with FDM masses ranging from $10^{-23}$~eV to $10^{-22}$~eV and adjust the box size by requiring that the large scale matter power spectrum follows the linear evolution.
By doing this, we can make sure the de Broglie wavelength is resolved, otherwise the evolution of large scale mass power spectrum will deviate from the linear evolution as demonstrated in Sec. \ref{TestOfSolver}.

\begin{figure}[htpb]
\vspace*{-0.3cm}
\begin{tabular}{c}
\includegraphics[width=0.5\textwidth]{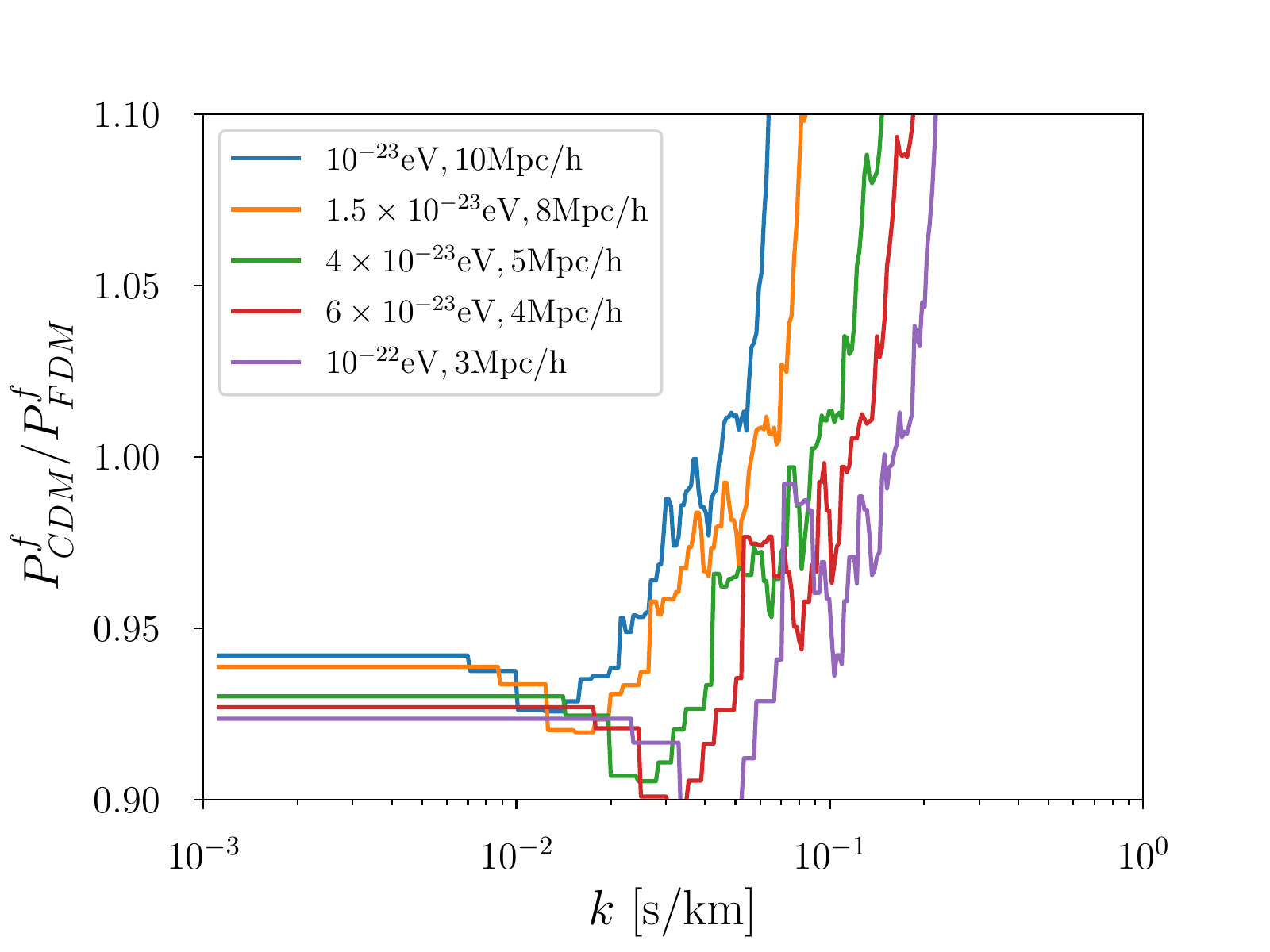} 
\end{tabular}
\caption{The 1D flux spectrum $P^f$ at $z=5$ with CDM
  dynamics divided by $P^f$ with FDM dynamics for realistic simulation with consistent FDM mass in the initial condition and the dynamics. Each line represents a simulation with different FDM mass and box size adjusted to resolve the de Broglie wavelength.}
\label{cdm}
\end{figure}

\begin{figure}[htpb]
\vspace*{-0.3cm}
\begin{tabular}{c}
\includegraphics[width=0.5\textwidth]{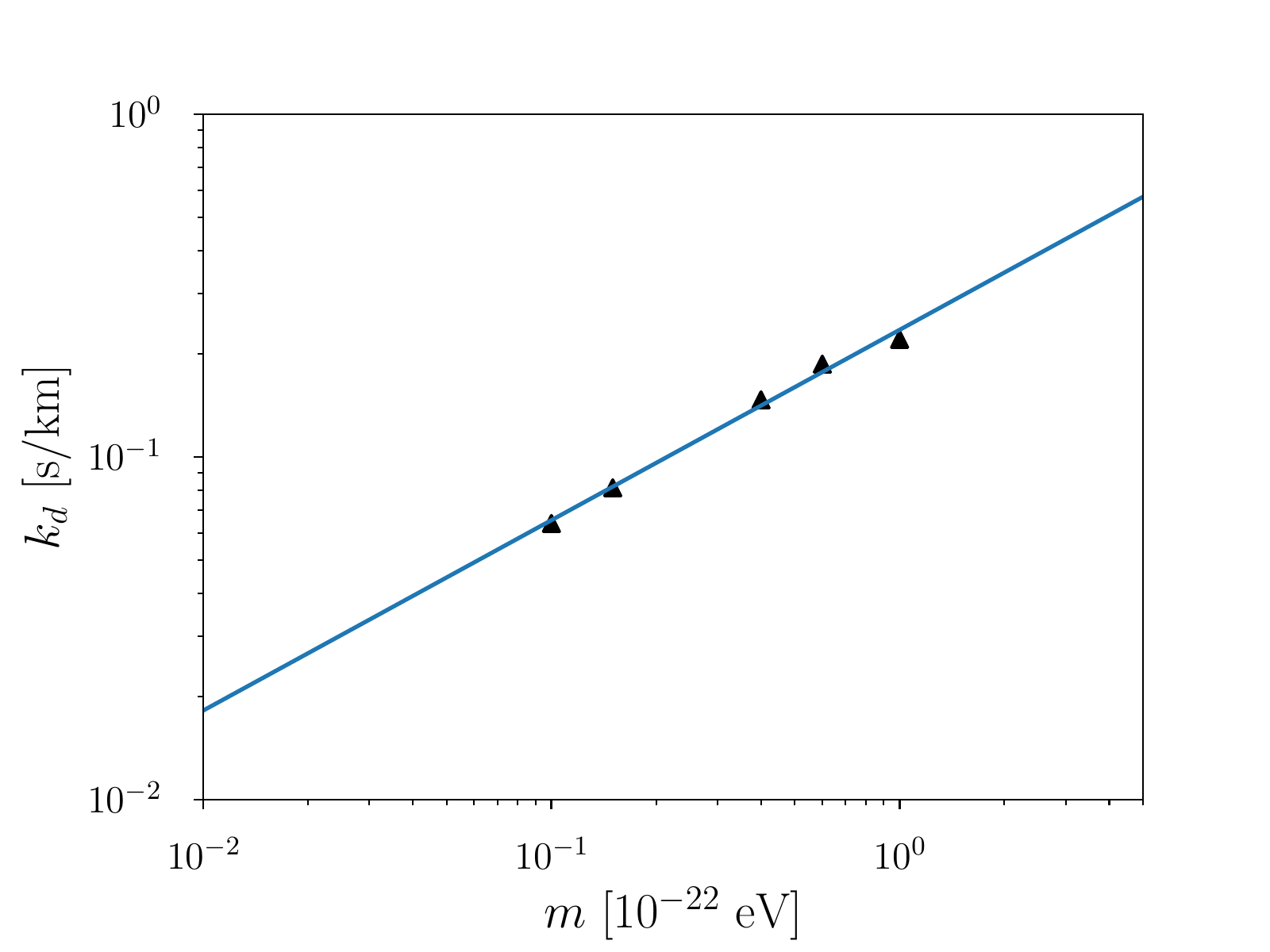} 
\end{tabular}
\caption{Relation between $k_d$ and FDM masses. The best fit line follows $k_{\rm d}= 0.23\; \mathrm{s/km}\left(m/10^{-22}\;\rm eV\right)^{0.56}$.}
\label{ext}
\end{figure}

Fig.\ref{cdm} shows the ratio of $P^f_{CDM}/P^f_{FDM}$ with each line representing a different FDM mass.
The ratio follows the same trend as the test simulations:
on large scales, $P^f$ from CDM and FDM dynamics are close to each other (differ less than $10\%$) and on small scales, $P^f_{CDM}$ overestimates $P^f_{FDM}$.
The scale $k$ at $z=5$ where $P^f_{CDM}$ starts to overestimate $P^f_{FDM}$ moves to smaller scales with increasing FDM mass.
\footnote{The fact that $P^f_{CMD}$ and $P^f_{FDM}$ have smaller deviations on large scales than the test simulations is consistent with what we discussed before. Fewer initial small scale fluctuations leave the system longer in the linear growth stage and yield better agreement.}
To quantify this effect, we define $k_d$ to be the smallest $k$ where $P^f_{CDM}$ exceeds $P^f_{FDM}$ by $10\%$.
Fig.\ref{ext} shows the the relation between the scale $k_d$ as a function of FDM mass.
Black triangles are the numerical data which closely follow a power law relation given by the blue line
\begin{equation}
	k_d \approx 0.23\; \mathrm{s/km}\left(\frac{m}{10^{-22}\;\rm eV}\right)^{0.56}\,.
\end{equation}

For general FDM masses, we can read from the power law relation the scales where $P^f_{CDM}$ from N-body simulations is a good approximation of $P^f_{FDM}$ at $z=5$.
For example, for FDM masses larger than $2\times 10^{-23}$~eV,
$P^f_{CDM}$ differs from $P^f_{FDM}$ by less than $10\%$ on our current observational scale of Lyman-alpha forest $k\lesssim 0.1$~s/km \citep{Croft:1997jf,Hui:1998hq,Croft:2000hs,McDonald:2004xn,Viel:2006yh}.
Therefore, one can avoid running expensive large scale FDM simulations
where the small de Broglie wavelength has to be resolved, but perform
relatively cheap N-body simulations to obtain good approximation of
$P^f_{FDM}$ and compare with observations. It is worth emphasizing
though that for other applications, such as investigating galactic
sub-structure, correctly simulating the FDM dynamics is crucial
(see discussion in Sec. \ref{TestOfSolver}).

\section{Discussion}
\label{discuss}

In this paper, we present numerical and perturbative
computations of the FDM model.

Perturbative calculations are carried out for both the fluid and wave 
formulations. For fluid perturbation theory, this is done up to third
order, focusing in particular on the one-loop mass power spectrum
(Appendix \ref{FPT}). Wave perturbation theory (Appendix
\ref{WPT}) has a more limited regime of validity compared to
fluid perturbation theory; namely the former requires
the physical length scale of interest be smaller than the de Broglie
wavelength (Eqn. (\ref{WPTrequire})) i.e. simply requiring small
density fluctuation $\delta$ is not sufficient.
Wave perturbation theory thus breaks down rather early in
cosmic history, even
when fluid perturbation theory is safely in the linear regime.
The wave description might still be useful for analytic understanding
in the highly nonlinear regime, especially in the context of complex
inteference structures--- a subject we hope to return in the
future.

On the numerical side, we perform cosmological simulations of FDM using both a Schr\"odinger-Poisson solver and a fluid solver.
The fluid solver integrates the Madelung formulation of the \sch equation and is able to produce the correct structures on large scales.
However, small scale oscillatory features (from wave interference) observed in the
Schr\"odinger-Poisson simulations are not reproduced by the fluid simulations.
In particular, the fluid solver fails to simulate the correct dynamics
when the local density vanishes due to the destructive interference of
waves, as demonstrated in Appendix \ref{testCollision}.
This is understandable because the quantum pressure in the fluid
formulation becomes ill-defined where the fluid density vanishes
(Eqn. (\ref{qpres})).
The failure of the fluid solver to give the correct quantum pressure
in regions of complex interference results in artificially enhanced
gravitational collapse, and over-predicts the mass power spectrum on
small scales.
Whether Lagrangian fluid codes like the SPH method
\citep{Nori:2018hud} suffer from the same problem is not clear, and we
recommend similar tests to those in Appendix
\ref{testCollision} be carried out.

Our Schr\"odinger-Poisson solver \texttt{SPoS}, on the other hand, is able to
simulate the correct dynamics on small scales. We obtain good agreement with third order perturbative calculations up to the mildly nonlinear scale.
However, \texttt{SPoS} is demanding in resolution
in that it requires the (typically small) de Broglie wavelength to be
resolved, even if one is primarily interested in structures on
much larger scales.

As an application, we compare the 1D Lyman-alpha flux power spectrum $P^f_{CDM}$  from N-body simulations with that computed from FDM dynamics,  $P^f_{FDM}$, using the same FDM initial conditions and determine the scales where $P^f_{CDM}$ is a good approximation to $P^f_{FDM}$.
Though $P^f$ from both CDM and FDM dynamics suffer from non-negligible finite box corrections, we find the ratio $P^f_{CDM}/P^f_{FDM}$ is reliable and independent of box size as long as the de Broglie wavelength is resolved.
The ratio $P^f_{CDM}/P^f_{FDM}$ is close to unity on large scales with deviation less than $10\%$ , but it eventually grows much larger on small scales.
We quantify the scales of agreement between the two flux power spectra
by measuring the smallest $k_d$ (largest length scale) where $P^f_{CDM}/P^f_{FDM}>1.1$, and our results suggest $k_d$ increases with FDM mass as $k_d\propto m^{0.56}$.
For FDM masses larger than $2\times 10^{-23}$~eV, $k_d>0.1$~s/km,
therefore $P^f_{CDM}$ is a good approximation to $P^f_{FDM}$  on our current observational scales of the Lyman-alpha forest $\lesssim 0.1$~s/km \citep{Croft:1997jf,Hui:1998hq,Croft:2000hs,McDonald:2004xn,Viel:2006yh}.

There have been a number of efforts to constrain FDM properties using
Lyman-alpha forest observations, including \citep{Irsic:2017yje} using
XQ-100 and HIRES/MIKE data, and \citep{Armengaud:2017nkf} using SDSS
data. Both groups employed N-body simulations with initial conditions
modified by FDM and use $P^f_{CDM}$ to approximate $P^f_{FDM}$.  Our
results confirm the validity of their approximation.
However, a number of astrophysical effects likely play a role in
interpreting the Lyman-alpha forest data, such as the spatial
fluctuations of the ionizing background and temperature at high
redshifts, and galactic winds at low redshifts. It is unclear if
current attempts to model these effects faithfully capture 
the range of possibilities (e.g. see discussion in \cite{Hui:2016ltb}).
More work remains to be done in this direction.

\appendix
\section{FDM with a Fluid Solver}
\label{FDMfluidSolver}

Our fluid solver for FDM dynamics is implemented in the ENZO code \citep{Bryan:2013hfa} by modifying the built-in 3D Zeus code \citep{Stone:1992wj,Stone:1992mq} to incorporate the quantum pressure.
The built-in Zeus method is an explicit method that solves the fluid transport on a Cartesian grid using an operator-split scheme.
The whole method contains two steps: \textit{source} and \textit{transport}.
In the source step, only velocities are updated due to the gravitational acceleration (and, in the non-FDM implementation, the pressure gradient terms).
Artificial viscosity, which is used in the original fluid solver to smooth shock fronts, is not needed for FDM dynamics, as the quantum pressure strongly disfavors sharp gradients. 
In the transport step, the mass and momentum fluxes are constructed and used to update the density and velocity through the conservative form of the fluid equations. 

The quantum pressure term for FDM is included in the source step as an extra term for the acceleration. 
The quantum pressure
$p=-\frac{1}{2}\nabla^2\log\rho-\frac{1}{4}\left(\nabla\log\rho\right)^2$ is
computed with a 4th-order accurate finite difference method.
Apart from the existing time step constraints already in ENZO, the time step must also respects the CFL condition for parabolic systems: $\Delta t \leq (\Delta x)^2ma^2/(6\hbar)$.

Below we describe a number of one-dimensional tests. 
All the tests are performed
without gravity and cosmological expansion (i.e. $a=1$).

One obvious worry is that the quantum pressure becomes
ill-defined when $\rho$ vanishes. Of course, numerically, the density
$\rho$ never vanishes exactly (nor does it go negative in all
cases we check); the question is whether the large quantum pressure
associated with regions of low density leads to unacceptable
inaccuracies (see Test 3 below). 

\subsection{A Test with a Gaussian Density Profile}
\label{testGaussian}

The evolution of a Gaussian wave packet admits a simple analytical solution for the 1D free Schr\"{o}dinger equation without an external potential:
\begin{equation}
	\psi(t,x) = \sqrt{\frac{1}{\alpha+i\frac{\hbar t}{m}}}\exp\left(-\frac{x^2}{2(\alpha+i\frac{\hbar t}{m})}\right)\,,
\end{equation}
where $\alpha$ is a constant and can be set to be real for simplicity.
The density and velocity follow from the wave function as
\begin{eqnarray}
	\rho &=& \frac{1}{\sqrt{\alpha^2+\frac{\hbar^2 t^2}{m^2}}}\exp\left(-\frac{\alpha x^2}{\alpha^2+\frac{\hbar^2 t^2}{m^2}}\right)\,,\\
	v &=& \frac{x}{\alpha^2+\frac{\hbar^2 t^2}{m^2}}\frac{\hbar^2 t}{m^2}\,.
\end{eqnarray}
\begin{figure}[htpb]
\vspace*{-0.3cm}
\begin{tabular}{c}
\includegraphics[width=0.5\textwidth]{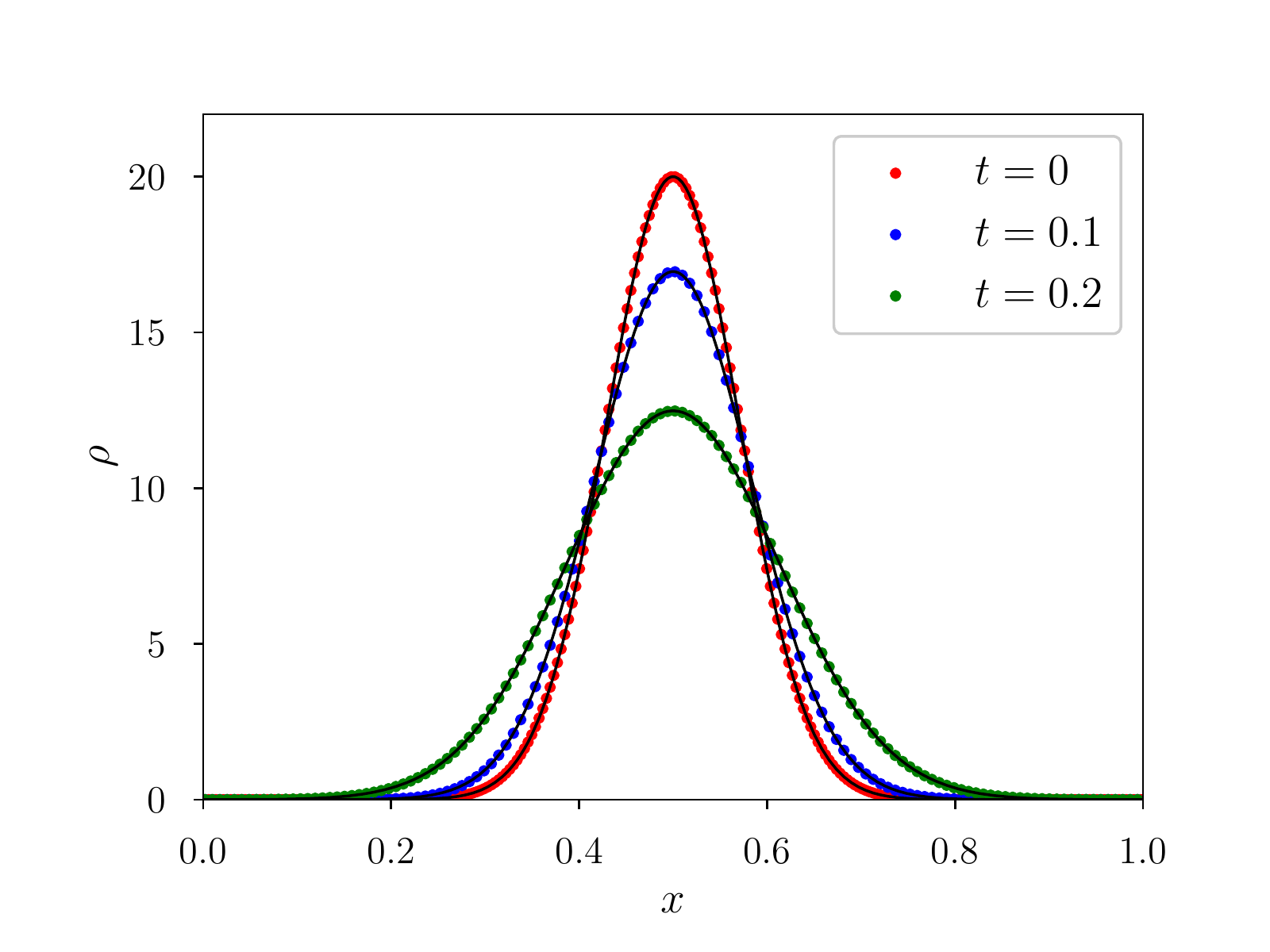} 
\end{tabular}
\caption{Comparison of analytical and numerical results for the
  density evolution of a Gaussian wave packet. Solid lines represent the
  exact solutions while dots represent the numerical results from a
  fluid simulation. The
  times shown are from top to bottom $t=0, 0.1, 0.2$ in code units (see text).}
\label{gaussian_rho}
\end{figure}

\begin{figure}[htpb]
\vspace*{-0.3cm}
\begin{tabular}{c}
\includegraphics[width=0.5\textwidth]{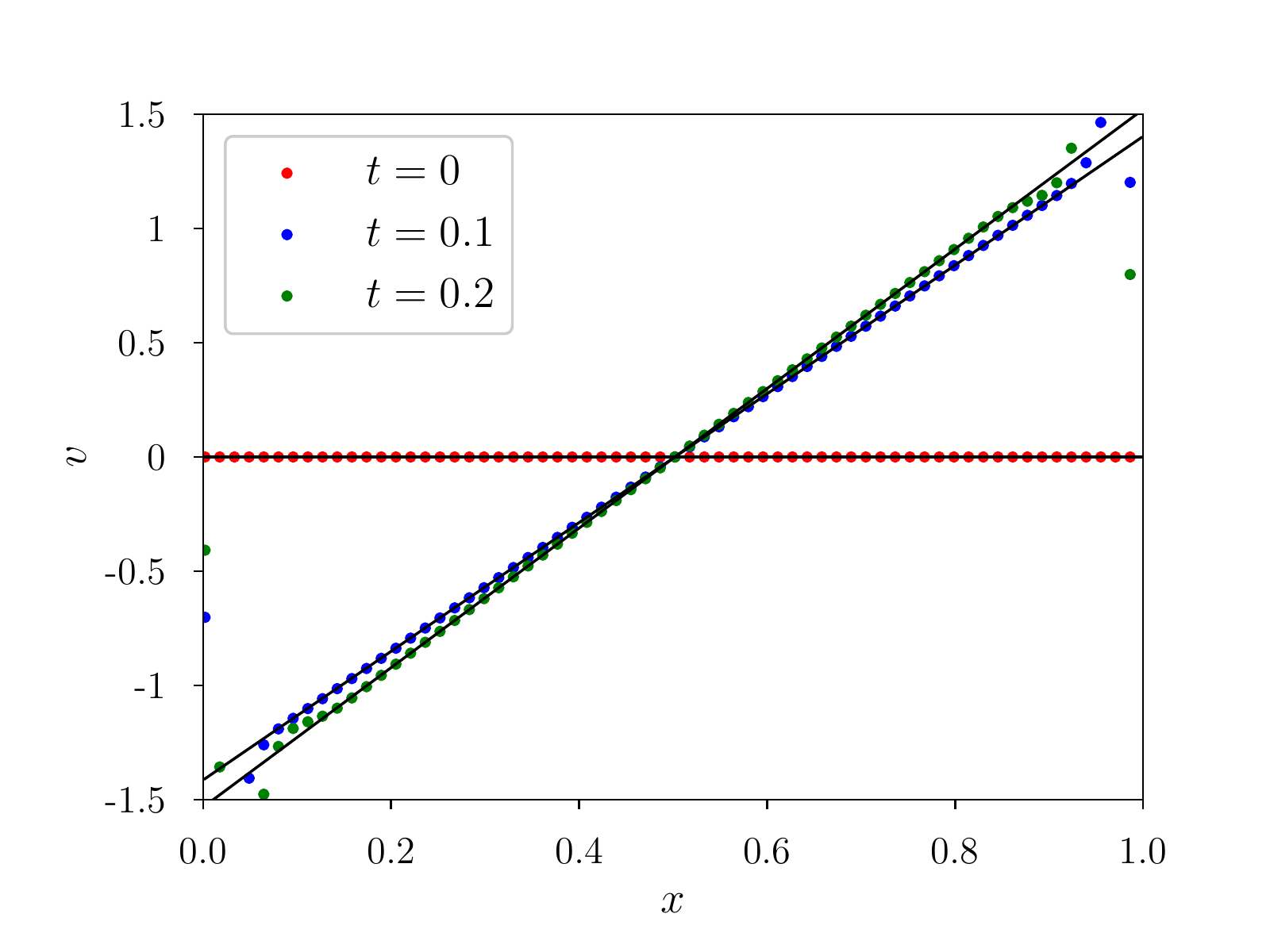} 
\end{tabular}
\caption{Comparison of analytical and numerical results for the
  velocity evolution of a Gaussian wave packet. Solid lines represent
  the exact solutions and dots represent the numerical results from a
  fluid simulation.
The flat line corresponds to $t=0$.
Of the tilted lines, the steeper one
corresponds to $t=0.2$ while the other is at $t=0.1$.}
\label{gaussian_v}
\end{figure}

Fig.(\ref{gaussian_rho}) and (\ref{gaussian_v}) show 
the comparison between analytical solutions and numerical results
(using the modified ENZO fluid code).
We start at $t=0$ where $v=0$ in the simulation box. 
The parameters are chosen to be $\alpha = 1/100$ and $\hbar/m=0.0626$ in the code's internal unit system. 
Solid lines represent analytical solutions and colored dots are numerical results at different time.
We can see the numerical values follow closely to the analytical solutions. 
The discrepancies of velocity near the boundaries are due to the periodic boundary condition we use for our finite box size.

\subsection{A Test with a Density Jump}
\label{testDensityJump}

One might worry that the fluid solver could miss non-trivial wave interference
effects. Here we demonstrate that the quantum pressure term, when
properly incorporated in a fluid code, is capable of reproducing
intricate interference patterns expected in certain solutions.
In particular, let us focus on a self-similar solution of the free
Schr\"odinger equation \cite{Hui:2016ltb}:
\begin{eqnarray}
	\psi(t,x)&=&\frac{\mathcal{A}}{2}+\mathcal{B}-\frac{\mathcal{A}}{2}(1\mp i)\mathcal{C}\left(x\sqrt{\frac{m}{\pi \hbar|t|}}\right)\nonumber\\
	&&-\frac{\mathcal{A}}{2}(1\pm i)\mathcal{S}\left(x\sqrt{\frac{m}{\pi \hbar|t|}}\right)\,,
\end{eqnarray}
where the upper/lower sign is for a positive/negative $t$ and $\mathcal{A}$, $\mathcal{B}$ are constants.
The functions $\mathcal{C}$ and $\mathcal{S}$ are the Fresnel integrals
\begin{eqnarray}
	\mathcal{C}(\theta) &=& \int_0^\theta \cos(\frac{\pi\xi^2}{2})\md \xi\,,\\
	\mathcal{S}(\theta) &=& \int_0^\theta \sin(\frac{\pi\xi^2}{2})\md \xi\,.
\end{eqnarray}

The density profile $\rho = |\psi|^2$ given by this solution asymptotes to $(\mathcal{A}+\mathcal{B})^2$ as $x\rightarrow-\infty$ and $\mathcal{B}^2$ as $x\rightarrow\infty$.
At $t = 0$, the density profile becomes a step function with constant value $(\mathcal{A}+\mathcal{B})^2$ and $\mathcal{B}^2$ for $x<0$ and $x>0$.
There is a singularity at $x=0$, where a sharp density jump exists, while the velocity remains regular and is zero everywhere.

\begin{figure}[htpb]
\vspace*{-0.3cm}
\begin{tabular}{c}
\includegraphics[width=0.5\textwidth]{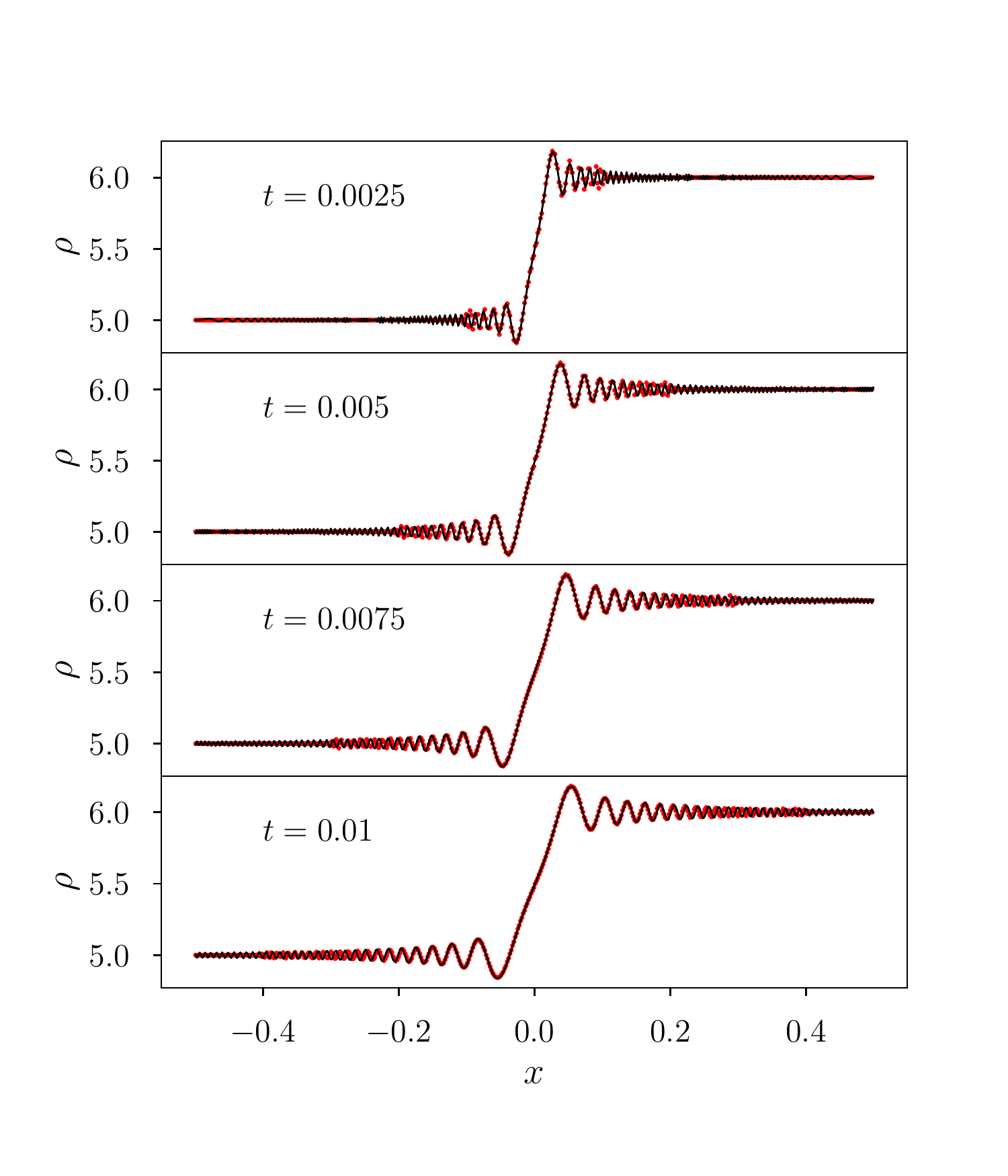} 
\end{tabular}
\caption{Snapshots of density evolution from the density jump. Solid (black)
  lines represent the exact solutions. The (red) dots represent
  numerical results from a fluid simulation.}
\label{fresnel_rho}
\end{figure}

The analytical evolution of the solution can be understood through the Fourier transform of the wave function.
The initial step function of density profile contains the waves of all wavelengths with the dispersion relation derived by the \sch equation $\omega\propto k^2$.
Those waves do not couple with each other and evolve independently.
Waves with short wavelength have larger speed but smaller amplitude.
They propagate faster and leave the longer wave modes behind, leading to an intricate interference pattern.

Fig.(\ref{fresnel_rho}) presents snapshots of the density
evolution starting from a step-function at $t=0$.
The constants are chosen such that $\hbar/m=0.0626$, $\mathcal{B}=\sqrt{6}$ and $\mathcal{A}=\sqrt{5}-\sqrt{6}$, so the initial density has value $6$ and $5$ on the right and left side of the origin.
The initial density jump at the origin launches waves which propagate away from it.
At the front of the wave propagation, the wavelength and amplitude of oscillations are getting smaller.
The numerical results agree well with the analytical results,
as long as the relevant oscillations are resolved. 

\subsection{A Failed Test: Collision of Two Wave Packets}
\label{testCollision}

A particular demanding test of the fluid formulation is to investigate
a case where the density vanishes in isolated places. The quantum
pressure $p$ becomes ill-defined and the fluid solver might not yield
the correct evolution. On the other hand, one might hope that
inaccuracies at isolated points do not affect significantly the
overall global evolution. This turns out not to be the case in this example.

The free \sch equation admits a solution where a single Gaussian wave
packet moves with momentum $\hbar k$
\begin{equation}
\psi(x,t) = \sqrt{\frac{\alpha}{\alpha + i\frac{\hbar}{m}t}}\exp\left[
-\frac{(x+x_0-ik\alpha)^2}{2(\alpha+i\frac{\hbar}{m}t)}
\right]
\exp\left(-\frac{\alpha k^2}{2}\right)\,.
\end{equation}
The linearity of the \sch equation allows two meeting wave packets to evolve on their own and pass
through each other, for instance:
\begin{eqnarray}
&& \psi(x,t) = \sqrt{\frac{\alpha}{\alpha + i\frac{\hbar}{m}t}}\exp\left[
-\frac{(x+x_0-ik\alpha)^2}{2(\alpha+i\frac{\hbar}{m}t)}
\right]
\exp\left(-\frac{\alpha k^2}{2}\right)\nonumber\\
&& \quad + \sqrt{\frac{\alpha}{\alpha + i\frac{\hbar}{m}t}}\exp\left[
-\frac{(x-x_0+ik\alpha)^2}{2(\alpha+i\frac{\hbar}{m}t)}
\right]
\exp\left(-\frac{\alpha k^2}{2}\right).
\end{eqnarray}
The density and velocity can be deduced in this situation
\begin{equation}
\label{rhovdef}
\rho = m|\psi|^2\,,\; v = \frac{\hbar}{2i m}\frac{\psi^*\nabla\psi-\psi\nabla\psi^*}{\rho}\,.
\end{equation}

\begin{figure}[htpb]
\vspace*{-0.3cm}
\begin{tabular}{c}
\includegraphics[width=0.5\textwidth]{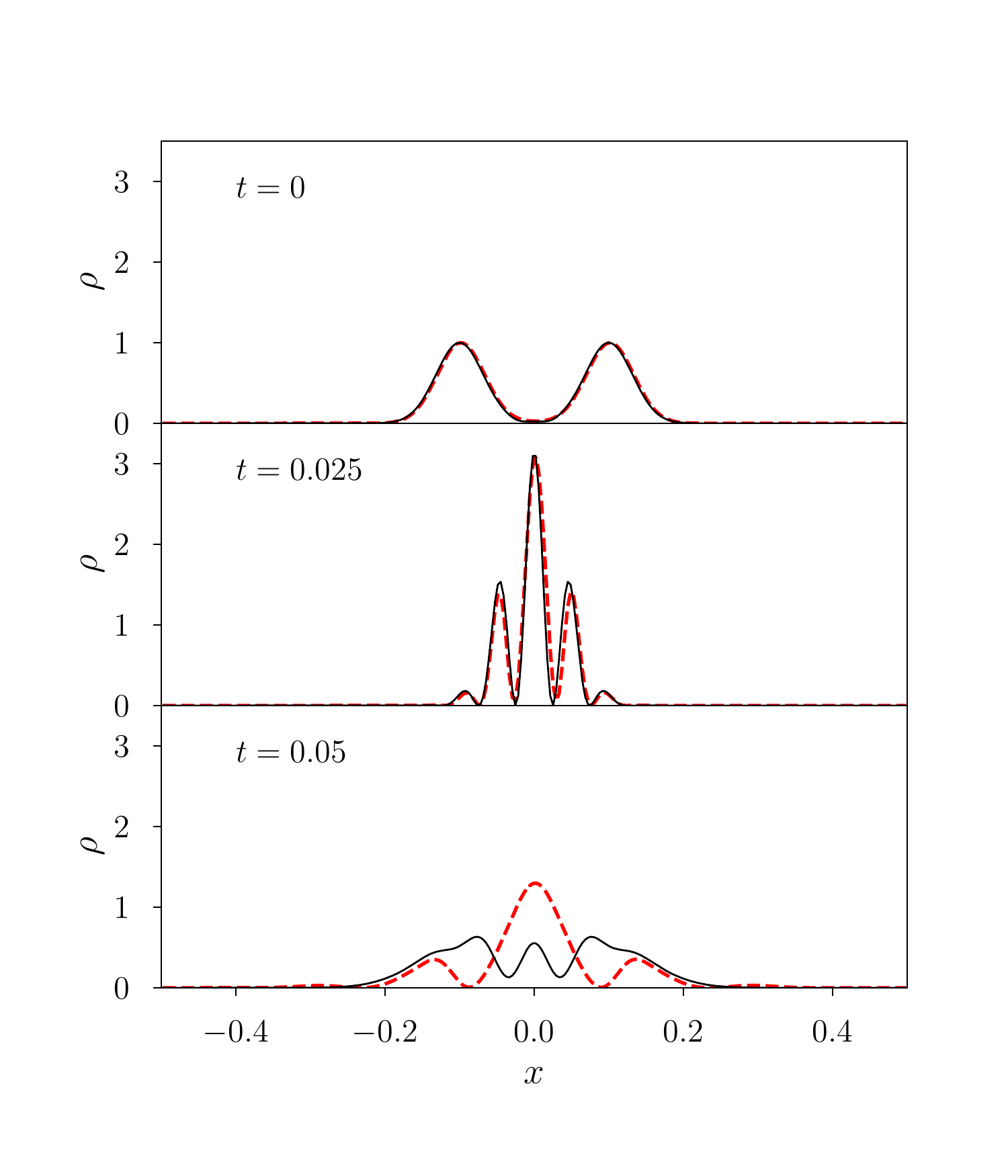} 
\end{tabular}
\caption{Snapshots of density evolution for the collision of two
  Gaussian wave packets. Solid (black) lines represent the exact
  solution. Dashed (red) lines represent the numerical results from
a fluid simulation. A wave simulation, on the other hand, reproduces
the exact solution very closely; the wave simulation results
essentially lie on top of the solid (black) lines.
}
\label{collision}
\end{figure}

Fig.\ref{collision} shows the comparison between the analytical and
numerical solutions (using the fluid code).
In this numerical test, we take $\alpha = 1/500$, $k = 20\pi$ and $x_0 = 0.1$.
As the two wave packets move towards each other, at some point the
interference between two waves results in vanishing density in certain locations.
In the snapshots, the numerical solution follows the analytical evolution very well until the vanishing density region from interference first appears (middle panel).
After that, the numerical solution deviates from the analytical
solution.

The local densities at the position of destructive interference in the fluid code are very small, but remain positive and never reach zero.
The error of density in those regions can not be smaller than machine error which effectively acts as a density floor.
On the other hand, the true quantum pressure (Eqn. \ref{qpres}) is proportional $\log\rho$ and diverges at the region of destructive interference diverges.
However small the error of the local density is, the error of the quantum pressure is always infinite where the true density is zero.
Therefore, the simulated dynamics deviates from the analytical solution after the destructive interference shows up.  We have verified that this result is independent of the spatial or temporal resolution employed.

Where the density is very low in the numerical simulations,
the quantum pressure is large and susceptible to sizable numerical errors.
The velocity suffers from a similar problem. 
One might think evolving the momentum density (as opposed to velocity)
could alleviate this problem. This is partly true: the momentum
density in fact vanishes in these places where the density vanishes.
The issue is that one still needs to advect the momentum density by a
suitable velocity field: the ill-defined nature of the velocity (or
its very large value) in such locations remains a problem.

Destructive interference like that
depicted in Fig.~\ref{collision}
might seem like a rather artificial situation, but it turns out
to be a fairly generic occurrence in the nonlinear regime (see
discussion in Sec.~\ref{Numericalcomputation}).
The fact that the fluid code fails to evolve correctly
past regions of destructive interference cause it to over-predict
the amount of power on small scales (see Fig. \ref{ps23} and \ref{slice_pic}  in 
Sec. \ref{Numericalcomputation}). 
One can circumvent this problem by switching to a wave code,
or by developing some kind of a hybrid code (fluid on large scales,
wave on small scales) which could combine the best features of both
formulations.
We have checked that the wave code successfully evolves past these
regions of vanishing density (a plot of the numerical solution against
the analytical one would not be too instructive, as they lie on top of
each other). 

\section{\label{FPT}Fluid Perturbation Theory}

The fluid formulation of the dynamics takes the form:
\begin{equation}
\partial_\eta \delta + \nabla_i [(1 + \delta) v^i] = 0 \, ,
\end{equation}
\begin{equation}
\label{EulerV}
\partial_\eta v^i + {\partial_\eta a\over a} v^i + v^j \nabla_j v^i = - \nabla_i \Phi +
  {1 \over 2 \bar m^2 a^2} \nabla_i {  \nabla^2 \sqrt{1 + \delta}
  \over \sqrt{1 + \delta} }  \, ,
\end{equation}
\begin{equation}
\label{nablaPoisson}
\nabla^2 \Phi = 4 \pi G a^2 \bar\rho \delta \, ,
\end{equation}
where $\delta$ is the overdensity defined by $\rho \equiv \bar\rho
(1+\delta)$, with $\bar\rho$ being the mean density. Here 
$\partial_\eta$ 
denotes derivative with respect to conformal time $\eta$. The
conformal time $\eta$ is related to proper time $t$ by $a \, d\eta =
dt$, where $a$ is the scale factor.
We introduce the symbol $\bar m \equiv m/\hbar$ to suppress
factors of $\hbar$. 
The last term of Eqn. (\ref{EulerV}) is often referred to as quantum
pressure, though it strictly speaking arises from a stress tensor that
has off-diagonal terms.


Assuming gradient flow, defining
$\Theta \equiv \nabla_i v^i$ and using $\delta(\bf{k})$ to denote the
Fourier transform of $\delta({\bf x})$ i.e.
\begin{eqnarray}
\delta({\bf x}) = \int {d^3 k \over (2\pi)^3} \delta ({\bf k}) e^{i
  {\bf k} \cdot {\bf x}}
\end{eqnarray}
Following standard perturbation theory techniques
(e.g. \cite{Fry:1983cj,Goroff:1986ep,Bernardeau:1992zw,Scoccimarro:1996jy,Bernardeau:2001qr}), 
the fluid equations can be rewritten as
\begin{eqnarray}
\label{MassconvPT}
&& \partial_\eta \delta ({\bf k}) + \Theta ({\bf k}) = 
- \int {d^3 k_1 d^3 k_2 \over
(2\pi)^3} 
\delta_D ({\bf k} - {\bf k}_{12}) \nonumber \\
&& \quad { {\bf k} \cdot {\bf
k}_2 \over k_2^2} 
\delta ({\bf k}_1 )\Theta ({\bf k}_2) \, ,
\end{eqnarray}
\begin{eqnarray}
\label{EulerPT}
&& \partial_\eta \Theta ({\bf k}) + {\partial_\eta a\over a} \Theta ({\bf k})
+ 4\pi G a^2 \bar\rho \delta({\bf k}) 
- {k^4 \over 4 a^2 \bar m^2} \delta({\bf k})
= \nonumber \\ && - \int {d^3 k_1 d^3 k_2 \over
(2\pi)^3} 
\delta_D ({\bf k} - {\bf k}_{12}) \Big[
{1\over 2} k^2 {{\bf k}_1 \cdot {\bf k}_2
\over k_1^2 k_2^2} \Theta ({\bf k}_1) \Theta ({\bf k}_2) \nonumber \\
&& \quad \quad + {k^4 \over 16 a^2 \bar m^2} \left(1 +  {k_1^2 + k_2^2 \over
  k^2} \right) \delta({\bf k}_1) \delta({\bf k}_2) \Big] \nonumber \\
&& + \int {d^3 k_1 d^3 k_2 d^3 k_3 \over
(2\pi)^6} 
\delta_D ({\bf k} - {\bf k}_{123}) \delta({\bf k}_1) \delta({\bf k}_2)
   \delta({\bf k}_3)  {k^4 \over 32 a^2 \bar m^2} \nonumber \\
&& \quad \quad 
\left[1 + {k_1^2 + k_2^2 + k_3^2 \over k^2} + {k_{12}^2 + k_{23}^2 +
   k_{31}^2 \over 3 k^2} \right] \nonumber \\
&& + ... \, ,
\end{eqnarray}
where the symbols ${\bf k}_{12}$ and ${\bf k}_{123}$ denote
respectively ${\bf k}_1 + {\bf k}_2$ and ${\bf k}_1 + {\bf k}_2 + {\bf
  k}_3$. The symbol $\delta_D$ represents the Dirac delta function.
The ellipsis on the right hand side of Eqn. (\ref{EulerPT})
represents higher order terms $O(\delta^4)$. 

For simplicity, we consider a matter-dominated flat universe
(a very good approximation at redshift $1000 > z > 1$)
i.e. $a = \bar H_0 {}^2 \eta^2 / 4$. Here, $\bar H_0$ is an effective
Hubble constant, and is related to the actual Hubble constant $H_0$ by
$\bar H_0 = H_0 \Omega_m {}_0 {}^{1/2}$ (where $\Omega_m {}_0$ is the
matter density today). 
It is helpful to define $\bar\eta \equiv \bar H_0 \eta$
and $\bar\Theta \equiv \Theta/\bar H_0$,
and rewrite the equations as:
\begin{eqnarray}
\label{masterEqt}
&& \partial_{\bar\eta} \Psi_a ({\bf k}) + \Omega_{ab}({\bf k}) \Psi_b ({\bf k}) 
= \nonumber \\
&& \int {d^3 k_1 d^3 k_2 \over (2\pi)^3} \delta_D ({\bf k} - {\bf
   k}_{12}) \gamma_{abc} ({\bf k}, {\bf
  k}_1, {\bf k}_2) \Psi_b ({\bf k}_1) \Psi_c ({\bf k}_2) \nonumber \\
&& + \int {d^3 k_1 d^3 k_2 d^3 k_3 \over (2\pi)^6} \delta_D ({\bf k} -
   {\bf k}_{123}) \Gamma_{abcd} ({\bf k}, {\bf
  k}_1, {\bf k}_2, {\bf k}_3) \nonumber \\
&& \quad \Psi_b ({\bf k}_1) \Psi_c ({\bf k}_2)
   \Psi_d ({\bf k}_3) \nonumber \\
&& + ...
\end{eqnarray}
where repeated indices e.g. $b, c$ are implicitly summed,
and all quantities are functions of time $\bar\eta$. The two-component
vector $\Psi$ represents:
\begin{eqnarray}
\left[\begin{array}{c} \Psi_1 \\ \Psi_2\end{array}\right] \equiv
\left[\begin{array}{c} \delta \\ \bar\Theta \end{array}\right] \, .
\end{eqnarray}
The matrix $\Omega ({\bf k})$ represents:
\begin{eqnarray}
\left[\begin{array}{cc} \Omega_{11} & \Omega_{12} \\ \Omega_{21} &
\Omega_{22}\end{array}\right] \equiv \left[\begin{array}{cc} 0 & 1 \\
{6\over \bar\eta {}^2} - {b_k^2 \over \bar\eta^4} &
{2\over \bar\eta} \end{array}\right] \, ,
\end{eqnarray}
where
\begin{equation}
b_k \equiv {2 k^2 \over \bar m\bar H_0 } \, .
\end{equation}
The matrix $\gamma_{abc} ({\bf k}, {\bf k}_1 , {\bf k}_2)$ is defined by:
\begin{equation}
\left[\begin{array}{cc} \gamma_{111} & \gamma_{112} \\ \gamma_{121} &
\gamma_{122} \end{array}\right]
\equiv - \left[\begin{array}{cc} 0 &
{{\bf k} \cdot {\bf k}_2 \over 2 k_2^2} \\ {{\bf k} \cdot {\bf k}_1 \over
2 k_1^2} & 0 \end{array}\right] \, ,
\end{equation}
\begin{equation}
\left[\begin{array}{cc} \gamma_{211} & \gamma_{212} \\ \gamma_{221} &
\gamma_{222} \end{array}\right]
\equiv - \left[\begin{array}{cc}
{b_k^2 \over 4
\bar\eta^4} \left( 1 + {k_1^2 + k_2^2 \over k^2} \right)&
0 \\ 0 & {k^2 {\bf k}_1 \cdot {\bf k}_2 \over 2 k_1^2 k_2^2} \end{array}\right] \, ,
\end{equation}
and all components of $\Gamma_{abcd} ({\bf k}, {\bf k}_1, {\bf k}_2,
{\bf k_3})$ 
vanish except for:
\begin{equation}
\Gamma_{2111} \equiv {b_k^2 \over
8 \bar\eta^4}  \Big( 1+ {k_1^2 + k_2^2 + k_3^2 \over k^2} + {k_{12}^2 +
  k_{23}^2 + k_{31}^2 \over 3 k^2} \Big)
\end{equation}

\begin{figure}[htpb]
\vspace*{-0.3cm}
\begin{tabular}{c}
\includegraphics[width=0.5\textwidth]{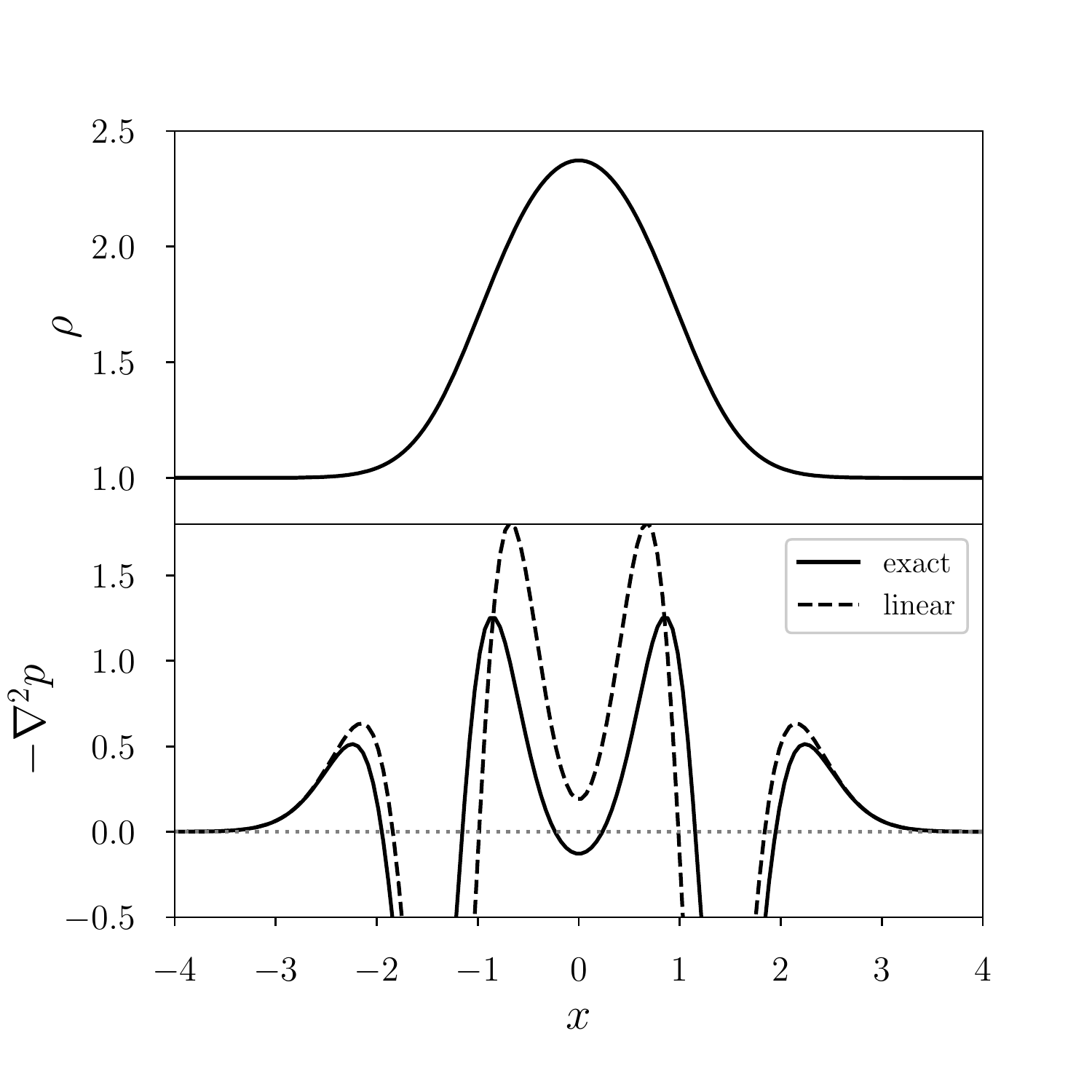} 
\end{tabular}
\caption{
$-\nabla^2 p$ for a density profile $\rho=[1+ (1+0.85 \exp x^2)^{-1}]^2$. The plot is drawn with $4\pi G$, $a$ and $\bar{m}$ all set to unity.
The exact value of $-\nabla^2 p$ is negative at the center, so the
quantum pressure actually enhances gravitational collapse (around the
center). However, the linear perturbative term is positive and acts to
slow down the collapse.
}
\label{fig:PertExp}
\end{figure}

Examining Eqn. (\ref{EulerPT}), the scale $k$ at which
the last two terms on the left balance each other
can be thought of as the Jeans scale i.e. when
$b_k^2 = 6 \bar\eta^2$:
\begin{equation}
k_{\rm Jeans} = {44.7\over {\rm Mpc}} 
\left( 6a
{\Omega_{m0} \over 0.3}\right)^{1/4}
\left( {H_0 \over 70 {\,\rm km/s}} 
{m \over 10^{-22} {\,\rm eV}} \right)^{1/2} \, .
\label{kJeans}
\end{equation}
The quantum pressure becomes important at
$k > k_{\rm Jeans}$. We stress that the Jeans scale is
a linear perturbation theory concept.
In fact, it is worth briefly commenting on how the quantum pressure
behaves beyond linear theory.
For a qualitative understanding, it's helpful to revert back to
position space in which the Euler equation (or the divergence thereof)
takes the form:
\begin{eqnarray}
&& \partial_\eta \Theta + 
{\partial_\eta a \over a} \Theta 
+ \partial_i (v^j \nabla_j v^i)
= - 4\pi G\bar\rho a^2 \delta \nonumber \\
&& \quad + {1\over 4 a^2 \bar m^2}
\nabla^2 \left( \nabla^2 \delta - {1\over 4} \nabla^2 \delta^2
- {1\over 2} \delta \nabla^2 \delta + ... \right) \, .
\end{eqnarray}
The first line of the equation is exactly the Euler equation in
the usual CDM case (pressureless fluid), where the term
on the right accounts for the effect of gravity -- it tends to
make $\Theta$ negative (i.e. convergent flow). 
The second line represents the effect of quantum pressure.
Its linear contribution (i.e. $\propto \nabla^2 \nabla^2 \delta$)
counteracts the effect of gravity. As an illustration,
consider a density peak with $\delta > 0$ such
as the one shown in Fig \ref{fig:PertExp}; the sign of the linear quantum pressure
term (i.e. $-\nabla^2 p \propto \nabla^2 \nabla^2 \delta$) 
is opposite to that of the gravity term (i.e. $-4\pi G \bar\rho a^2
\delta$) at the top of the peak.
This is precisely the Jeans phenomenon in linear theory.
{\it On the other hand, the quadratic terms from the quantum pressure acts in the same direction as gravity!} 
As demonstrated in Fig. \ref{fig:PertExp}, the linear order of
$-\nabla^2 p$ (dashed line) is positive at the center which
counteracts the gravity. However, the exact evaluation of nonlinear
$-\nabla^2 p$ (solid line) is actually negative and enhances the
gravitational collapse (close to the density peak).
The density profile chosen is admittedly artificial, but it serves
to illustrate the point that quantum pressure needs not always oppose
fluctuation growth. Thus while the fuzziness of dark matter leads
to a suppression of 
the {\it linear} power on scales smaller than the Jeans scale (see
below), the reverse could happen in the {\it nonlinear} regime.
This is borne out by numerical simulations, showing that nonlinear halos often have large density fluctuations due to the accidental
interference of waves (references). In this section, we have
a relatively narrow goal: compute the one-loop perturbative
corrections to the linear (tree) power spectrum, which
requires following the perturbation evolution up to cubic order. 
We will see that the quantum pressure in the end leads to a suppression of
power for this particular quantity.

Eqn. (\ref{masterEqt}) can be solved perturbatively, assuming
small perturbations.
Suppose $\Psi_a = \Psi_a^{(1)} + \Psi_a^{(2)} +
\Psi_a^{(3)} + ...$ where $\Psi_a^{(2)}$ is of the same order as
$[\Psi_a^{(1)}]^2$, etc. The first order (linear) solution is
\begin{equation}
\left[ \begin{array}{c} \Psi_1^{(1)} ({\bf k}, \bar\eta) \\
\Psi_2^{(1)} ({\bf k}, \bar\eta) \end{array}\right]
= \left[ \begin{array}{c} D_k (\bar\eta) \\ -\partial_{\bar\eta} D_k
           (\bar\eta) \end{array} \right] \delta^{(1)} ({\bf k},
       \bar\eta_{\rm in}) \, ,
\end{equation}
where we have made the time dependence explicit,
and $\bar\eta_{\rm in}$ is some early time (that can be sent to $0$). The
linear growth factor $D_k (\bar\eta)$ satisfies the equation:
\begin{equation}
\label{linearDeqt}
\partial_{\bar\eta}^2 D_k + {2\over \bar\eta} \partial_{\bar\eta} D_k
- \left( {6 \over \bar\eta^2} - {b_k^2 \over \bar\eta^4} \right) D_k =
0 \, ,
\end{equation}
and is given by:
\begin{equation}
D_k (\bar\eta)_ = \sqrt{\bar\eta_{\rm in} \over \bar\eta} 
{J_{-5/2} (b_k / \bar\eta) \over J_{-5/2} (b_k / \bar\eta_{\rm in})} \, ,
\end{equation}
where $J_{-5/2}$ is the Bessel function:
\begin{equation}
J_{-5/2} (z) \equiv \sqrt{2 \over \pi z}
\left[ {3 {\,\rm cos} z \over z^2} + {3 {\,\rm sin} z \over z} 
- {\rm cos} z \right] \, .
\end{equation}
Replacing $J_{-5/2}$ by $J_{5/2}$ would have given the decaying mode,
with
\begin{equation}
J_{5/2} (z) \equiv \sqrt{2 \over \pi z}
\left[ {3 {\,\rm sin} z \over z^2} - {3 {\,\rm cos} z \over z} 
- {\rm sin} z \right] \, .
\end{equation}
The small $z$ limit of the Bessel functions is:
\begin{equation}
J_{-5/2} (z) \rightarrow \sqrt{2\over \pi} {3 \over z^{5/2}} \, , \,
J_{5/2} (z) \rightarrow \sqrt{2\over \pi} {z^{5/2} \over 15} \, ,
\end{equation}
and thus $D_k \propto \bar\eta^2$ in the small $k$ limit, as expected.
The argument of the Bessel function $b_k/\bar\eta$ determines
the importance of quantum pressure: when it is large
(i.e. $k \gsim k_{\rm Jeans}$), quantum
pressure is important and the growth factor $D_k (\bar\eta)$ 
is oscillatory with a decaying amplitude.

For an accurate evaluation of the Bessel function,
it is best not to use these explicit expressions but rather use
recurrence relations (see Numerical Recipe).
Going beyond linear theory requires a solution
to Eqn. (\ref{linearDeqt}) with a non-vanishing right hand side
i.e. the following equation for $u(\bar\eta)$ for some general source $g(\bar\eta)$:
\begin{equation}
\partial_{\bar\eta}^2 u(\bar\eta) + {2\over \bar\eta} \partial_{\bar\eta} u(\bar\eta)
-
\left( {6 \over \bar\eta^2} - {b_k^2 \over \bar\eta^4} \right)
u(\bar\eta) = g(\bar\eta)
\end{equation}
can be solved by (up to the addition of a homogenous solution):
\begin{equation}
u(\bar\eta) = \int_{\bar\eta_{in}}^{\bar\eta} ds g(s) G_k (s,
\bar\eta) \, ,
\end{equation}
where $G_k (s, \bar\eta)$ is the propagator:
\begin{eqnarray}
&& G_k (s, \bar\eta) = {\pi s^2 \over 2}
\Big[ s^{-1/2} J_{5/2} (b_k/s) \bar\eta^{-1/2} J_{-5/2} (b_k/\bar\eta)
\nonumber \\
&& \quad \quad - \bar\eta^{-1/2} J_{5/2} (b_k/\bar\eta) s^{-1/2} J_{-5/2}
   (b_k/s) \Big] \, .
\end{eqnarray}
It is useful to note that $G_k (\bar\eta, \bar\eta) = 0$
and:
\begin{eqnarray}
&& \partial_{\bar\eta} G_k (s, \bar\eta) \Big|_{s = \bar\eta} = 1
  \nonumber \\
&& \partial_s G_k (s, \bar\eta) \Big|_{s = \bar\eta} = -1 
\nonumber \\
&& G_{k\rightarrow 0} (s, \bar\eta) = {1\over 5}
\left({\bar\eta^2 \over s} - {s^4 \over \bar\eta^3}\right)\, .
\end{eqnarray}
With these definitions in place, Eqn. (\ref{linearDeqt}) has
the following formal solution:
\begin{eqnarray}
\label{linearFormal}
&& \Psi_a ({\bf k}, \bar\eta) = \Psi_a^{(1)} ({\bf k}, \bar\eta)
\nonumber \\ && + \int {d^3 k_1 d^3 k_2 \over (2\pi)^3} 
\delta_D ({\bf k} - {\bf k}_{12}) \nonumber \\ && \quad \int_{\bar\eta_{\rm
  in}}^{\bar\eta}
ds \, W_{abc} ({\bf k}, {\bf k}_1, {\bf k}_2; s, \bar\eta) \Psi_b ({\bf
  k}_1, s) \Psi_c({\bf k}_2, s) \nonumber \\
&& + \int {d^3 k_1 d^3 k_2 d^3 k_3 \over (2\pi)^6} 
\delta_D ({\bf k} - {\bf k}_{123}) \nonumber \\
&& \quad \int_{\bar\eta_{\rm
  in}}^{\bar\eta} ds \, U_{abcd} ({\bf k}, {\bf k}_1, {\bf k}_2, {\bf k}_3;
  s, \bar\eta) \nonumber \\ && \quad \quad \quad
\Psi_b ({\bf k}_1, s) \Psi_c ({\bf k}_2, s) \Psi_d ({\bf
   k}_3, s) \nonumber \\
&& + ... \, ,
\end{eqnarray}
where the ellipsis represents terms with products of four $\Psi$'s or
more. The kernels $W$ and $U$ are defined by:
\begin{eqnarray}
&& W_{1bc}({\bf k}, {\bf k}_1, {\bf k}_2; s, \bar\eta) \equiv 
\Big( - \gamma_{1bc} ({\bf k}, {\bf
  k}_1, {\bf k}_2; s) \partial_s \nonumber \\
&& -\gamma_{2bc} ({\bf k}, {\bf k}_1, {\bf k}_2; s) + {2\over s}
  \gamma_{1bc} ({\bf k}, {\bf k}_1, {\bf k}_2; s) \Big) G_k (s,
  \bar\eta) \, , \nonumber \\
&& W_{2bc}({\bf k}, {\bf k}_1, {\bf k}_2; s, \bar\eta) \equiv
\Big(\gamma_{2bc} ({\bf k}, {\bf k}_1, {\bf k}_2; s) \nonumber \\
&& - {2\over s} 
\gamma_{1bc} ({\bf k}, {\bf k}_1, {\bf k}_2; s) 
+ \gamma_{1bc} ({\bf k}, {\bf k}_1, {\bf k}_2; s) \partial_s \Big)
\partial_{\bar\eta} G_k (s, \bar\eta) \, , \nonumber \\
&& U_{1bcd} ({\bf k}, {\bf k}_1, {\bf k}_2, {\bf k}_3; s, \bar\eta)
= \nonumber \\
&& \quad \quad - \Gamma_{2bcd} ({\bf k}, {\bf k}_1, {\bf k}_2, {\bf k}_3; s) G_k(s,
   \bar\eta) \, , \nonumber \\
&& U_{2bcd} ({\bf k}, {\bf k}_1, {\bf k}_2, {\bf k}_3; s, \bar\eta)
= \nonumber \\
&& \quad \quad \Gamma_{2bcd} ({\bf k}, {\bf k}_1, {\bf k}_2, {\bf
   k}_3; s) \partial_{\bar\eta} G_k(s,
   \bar\eta) \, , 
\end{eqnarray}
where the implicit time dependence of $\gamma$ and $\Gamma$
is made explicit for clarity.
Eqn. (\ref{linearFormal}) can be used to obtain the perturbative
solution iteratively: plugging in $\Psi = \Psi^{(1)}$ to the right
gives $\Psi^{(1)} + \Psi^{(2)}$ (ignoring third order terms and higher); plugging in
$\Psi = \Psi^{(1)} + \Psi^{(2)}$ to the right gives 
$\Psi^{(1)} + \Psi^{(2)} + \Psi^{(3)}$ (ignoring fourth order terms
and higher).

We are primarily interested in the density (equal-time) power spectrum $P(k,\bar\eta)$:
\begin{equation}
\langle \Psi_1 ({\bf k}, \bar\eta) \Psi_1 ({\bf k}', \bar\eta) \rangle
  \equiv (2\pi)^3 \delta_D ({\bf k} + {\bf k}') P (k,\bar\eta) \, .
\end{equation}
The linear (or tree) power spectrum $P^{(11)}$ is given by
\begin{equation}
\langle \Psi_1^{(1)} ({\bf k}, \bar\eta) \Psi_1^{(1)} ({\bf k}', \bar\eta) \rangle
  \equiv (2\pi)^3 \delta_D ({\bf k} + {\bf k}') P^{(11)} (k,\bar\eta)
  \, ,
\end{equation}
and can be expressed in terms of the initial tree power times the
growth factors:
\begin{equation}
P^{(11)} (k, \bar\eta) = P^{(11)} (k, \bar\eta_{\rm in})
D_k(\bar\eta)^2/D_k(\bar\eta_{\rm in})^2 \, .
\end{equation}

The tree power spectrum at redshift $z=5$ for
an FDM particle mass of $10^{-23}$ eV is shown in the
lower panel of Fig. \ref{pcomparefig} (the lower dashed line).
The tree power spectrum at the same redshift
and with the same initial condition but CDM dynamics is shown
as the lower solid line in the same panel.
The fact that they are very similar is due to the fact that
the growth factor $D_k$ does not differ a whole lot between
FDM and CDM dynamics for $k < 10$ h/Mpc. 
(Recall from Eqn. (\ref{kJeans}) that 
$k_{\rm Jeans} \sim 14$/Mpc for a mass of $10^{-23}$ eV.) 
In other words, the suppression of power seen in
Fig. \ref{pcomparefig} at high momenta has nothing
to do with the low redshift FDM dynamics; it's all in the initial
condition. Indeed, the scale at which the initial power
spectrum is suppressed by a factor of two is:
\begin{equation}
k_{1/2} = 1.62 {\rm /Mpc} \left( \frac{m}{ 10^{-23} {\,\rm eV} }\right)^{4/9} \, .
\end{equation}
This is substantially smaller (i.e. larger scale) compared to
$k_{\rm Jeans}$ at $z=5$. Thus, based on linear perturbation theory
alone, we do not expect CDM and FDM dynamics to give vastly
different power spectra starting from the same initial condition.
The question is whether this persists at higher order.

There are two lowest order perturbative corrections (the so called
loop corrections) $P^{(22)}$ and $P^{(13)}$, i.e.
\begin{equation}
P (k, \bar\eta) = P^{(11)} (k,\bar\eta) + P^{1-{\rm loop}} (k,
\bar\eta) \, ,
\end{equation}
with $P^{1-{\rm loop}} \equiv P^{(22)} + P^{(13)}$, are given by:
\begin{equation}
\langle \Psi_1^{(2)} ({\bf k}, \bar\eta) \Psi_1^{(2)} ({\bf k}', \bar\eta) \rangle
  \equiv (2\pi)^3 \delta_D ({\bf k} + {\bf k}') P^{(22)} (k.\bar\eta) \, ,
\end{equation}
\begin{eqnarray}
\langle \Psi_1^{(1)} ({\bf k}, \bar\eta) \Psi_1^{(3)} ({\bf k}',
  \bar\eta) \rangle
+ \langle \Psi_1^{(3)} ({\bf k}, \bar\eta) \Psi_1^{(1)} ({\bf k}',
  \bar\eta) \rangle
\nonumber \\ \equiv (2\pi)^3 \delta_D ({\bf k} + {\bf k}') P^{(13)}
  (k,\bar\eta) \, .
\end{eqnarray}

It can be shown the loop correction $P^{(22)}$ takes the form:
\begin{eqnarray}
\label{P22}
&& P^{(22)} (k, \bar\eta) = 2 \int {d^3 k_1 d^3 k_2 \over (2\pi)^3}
\delta_D ({\bf k} - {\bf k}_{12}) F_{\rm II} ({\bf k}_1, {\bf k}_2, \bar\eta)
  {}^2 \nonumber \\
&& \quad \quad P^{(11)} (k_1, \bar\eta) P^{(11)} (k_2, \bar\eta) \, ,
\end{eqnarray}
where the kernel $F_{\rm II}$ is defined by
\begin{eqnarray}
F_{\rm II} ({\bf k}_1 , {\bf k}_2, \bar\eta) \equiv
\int_{\bar\eta_{\rm in}}^{\bar\eta} ds \, W_{1bc} ({\bf k}_{12}, {\bf k}_1,
  {\bf k}_2; s, \bar\eta) 
\nonumber \\ f_b (k_1 , s, \bar\eta) f_c (k_2 , s, \bar\eta) \, ,
\end{eqnarray}
with $f_b, f_c$ defined by
\begin{eqnarray}
&& f_1 (k, s, \bar\eta) \equiv D_k (s) / D_k (\bar\eta) \nonumber \\
&& f_2 (k, s, \bar\eta) \equiv -\partial_s D_k (s) / D_k (\bar\eta) \, .
\end{eqnarray}
This $F_{\rm II}$ is often referred to as $F_2$ in the literature (or
the symmetrized version thereof) and matches
that in the $k\rightarrow 0$ limit (or $\bar m \rightarrow \infty$).
Numerically, it is advantageous to compute $P^{(22)}$ by
rewriting the integral in Eqn. (\ref{P22}) as:
\begin{eqnarray}
\label{P22integral}
&& P^{(22)} (k, \bar\eta) = 4 \int_{k_1 \le |{\bf k} - {\bf k_1}|} {d^3 k_1 \over (2\pi)^3}
F_{\rm II} ({\bf k}_1, {\bf k} - {\bf k_1}, \bar\eta)
  {}^2 \nonumber \\
&& \quad \quad P^{(11)} (k_1, \bar\eta) P^{(11)} (|{\bf k} - {\bf k}_1|, \bar\eta) \, ,
\end{eqnarray}
where the restriction of $k_1 \le |{\bf k} - {\bf k_1}|$ implies a
restriction on the angle between ${\bf k}$ and ${\bf k}_1$ if $k_1 >
k/2$. 
This way of integrating avoids divergent terms in $F$ that go
as $k^2/|{\bf k} - {\bf k}_1|^2$ when $|{\bf k} - {\bf k}_1|$ becomes
small \cite{Carrasco:2013sva}.
Terms that diverge as $k^2/k_1^2$ as $k_1$ becomes small are
harmless because of the integration measure $d^3 k_1$. 
(In fact, even $O(k/k_1)$ terms in $F$ can be avoided by 
a further rewriting.
) It is also to note that $F_{II} ({\bf k}_1 , {\bf k}_2, \bar\eta) = 0$
if ${\bf k}_1 + {\bf k}_2 = 0$,
because the corresponding kernels $\gamma_{abc}$'s vanish in that case.

The other loop correction to the power spectrum $P^{(13)}$,
which is of the same order as $P^{(22)}$, is given by
\begin{eqnarray}
\label{P13integral}
&& P^{(13)} (k, \bar\eta) = \int {d^3 k_2 \over (2\pi)^3}
\Big( 4 F_{\rm III} ({\bf k}_2, -{\bf k}_2, {\bf k}, \bar\eta)
+ 
\nonumber \\
&& 
+ 2 H_{\rm III} ({\bf
   k}, {\bf k}_2, \bar\eta) \Big) 
P^{(11)} (k,
  \bar\eta) P^{(11)} (k_2, \bar\eta) \,  ,\nonumber \\
\end{eqnarray}
where the kernel $F_{\rm III}$ is defined by
\begin{eqnarray}
&& F_{\rm III} ({\bf k}_2, {\bf k}_1', {\bf k}_2', \bar\eta) \equiv
2 \int_0^{\bar\eta} ds \, W_{1bc} ({\bf k}_{12}, {\bf k}_1, {\bf k}_2; s,
  \bar\eta) \nonumber \\
&& \quad \int_0^s ds' \, W_{bb'c'} ({\bf k}_1, {\bf k}_1' , {\bf k}_2';
  s', s) f_{b'} (k_1', s', \bar\eta) \nonumber \\
&& \quad \quad f_{c'} (k_2', s', \bar\eta) f_c(k_2, s, \bar\eta) \, ,
\end{eqnarray}
where ${\bf k}_1 = {\bf k}_1' + {\bf k}_2'$. Note that
$F_{\rm III} ({\bf k}_2, {\bf k}_1' , {\bf k}_2' , \bar\eta) = 0$
if ${\bf k}_1' + {\bf k}_2' = 0$ because $W_{\rm abc} ({\bf k}_{12},
{\bf k}_1, {\bf k}_2; s, \bar\eta) = 0$ in the case ${\bf k}_{12} = 0$. 
This is why a term that goes like $F_{\rm III} ({\bf k}, {\bf k}_2,
-{\bf k}_2, \bar\eta)$ is absent from Eqn. (\ref{P13integral}).
A common convention in the literature is to use a fully symmetrized
version of $F_{\rm III}$ in which case such a term would be present.
\footnote{In other words, the general structure should
be to replace $4 F_{\rm III}$ in Eqn. (\ref{P13integral}) by
$2 F_{\rm III} ({\bf k}_2, -{\bf k}_2, {\bf k})
+ 2 F_{\rm III} ({\bf k}_2, {\bf k}, -{\bf k}_2)
+ 2 F_{\rm III} ({\bf k}, {\bf k}_2, -{\bf k}_2)$.
The last contribution vanishes as argued above.
The second term matches the first i.e.
$F_{\rm III} ({\bf k}_2, -{\bf k}_2, {\bf k})
= F_{\rm III} ({\bf k}_2, {\bf k}, -{\bf k}_2)$, which is
guaranteed by the fact that
$W_{bb'c'} ({\bf k}_1, {\bf k}_1' , {\bf k}_2') = W_{bc'b'} ({\bf
  k}_1, {\bf k}_2' ,  {\bf k}_1')$. If one were to use a fully
symmetrized version of the kernel, one would have
replaced $4 F_{\rm III}$ by $6 F_{\rm III}^{\rm symm.}$.}

The kernel $H_{\rm III}$ is given by:
\begin{eqnarray}
&& H_{\rm III} ({\bf k}, {\bf k}_2, \bar\eta) \equiv f_a(k, \bar\eta, \bar\eta)
\int_{\bar\eta_{in}}^{\bar\eta} ds \nonumber \\
&& \big(U_{abcd} ({\bf k}, {\bf k}_2,
  -{\bf k}_2, {\bf k}; s, \bar\eta) 
f_b(k_2, s, \bar\eta) f_c(k_2, s,
  \bar\eta) f_d(k, s, \bar\eta) \nonumber \\
&& + U_{abcd} ({\bf k}, {\bf k}_2, {\bf k}, -{\bf k}_2; s, \bar\eta)
f_b(k_2, s, \bar\eta) f_c(k, s, \bar\eta) f_d(k_2, s, \bar\eta)
   \nonumber \\
&& + U_{abcd} ({\bf k}, {\bf k}, {\bf k}_2, -{\bf k}_2; s, \bar\eta)
f_b(k, s, \bar\eta) f_c(k_2, s, \bar\eta) f_d(k_2, s, \bar\eta) \Big)
   \, .
   \nonumber \\
\end{eqnarray}

Fig. \ref{pcomparefig} lower panel shows the total
power ($P^{11} + P^{22} + P^{13}$) computed this way:
the upper solid line uses CDM dynamics while the upper dashed line
uses FDM dynamics (same initial condition). 
The power is enhanced at high momenta compared
with the tree power spectrum (the lower solid/dashed lines).
The power at $k \gsim k_{1/2}$ comes almost entirely
from transferred power from large scales (low momenta).
Just as in the case of the tree power, CDM vs FDM dynamics
does not appear to make a big difference in the total power.
The total power with FDM dynamics is slightly lower than that
with CDM dynamics, except possibly at $k \sim 10$ h/Mpc, though
at that point, the power spectrum is rather low in any case.

Fig. \ref{pcomparefig} upper panel compares the total power
in two cases (both with FDM dynamics): the dashed line
is the same as the black dashed line in the lower panel; the dotted line shows the
power computed using Eqn. (\ref{P22integral}) and (\ref{P13integral})
with an infrared cut-off of the loop momentum at $2\pi/10$ h/Mpc.
The latter is chosen to mimic the effect of having a finite simulation
box of size $10$ Mpc/h. One can see the transferred power from large
scales is diminished, because the relevant large scale modes are
absent in a finite box.

Lastly, let's note that the bispectrum $B$, defined by
\begin{eqnarray}
&& \langle \Psi_1 ({\bf k_1}, \bar\eta) \Psi_1 ({\bf k_2}, \bar\eta) \Psi_1 ({\bf k_3}, \bar\eta) 
\equiv \nonumber \\
&& (2\pi)^3 \delta_D ({\bf k_1} + {\bf k_2} + {\bf k_3}) B({\bf k_1},
   {\bf k_2},
   {\bf k_3}, \bar\eta)  \, ,
\end{eqnarray}
is given, to the lowest order in perturbation theory, by:
\begin{eqnarray}
&& B({\bf k_1}, {\bf k_2}, {\bf k_3}, \bar\eta) = \nonumber \\
&& \quad \quad 2 F_{II}({\bf k_1}, {\bf k_2},
  \bar\eta) P^{(11)} (k_1, \bar\eta) P^{(11)} (k_2, \bar\eta) +
  \nonumber \\
&& \quad \quad 2 F_{II}({\bf k_2}, {\bf k_3},
  \bar\eta) P^{(11)} (k_2, \bar\eta) P^{(11)} (k_3, \bar\eta) +
  \nonumber \\
&& \quad \quad 2 F_{II}({\bf k_3}, {\bf k_1},
  \bar\eta) P^{(11)} (k_3, \bar\eta) P^{(11)} (k_1, \bar\eta) \, .
\end{eqnarray}

\section{\label{WPT}Wave Perturbation Theory}

Instead of the fluid equations, one could instead consider the wave
formulation:
\begin{equation}
i\left(\partial_t \psi + {3\over 2} H\psi\right) =
\left( - {\nabla^2 \over 2\bar m a^2} + \bar m \Phi \right)\psi \, ,
\end{equation}
supplemented by the Poisson equation (\ref{nablaPoisson}). 
Here, $t$ is the proper time, and $H$ is the Hubble constant
$\partial_t a / a$. 

The density $\delta$ is related to $\psi$ by
\begin{equation}
\delta = {|\psi |^2 - |\bar\psi|^2 \over |\bar\psi|^2} \, ,
\end{equation}
where $\bar\psi$ is the cosmic mean and can be taken to be real.
The velocity ${\boldsymbol{v}}$ can be deduced from
\begin{equation}
(1 + \delta){\boldsymbol{v}} = {1\over 2 i a \bar m |\bar\psi|^2} \left( \psi^* 
{\mbox{\boldmath$\nabla$}}
  \psi - \psi {\mbox{\boldmath$\nabla$}} \psi^* \right) \, .
\end{equation}

Let us constrast wave perturbation theory with fluid perturbation
theory. For the latter, we assume $\delta$ is small, in which
case the linearized mass conservation equation tells us, in momentum space:
\begin{equation}
v \sim k^{-1} (\partial_\eta D_k / D_k) \delta \sim {aH \over k} \delta
\end{equation}
where, in the last equality, we assume the growth factor $D_k \sim a$,
appropriate for low momenta. Thus, as long as the wavelength of
interest is sub-Hubble ($k > aH$) which is almost always the case, 
a small $\delta$ implies a small $v$ as well. 

For wave perturbation theory, we assume $\delta\psi \equiv \psi - \bar\psi$ is small.
It turns out this breaks down earlier (at higher redshift) than 
the break down of small $\delta$ and $v$. 
Let us relate $\delta$ and $v$ to the first order perturbation in $\psi$:
\begin{equation}
\delta = (\delta\psi + \delta\psi^*)/ \bar\psi \, ,
\end{equation}
and
\begin{equation}
{\boldsymbol{v}} = {1\over 2ia \bar m \bar\psi} {\mbox{\boldmath$\nabla$}}
(\delta\psi - \delta\psi^*)
\end{equation}
The smallness of $\delta$ is thus equivalent to
the smallness of $(\delta \psi + \delta\psi^*)/\bar\psi$.
It does not automatically guarantee that 
$(\delta\psi - \delta\psi^*)/\bar\psi$ is also small.
In momentum space, $(\delta\psi - \delta\psi^*)/\bar\psi$
can be written as
\begin{equation}
\label{WPTrequire}
{\delta\psi - \delta\psi^* \over \bar\psi}
\sim {a \bar m v \over k} \sim {a^2 \bar m H \over k^2} \delta \, ,
\end{equation}
where we have used linearized mass conservation in the
last equality. We see that the smallness of 
$(\delta\psi - \delta\psi^*)/\bar\psi$ is much more demanding than
the smallness of $v$: we need $a^2 \bar m H < k^2$ which is roughly
$k_{\rm Jeans} < k$. In other words, for wave perturbation theory to
work, it is not sufficient for $\delta$ to be small;
one needs $(k_{\rm Jeans}/k)^2 \delta$ to be small.
If the length scale of interest is above Jeans scale $k < k_{\rm
  Jeans}$, one would need a {\it very} small $\delta$ to compensate
(i.e. by going to an earlier time).
\footnote{We thank Michael Landry and Alberto Nicolis 
for discussions regarding wave
perturbation theory.}

Thus it appears wave perturbation theory has a rather limited
range of applicability. A corollary of the above reasoning is
that a numerical simulation of the wave equations is also
very demanding on resolution: suppose one is interested in
simulating structure growth on some large length scale (small $k \sim k_0$).
The typical velocity on such a scale is $v \sim (a H / k_0)
\delta$. Accounting for the typical power spectrum for $\delta$,
this velocity generally becomes larger as $k_0$ becomes smaller.
The velocity shows up in a wave simulation as the gradient of the
phase of $\psi$. It is  important that the wave simulation resolves
the de Broglie wavelength associated with this velocity
i.e. $k_{\rm resolution} \sim a \bar m v$. Thus, a
wave simulation requires high resolution even if the length scale
of interest is large.

The wave description is perhaps more useful in
the highly nonlinear regime, where $\delta$ is large
but $v$ remains modest. The velocity information
is repackaged into the phase of the $\psi$ in such
a way that the wave equation is linear in the absence of gravity.
Contrast this with the fluid equations which are nonlinear even
if gravity is turned off. How to take advantage of the fluid-wave
mapping to gain insight into nonlinear clustering is a topic
for further investigation.

\acknowledgements

We thank Marco Baldi, Zoltan Haiman, Vid Ir\u si\u c, Michael Landry,
Alberto Nicolis, Jerry
Ostriker, Massimo Pietroni, Roman Scoccimarro, Sergey Sibiriyakov and
Matteo Viel for useful discussions.
We gratefully recognize computational resources provided by NSF 
XSEDE through grant number TGMCA99S024, the NASA High-End 
Computing Program through the NASA Advanced Supercomputing 
Division at Ames Research Center, Columbia University, and the 
Flatiron Institute.  
This work made significant use of many open source software packages, 
including yt, Enzo, Python, IPython, NumPy, SciPy, Matplotlib, HDF5, h5py. 
These are products of collaborative effort by many independent developers 
from numerous institutions around the world. Their commitment to open 
science has helped to make this work possible.
G.L.B. acknowledges support from the NSF (AST-161955) and NASA (NNX15AB20G).
L.H. and X.L. acknowledge support from NASA (NXX16AB27G) and DOE
(DE-SC011941), and thank Andy Cohen, Henry Tye and the Institute for
Advanced Study at the Hong Kong University of Science and Technology for
hospitality.

\bibliography{ms}

\begin{thebibliography}{43}%
\makeatletter
\providecommand \@ifxundefined [1]{%
 \@ifx{#1\undefined}
}%
\providecommand \@ifnum [1]{%
 \ifnum #1\expandafter \@firstoftwo
 \else \expandafter \@secondoftwo
 \fi
}%
\providecommand \@ifx [1]{%
 \ifx #1\expandafter \@firstoftwo
 \else \expandafter \@secondoftwo
 \fi
}%
\providecommand \natexlab [1]{#1}%
\providecommand \enquote  [1]{``#1''}%
\providecommand \bibnamefont  [1]{#1}%
\providecommand \bibfnamefont [1]{#1}%
\providecommand \citenamefont [1]{#1}%
\providecommand \href@noop [0]{\@secondoftwo}%
\providecommand \href [0]{\begingroup \@sanitize@url \@href}%
\providecommand \@href[1]{\@@startlink{#1}\@@href}%
\providecommand \@@href[1]{\endgroup#1\@@endlink}%
\providecommand \@sanitize@url [0]{\catcode `\\12\catcode `\$12\catcode
  `\&12\catcode `\#12\catcode `\^12\catcode `\_12\catcode `\%12\relax}%
\providecommand \@@startlink[1]{}%
\providecommand \@@endlink[0]{}%
\providecommand \url  [0]{\begingroup\@sanitize@url \@url }%
\providecommand \@url [1]{\endgroup\@href {#1}{\urlprefix }}%
\providecommand \urlprefix  [0]{URL }%
\providecommand \Eprint [0]{\href }%
\providecommand \doibase [0]{http://dx.doi.org/}%
\providecommand \selectlanguage [0]{\@gobble}%
\providecommand \bibinfo  [0]{\@secondoftwo}%
\providecommand \bibfield  [0]{\@secondoftwo}%
\providecommand \translation [1]{[#1]}%
\providecommand \BibitemOpen [0]{}%
\providecommand \bibitemStop [0]{}%
\providecommand \bibitemNoStop [0]{.\EOS\space}%
\providecommand \EOS [0]{\spacefactor3000\relax}%
\providecommand \BibitemShut  [1]{\csname bibitem#1\endcsname}%
\let\auto@bib@innerbib\@empty
\bibitem [{\citenamefont {Baldeschi}\ \emph {et~al.}(1983)\citenamefont
  {Baldeschi}, \citenamefont {Ruffini},\ and\ \citenamefont
  {Gelmini}}]{Baldeschi:1983mq}%
  \BibitemOpen
  \bibfield  {author} {\bibinfo {author} {\bibfnamefont {M.~R.}\ \bibnamefont
  {Baldeschi}}, \bibinfo {author} {\bibfnamefont {R.}~\bibnamefont {Ruffini}},
  \ and\ \bibinfo {author} {\bibfnamefont {G.~B.}\ \bibnamefont {Gelmini}},\
  }\href {\doibase 10.1016/0370-2693(83)90688-3} {\bibfield  {journal}
  {\bibinfo  {journal} {Phys. Lett.}\ }\textbf {\bibinfo {volume} {122B}},\
  \bibinfo {pages} {221} (\bibinfo {year} {1983})}\BibitemShut {NoStop}%
\bibitem [{\citenamefont {Hu}\ \emph {et~al.}(2000)\citenamefont {Hu},
  \citenamefont {Barkana},\ and\ \citenamefont {Gruzinov}}]{Hu:2000ke}%
  \BibitemOpen
  \bibfield  {author} {\bibinfo {author} {\bibfnamefont {W.}~\bibnamefont
  {Hu}}, \bibinfo {author} {\bibfnamefont {R.}~\bibnamefont {Barkana}}, \ and\
  \bibinfo {author} {\bibfnamefont {A.}~\bibnamefont {Gruzinov}},\ }\href
  {\doibase 10.1103/PhysRevLett.85.1158} {\bibfield  {journal} {\bibinfo
  {journal} {Phys. Rev. Lett.}\ }\textbf {\bibinfo {volume} {85}},\ \bibinfo
  {pages} {1158} (\bibinfo {year} {2000})},\ \Eprint
  {http://arxiv.org/abs/astro-ph/0003365} {arXiv:astro-ph/0003365 [astro-ph]}
  \BibitemShut {NoStop}%
\bibitem [{\citenamefont {Lesgourgues}\ \emph {et~al.}(2002)\citenamefont
  {Lesgourgues}, \citenamefont {Arbey},\ and\ \citenamefont
  {Salati}}]{Lesgourgues:2002hk}%
  \BibitemOpen
  \bibfield  {author} {\bibinfo {author} {\bibfnamefont {J.}~\bibnamefont
  {Lesgourgues}}, \bibinfo {author} {\bibfnamefont {A.}~\bibnamefont {Arbey}},
  \ and\ \bibinfo {author} {\bibfnamefont {P.}~\bibnamefont {Salati}},\ }\href
  {\doibase 10.1016/S1387-6473(02)00247-6} {\bibfield  {journal} {\bibinfo
  {journal} {New Astron. Rev.}\ }\textbf {\bibinfo {volume} {46}},\ \bibinfo
  {pages} {791} (\bibinfo {year} {2002})}\BibitemShut {NoStop}%
\bibitem [{\citenamefont {Amendola}\ and\ \citenamefont
  {Barbieri}(2006)}]{Amendola:2005ad}%
  \BibitemOpen
  \bibfield  {author} {\bibinfo {author} {\bibfnamefont {L.}~\bibnamefont
  {Amendola}}\ and\ \bibinfo {author} {\bibfnamefont {R.}~\bibnamefont
  {Barbieri}},\ }\href {\doibase 10.1016/j.physletb.2006.08.069} {\bibfield
  {journal} {\bibinfo  {journal} {Phys. Lett.}\ }\textbf {\bibinfo {volume}
  {B642}},\ \bibinfo {pages} {192} (\bibinfo {year} {2006})},\ \Eprint
  {http://arxiv.org/abs/hep-ph/0509257} {arXiv:hep-ph/0509257 [hep-ph]}
  \BibitemShut {NoStop}%
\bibitem [{\citenamefont {Chavanis}(2011)}]{Chavanis:2011zi}%
  \BibitemOpen
  \bibfield  {author} {\bibinfo {author} {\bibfnamefont {P.-H.}\ \bibnamefont
  {Chavanis}},\ }\href {\doibase 10.1103/PhysRevD.84.043531} {\bibfield
  {journal} {\bibinfo  {journal} {Phys. Rev.}\ }\textbf {\bibinfo {volume}
  {D84}},\ \bibinfo {pages} {043531} (\bibinfo {year} {2011})},\ \Eprint
  {http://arxiv.org/abs/1103.2050} {arXiv:1103.2050 [astro-ph.CO]} \BibitemShut
  {NoStop}%
\bibitem [{\citenamefont {Arvanitaki}\ \emph {et~al.}(2010)\citenamefont
  {Arvanitaki}, \citenamefont {Dimopoulos}, \citenamefont {Dubovsky},
  \citenamefont {Kaloper},\ and\ \citenamefont
  {March-Russell}}]{Arvanitaki:2009fg}%
  \BibitemOpen
  \bibfield  {author} {\bibinfo {author} {\bibfnamefont {A.}~\bibnamefont
  {Arvanitaki}}, \bibinfo {author} {\bibfnamefont {S.}~\bibnamefont
  {Dimopoulos}}, \bibinfo {author} {\bibfnamefont {S.}~\bibnamefont
  {Dubovsky}}, \bibinfo {author} {\bibfnamefont {N.}~\bibnamefont {Kaloper}}, \
  and\ \bibinfo {author} {\bibfnamefont {J.}~\bibnamefont {March-Russell}},\
  }\href {\doibase 10.1103/PhysRevD.81.123530} {\bibfield  {journal} {\bibinfo
  {journal} {Phys. Rev.}\ }\textbf {\bibinfo {volume} {D81}},\ \bibinfo {pages}
  {123530} (\bibinfo {year} {2010})},\ \Eprint {http://arxiv.org/abs/0905.4720}
  {arXiv:0905.4720 [hep-th]} \BibitemShut {NoStop}%
\bibitem [{\citenamefont {Marsh}(2016)}]{Marsh:2015xka}%
  \BibitemOpen
  \bibfield  {author} {\bibinfo {author} {\bibfnamefont {D.~J.~E.}\
  \bibnamefont {Marsh}},\ }\href {\doibase 10.1016/j.physrep.2016.06.005}
  {\bibfield  {journal} {\bibinfo  {journal} {Phys. Rept.}\ }\textbf {\bibinfo
  {volume} {643}},\ \bibinfo {pages} {1} (\bibinfo {year} {2016})},\ \Eprint
  {http://arxiv.org/abs/1510.07633} {arXiv:1510.07633 [astro-ph.CO]}
  \BibitemShut {NoStop}%
\bibitem [{\citenamefont {Hui}\ \emph {et~al.}(2017)\citenamefont {Hui},
  \citenamefont {Ostriker}, \citenamefont {Tremaine},\ and\ \citenamefont
  {Witten}}]{Hui:2016ltb}%
  \BibitemOpen
  \bibfield  {author} {\bibinfo {author} {\bibfnamefont {L.}~\bibnamefont
  {Hui}}, \bibinfo {author} {\bibfnamefont {J.~P.}\ \bibnamefont {Ostriker}},
  \bibinfo {author} {\bibfnamefont {S.}~\bibnamefont {Tremaine}}, \ and\
  \bibinfo {author} {\bibfnamefont {E.}~\bibnamefont {Witten}},\ }\href
  {\doibase 10.1103/PhysRevD.95.043541} {\bibfield  {journal} {\bibinfo
  {journal} {Phys. Rev.}\ }\textbf {\bibinfo {volume} {D95}},\ \bibinfo {pages}
  {043541} (\bibinfo {year} {2017})},\ \Eprint
  {http://arxiv.org/abs/1610.08297} {arXiv:1610.08297 [astro-ph.CO]}
  \BibitemShut {NoStop}%
\bibitem [{\citenamefont {Svrcek}\ and\ \citenamefont
  {Witten}(2006)}]{Svrcek:2006yi}%
  \BibitemOpen
  \bibfield  {author} {\bibinfo {author} {\bibfnamefont {P.}~\bibnamefont
  {Svrcek}}\ and\ \bibinfo {author} {\bibfnamefont {E.}~\bibnamefont
  {Witten}},\ }\href {\doibase 10.1088/1126-6708/2006/06/051} {\bibfield
  {journal} {\bibinfo  {journal} {JHEP}\ }\textbf {\bibinfo {volume} {06}},\
  \bibinfo {pages} {051} (\bibinfo {year} {2006})},\ \Eprint
  {http://arxiv.org/abs/hep-th/0605206} {arXiv:hep-th/0605206 [hep-th]}
  \BibitemShut {NoStop}%
\bibitem [{\citenamefont {Schive}\ \emph
  {et~al.}(2014{\natexlab{a}})\citenamefont {Schive}, \citenamefont {Chiueh},\
  and\ \citenamefont {Broadhurst}}]{Schive:2014dra}%
  \BibitemOpen
  \bibfield  {author} {\bibinfo {author} {\bibfnamefont {H.-Y.}\ \bibnamefont
  {Schive}}, \bibinfo {author} {\bibfnamefont {T.}~\bibnamefont {Chiueh}}, \
  and\ \bibinfo {author} {\bibfnamefont {T.}~\bibnamefont {Broadhurst}},\
  }\href {\doibase 10.1038/nphys2996} {\bibfield  {journal} {\bibinfo
  {journal} {Nature Phys.}\ }\textbf {\bibinfo {volume} {10}},\ \bibinfo
  {pages} {496} (\bibinfo {year} {2014}{\natexlab{a}})},\ \Eprint
  {http://arxiv.org/abs/1406.6586} {arXiv:1406.6586 [astro-ph.GA]} \BibitemShut
  {NoStop}%
\bibitem [{\citenamefont {Schive}\ \emph
  {et~al.}(2014{\natexlab{b}})\citenamefont {Schive}, \citenamefont {Liao},
  \citenamefont {Woo}, \citenamefont {Wong}, \citenamefont {Chiueh},
  \citenamefont {Broadhurst},\ and\ \citenamefont {Hwang}}]{Schive:2014hza}%
  \BibitemOpen
  \bibfield  {author} {\bibinfo {author} {\bibfnamefont {H.-Y.}\ \bibnamefont
  {Schive}}, \bibinfo {author} {\bibfnamefont {M.-H.}\ \bibnamefont {Liao}},
  \bibinfo {author} {\bibfnamefont {T.-P.}\ \bibnamefont {Woo}}, \bibinfo
  {author} {\bibfnamefont {S.-K.}\ \bibnamefont {Wong}}, \bibinfo {author}
  {\bibfnamefont {T.}~\bibnamefont {Chiueh}}, \bibinfo {author} {\bibfnamefont
  {T.}~\bibnamefont {Broadhurst}}, \ and\ \bibinfo {author} {\bibfnamefont
  {W.~Y.~P.}\ \bibnamefont {Hwang}},\ }\href {\doibase
  10.1103/PhysRevLett.113.261302} {\bibfield  {journal} {\bibinfo  {journal}
  {Phys. Rev. Lett.}\ }\textbf {\bibinfo {volume} {113}},\ \bibinfo {pages}
  {261302} (\bibinfo {year} {2014}{\natexlab{b}})},\ \Eprint
  {http://arxiv.org/abs/1407.7762} {arXiv:1407.7762 [astro-ph.GA]} \BibitemShut
  {NoStop}%
\bibitem [{\citenamefont {Mocz}\ and\ \citenamefont
  {Succi}(2015)}]{Mocz:2015sda}%
  \BibitemOpen
  \bibfield  {author} {\bibinfo {author} {\bibfnamefont {P.}~\bibnamefont
  {Mocz}}\ and\ \bibinfo {author} {\bibfnamefont {S.}~\bibnamefont {Succi}},\
  }\href {\doibase 10.1103/PhysRevE.91.053304} {\bibfield  {journal} {\bibinfo
  {journal} {Phys. Rev.}\ }\textbf {\bibinfo {volume} {E91}},\ \bibinfo {pages}
  {053304} (\bibinfo {year} {2015})},\ \Eprint
  {http://arxiv.org/abs/1503.03869} {arXiv:1503.03869 [physics.comp-ph]}
  \BibitemShut {NoStop}%
\bibitem [{\citenamefont {Schwabe}\ \emph {et~al.}(2016)\citenamefont
  {Schwabe}, \citenamefont {Niemeyer},\ and\ \citenamefont
  {Engels}}]{Schwabe:2016rze}%
  \BibitemOpen
  \bibfield  {author} {\bibinfo {author} {\bibfnamefont {B.}~\bibnamefont
  {Schwabe}}, \bibinfo {author} {\bibfnamefont {J.~C.}\ \bibnamefont
  {Niemeyer}}, \ and\ \bibinfo {author} {\bibfnamefont {J.~F.}\ \bibnamefont
  {Engels}},\ }\href {\doibase 10.1103/PhysRevD.94.043513} {\bibfield
  {journal} {\bibinfo  {journal} {Phys. Rev.}\ }\textbf {\bibinfo {volume}
  {D94}},\ \bibinfo {pages} {043513} (\bibinfo {year} {2016})},\ \Eprint
  {http://arxiv.org/abs/1606.05151} {arXiv:1606.05151 [astro-ph.CO]}
  \BibitemShut {NoStop}%
\bibitem [{\citenamefont {Veltmaat}\ and\ \citenamefont
  {Niemeyer}(2016)}]{Veltmaat:2016rxo}%
  \BibitemOpen
  \bibfield  {author} {\bibinfo {author} {\bibfnamefont {J.}~\bibnamefont
  {Veltmaat}}\ and\ \bibinfo {author} {\bibfnamefont {J.~C.}\ \bibnamefont
  {Niemeyer}},\ }\href {\doibase 10.1103/PhysRevD.94.123523} {\bibfield
  {journal} {\bibinfo  {journal} {Phys. Rev.}\ }\textbf {\bibinfo {volume}
  {D94}},\ \bibinfo {pages} {123523} (\bibinfo {year} {2016})},\ \Eprint
  {http://arxiv.org/abs/1608.00802} {arXiv:1608.00802 [astro-ph.CO]}
  \BibitemShut {NoStop}%
\bibitem [{\citenamefont {Mocz}\ \emph {et~al.}(2017)\citenamefont {Mocz},
  \citenamefont {Vogelsberger}, \citenamefont {Robles}, \citenamefont {Zavala},
  \citenamefont {Boylan-Kolchin},\ and\ \citenamefont
  {Hernquist}}]{Mocz:2017wlg}%
  \BibitemOpen
  \bibfield  {author} {\bibinfo {author} {\bibfnamefont {P.}~\bibnamefont
  {Mocz}}, \bibinfo {author} {\bibfnamefont {M.}~\bibnamefont {Vogelsberger}},
  \bibinfo {author} {\bibfnamefont {V.}~\bibnamefont {Robles}}, \bibinfo
  {author} {\bibfnamefont {J.}~\bibnamefont {Zavala}}, \bibinfo {author}
  {\bibfnamefont {M.}~\bibnamefont {Boylan-Kolchin}}, \ and\ \bibinfo {author}
  {\bibfnamefont {L.}~\bibnamefont {Hernquist}},\ }\href {\doibase
  10.1093/mnras/stx1887} {\bibfield  {journal} {\bibinfo  {journal} {Mon. Not.
  Roy. Astron. Soc.}\ }\textbf {\bibinfo {volume} {471}},\ \bibinfo {pages}
  {4559} (\bibinfo {year} {2017})},\ \Eprint {http://arxiv.org/abs/1705.05845}
  {arXiv:1705.05845 [astro-ph.CO]} \BibitemShut {NoStop}%
\bibitem [{\citenamefont {Lin}\ \emph {et~al.}(2018)\citenamefont {Lin},
  \citenamefont {Schive}, \citenamefont {Wong},\ and\ \citenamefont
  {Chiueh}}]{Lin:2018whl}%
  \BibitemOpen
  \bibfield  {author} {\bibinfo {author} {\bibfnamefont {S.-C.}\ \bibnamefont
  {Lin}}, \bibinfo {author} {\bibfnamefont {H.-Y.}\ \bibnamefont {Schive}},
  \bibinfo {author} {\bibfnamefont {S.-K.}\ \bibnamefont {Wong}}, \ and\
  \bibinfo {author} {\bibfnamefont {T.}~\bibnamefont {Chiueh}},\ }\href
  {\doibase 10.1103/PhysRevD.97.103523} {\bibfield  {journal} {\bibinfo
  {journal} {Phys. Rev.}\ }\textbf {\bibinfo {volume} {D97}},\ \bibinfo {pages}
  {103523} (\bibinfo {year} {2018})},\ \Eprint
  {http://arxiv.org/abs/1801.02320} {arXiv:1801.02320 [astro-ph.CO]}
  \BibitemShut {NoStop}%
\bibitem [{\citenamefont {Veltmaat}\ \emph {et~al.}(2018)\citenamefont
  {Veltmaat}, \citenamefont {Niemeyer},\ and\ \citenamefont
  {Schwabe}}]{Veltmaat:2018dfz}%
  \BibitemOpen
  \bibfield  {author} {\bibinfo {author} {\bibfnamefont {J.}~\bibnamefont
  {Veltmaat}}, \bibinfo {author} {\bibfnamefont {J.~C.}\ \bibnamefont
  {Niemeyer}}, \ and\ \bibinfo {author} {\bibfnamefont {B.}~\bibnamefont
  {Schwabe}},\ }\href {\doibase 10.1103/PhysRevD.98.043509} {\bibfield
  {journal} {\bibinfo  {journal} {Phys. Rev.}\ }\textbf {\bibinfo {volume}
  {D98}},\ \bibinfo {pages} {043509} (\bibinfo {year} {2018})},\ \Eprint
  {http://arxiv.org/abs/1804.09647} {arXiv:1804.09647 [astro-ph.CO]}
  \BibitemShut {NoStop}%
\bibitem [{\citenamefont {Nori}\ and\ \citenamefont
  {Baldi}(2018)}]{Nori:2018hud}%
  \BibitemOpen
  \bibfield  {author} {\bibinfo {author} {\bibfnamefont {M.}~\bibnamefont
  {Nori}}\ and\ \bibinfo {author} {\bibfnamefont {M.}~\bibnamefont {Baldi}},\
  }\href@noop {} {\  (\bibinfo {year} {2018})},\ \Eprint
  {http://arxiv.org/abs/1801.08144} {arXiv:1801.08144 [astro-ph.CO]}
  \BibitemShut {NoStop}%
\bibitem [{\citenamefont {Madelung}(1927)}]{Madelung1927}%
  \BibitemOpen
  \bibfield  {author} {\bibinfo {author} {\bibfnamefont {E.}~\bibnamefont
  {Madelung}},\ }\href {\doibase 10.1007/BF01400372} {\bibfield  {journal}
  {\bibinfo  {journal} {Zeitschrift f{\"u}r Physik}\ }\textbf {\bibinfo
  {volume} {40}},\ \bibinfo {pages} {322} (\bibinfo {year} {1927})}\BibitemShut
  {NoStop}%
\bibitem [{\citenamefont {{Feynman}}(1965)}]{1965flp..book.....F}%
  \BibitemOpen
  \bibfield  {author} {\bibinfo {author} {\bibfnamefont {R.~P.}\ \bibnamefont
  {{Feynman}}},\ }\href@noop {} {\emph {\bibinfo {title} {Reading, Ma.:
  Addison-Wesley, 1965, edited by Feynman, Richard P.; Leighton, Robert B.;
  Sands, Matthew}}}\ (\bibinfo {year} {1965})\BibitemShut {NoStop}%
\bibitem [{\citenamefont {Mocz}\ \emph {et~al.}(2018)\citenamefont {Mocz},
  \citenamefont {Lancaster}, \citenamefont {Fialkov}, \citenamefont {Becerra},\
  and\ \citenamefont {Chavanis}}]{Mocz:2018ium}%
  \BibitemOpen
  \bibfield  {author} {\bibinfo {author} {\bibfnamefont {P.}~\bibnamefont
  {Mocz}}, \bibinfo {author} {\bibfnamefont {L.}~\bibnamefont {Lancaster}},
  \bibinfo {author} {\bibfnamefont {A.}~\bibnamefont {Fialkov}}, \bibinfo
  {author} {\bibfnamefont {F.}~\bibnamefont {Becerra}}, \ and\ \bibinfo
  {author} {\bibfnamefont {P.-H.}\ \bibnamefont {Chavanis}},\ }\href {\doibase
  10.1103/PhysRevD.97.083519} {\bibfield  {journal} {\bibinfo  {journal} {Phys.
  Rev.}\ }\textbf {\bibinfo {volume} {D97}},\ \bibinfo {pages} {083519}
  (\bibinfo {year} {2018})},\ \Eprint {http://arxiv.org/abs/1801.03507}
  {arXiv:1801.03507 [astro-ph.CO]} \BibitemShut {NoStop}%
\bibitem [{\citenamefont {Zhang}\ \emph {et~al.}(2017)\citenamefont {Zhang},
  \citenamefont {Kuo}, \citenamefont {Liu}, \citenamefont {Tsai}, \citenamefont
  {Cheung},\ and\ \citenamefont {Chu}}]{Zhang:2017chj}%
  \BibitemOpen
  \bibfield  {author} {\bibinfo {author} {\bibfnamefont {J.}~\bibnamefont
  {Zhang}}, \bibinfo {author} {\bibfnamefont {J.-L.}\ \bibnamefont {Kuo}},
  \bibinfo {author} {\bibfnamefont {H.}~\bibnamefont {Liu}}, \bibinfo {author}
  {\bibfnamefont {Y.-L.~S.}\ \bibnamefont {Tsai}}, \bibinfo {author}
  {\bibfnamefont {K.}~\bibnamefont {Cheung}}, \ and\ \bibinfo {author}
  {\bibfnamefont {M.-C.}\ \bibnamefont {Chu}},\ }\href@noop {} {\  (\bibinfo
  {year} {2017})},\ \Eprint {http://arxiv.org/abs/1708.04389} {arXiv:1708.04389
  [astro-ph.CO]} \BibitemShut {NoStop}%
\bibitem [{\citenamefont {Iršič}\ \emph {et~al.}(2017)\citenamefont
  {Iršič}, \citenamefont {Viel}, \citenamefont {Haehnelt}, \citenamefont
  {Bolton},\ and\ \citenamefont {Becker}}]{Irsic:2017yje}%
  \BibitemOpen
  \bibfield  {author} {\bibinfo {author} {\bibfnamefont {V.}~\bibnamefont
  {Iršič}}, \bibinfo {author} {\bibfnamefont {M.}~\bibnamefont {Viel}},
  \bibinfo {author} {\bibfnamefont {M.~G.}\ \bibnamefont {Haehnelt}}, \bibinfo
  {author} {\bibfnamefont {J.~S.}\ \bibnamefont {Bolton}}, \ and\ \bibinfo
  {author} {\bibfnamefont {G.~D.}\ \bibnamefont {Becker}},\ }\href {\doibase
  10.1103/PhysRevLett.119.031302} {\bibfield  {journal} {\bibinfo  {journal}
  {Phys. Rev. Lett.}\ }\textbf {\bibinfo {volume} {119}},\ \bibinfo {pages}
  {031302} (\bibinfo {year} {2017})},\ \Eprint
  {http://arxiv.org/abs/1703.04683} {arXiv:1703.04683 [astro-ph.CO]}
  \BibitemShut {NoStop}%
\bibitem [{\citenamefont {Armengaud}\ \emph {et~al.}(2017)\citenamefont
  {Armengaud}, \citenamefont {Palanque-Delabrouille}, \citenamefont {Yèche},
  \citenamefont {Marsh},\ and\ \citenamefont {Baur}}]{Armengaud:2017nkf}%
  \BibitemOpen
  \bibfield  {author} {\bibinfo {author} {\bibfnamefont {E.}~\bibnamefont
  {Armengaud}}, \bibinfo {author} {\bibfnamefont {N.}~\bibnamefont
  {Palanque-Delabrouille}}, \bibinfo {author} {\bibfnamefont {C.}~\bibnamefont
  {Yèche}}, \bibinfo {author} {\bibfnamefont {D.~J.~E.}\ \bibnamefont
  {Marsh}}, \ and\ \bibinfo {author} {\bibfnamefont {J.}~\bibnamefont {Baur}},\
  }\href {\doibase 10.1093/mnras/stx1870} {\bibfield  {journal} {\bibinfo
  {journal} {Mon. Not. Roy. Astron. Soc.}\ }\textbf {\bibinfo {volume} {471}},\
  \bibinfo {pages} {4606} (\bibinfo {year} {2017})},\ \Eprint
  {http://arxiv.org/abs/1703.09126} {arXiv:1703.09126 [astro-ph.CO]}
  \BibitemShut {NoStop}%
\bibitem [{\citenamefont {Nori}\ \emph {et~al.}(2018)\citenamefont {Nori},
  \citenamefont {Murgia}, \citenamefont {Iršič}, \citenamefont {Baldi},\ and\
  \citenamefont {Viel}}]{Nori:2018pka}%
  \BibitemOpen
  \bibfield  {author} {\bibinfo {author} {\bibfnamefont {M.}~\bibnamefont
  {Nori}}, \bibinfo {author} {\bibfnamefont {R.}~\bibnamefont {Murgia}},
  \bibinfo {author} {\bibfnamefont {V.}~\bibnamefont {Iršič}}, \bibinfo
  {author} {\bibfnamefont {M.}~\bibnamefont {Baldi}}, \ and\ \bibinfo {author}
  {\bibfnamefont {M.}~\bibnamefont {Viel}},\ }\href@noop {} {\  (\bibinfo
  {year} {2018})},\ \Eprint {http://arxiv.org/abs/1809.09619} {arXiv:1809.09619
  [astro-ph.CO]} \BibitemShut {NoStop}%
\bibitem [{\citenamefont {Bryan}\ \emph {et~al.}(2014)\citenamefont {Bryan}
  \emph {et~al.}}]{Bryan:2013hfa}%
  \BibitemOpen
  \bibfield  {author} {\bibinfo {author} {\bibfnamefont {G.~L.}\ \bibnamefont
  {Bryan}} \emph {et~al.} (\bibinfo {collaboration} {ENZO}),\ }\href {\doibase
  10.1088/0067-0049/211/2/19} {\bibfield  {journal} {\bibinfo  {journal}
  {Astrophys. J. Suppl.}\ }\textbf {\bibinfo {volume} {211}},\ \bibinfo {pages}
  {19} (\bibinfo {year} {2014})},\ \Eprint {http://arxiv.org/abs/1307.2265}
  {arXiv:1307.2265 [astro-ph.IM]} \BibitemShut {NoStop}%
\bibitem [{\citenamefont {{Turk}}\ \emph {et~al.}(2011)\citenamefont {{Turk}},
  \citenamefont {{Smith}}, \citenamefont {{Oishi}}, \citenamefont {{Skory}},
  \citenamefont {{Skillman}}, \citenamefont {{Abel}},\ and\ \citenamefont
  {{Norman}}}]{yt}%
  \BibitemOpen
  \bibfield  {author} {\bibinfo {author} {\bibfnamefont {M.~J.}\ \bibnamefont
  {{Turk}}}, \bibinfo {author} {\bibfnamefont {B.~D.}\ \bibnamefont {{Smith}}},
  \bibinfo {author} {\bibfnamefont {J.~S.}\ \bibnamefont {{Oishi}}}, \bibinfo
  {author} {\bibfnamefont {S.}~\bibnamefont {{Skory}}}, \bibinfo {author}
  {\bibfnamefont {S.~W.}\ \bibnamefont {{Skillman}}}, \bibinfo {author}
  {\bibfnamefont {T.}~\bibnamefont {{Abel}}}, \ and\ \bibinfo {author}
  {\bibfnamefont {M.~L.}\ \bibnamefont {{Norman}}},\ }\href {\doibase
  10.1088/0067-0049/192/1/9} {\bibfield  {journal} {\bibinfo  {journal} {ApJS}\
  }\textbf {\bibinfo {volume} {192}},\ \bibinfo {eid} {9} (\bibinfo {year}
  {2011})},\ \Eprint {http://arxiv.org/abs/1011.3514} {arXiv:1011.3514
  [astro-ph.IM]} \BibitemShut {NoStop}%
\bibitem [{\citenamefont {Croft}\ \emph {et~al.}(1998)\citenamefont {Croft},
  \citenamefont {Weinberg}, \citenamefont {Katz},\ and\ \citenamefont
  {Hernquist}}]{Croft:1997jf}%
  \BibitemOpen
  \bibfield  {author} {\bibinfo {author} {\bibfnamefont {R.~A.~C.}\
  \bibnamefont {Croft}}, \bibinfo {author} {\bibfnamefont {D.~H.}\ \bibnamefont
  {Weinberg}}, \bibinfo {author} {\bibfnamefont {N.}~\bibnamefont {Katz}}, \
  and\ \bibinfo {author} {\bibfnamefont {L.}~\bibnamefont {Hernquist}},\ }\href
  {\doibase 10.1086/305289} {\bibfield  {journal} {\bibinfo  {journal}
  {Astrophys. J.}\ }\textbf {\bibinfo {volume} {495}},\ \bibinfo {pages} {44}
  (\bibinfo {year} {1998})},\ \Eprint {http://arxiv.org/abs/astro-ph/9708018}
  {arXiv:astro-ph/9708018 [astro-ph]} \BibitemShut {NoStop}%
\bibitem [{\citenamefont {Hui}(1999)}]{Hui:1998hq}%
  \BibitemOpen
  \bibfield  {author} {\bibinfo {author} {\bibfnamefont {L.}~\bibnamefont
  {Hui}},\ }\href {\doibase 10.1086/307134} {\bibfield  {journal} {\bibinfo
  {journal} {Astrophys. J.}\ }\textbf {\bibinfo {volume} {516}},\ \bibinfo
  {pages} {519} (\bibinfo {year} {1999})},\ \Eprint
  {http://arxiv.org/abs/astro-ph/9807068} {arXiv:astro-ph/9807068 [astro-ph]}
  \BibitemShut {NoStop}%
\bibitem [{\citenamefont {Croft}\ \emph {et~al.}(2002)\citenamefont {Croft},
  \citenamefont {Weinberg}, \citenamefont {Bolte}, \citenamefont {Burles},
  \citenamefont {Hernquist}, \citenamefont {Katz}, \citenamefont {Kirkman},\
  and\ \citenamefont {Tytler}}]{Croft:2000hs}%
  \BibitemOpen
  \bibfield  {author} {\bibinfo {author} {\bibfnamefont {R.~A.~C.}\
  \bibnamefont {Croft}}, \bibinfo {author} {\bibfnamefont {D.~H.}\ \bibnamefont
  {Weinberg}}, \bibinfo {author} {\bibfnamefont {M.}~\bibnamefont {Bolte}},
  \bibinfo {author} {\bibfnamefont {S.}~\bibnamefont {Burles}}, \bibinfo
  {author} {\bibfnamefont {L.}~\bibnamefont {Hernquist}}, \bibinfo {author}
  {\bibfnamefont {N.}~\bibnamefont {Katz}}, \bibinfo {author} {\bibfnamefont
  {D.}~\bibnamefont {Kirkman}}, \ and\ \bibinfo {author} {\bibfnamefont
  {D.}~\bibnamefont {Tytler}},\ }\href {\doibase 10.1086/344099} {\bibfield
  {journal} {\bibinfo  {journal} {Astrophys. J.}\ }\textbf {\bibinfo {volume}
  {581}},\ \bibinfo {pages} {20} (\bibinfo {year} {2002})},\ \Eprint
  {http://arxiv.org/abs/astro-ph/0012324} {arXiv:astro-ph/0012324 [astro-ph]}
  \BibitemShut {NoStop}%
\bibitem [{\citenamefont {McDonald}\ \emph {et~al.}(2005)\citenamefont
  {McDonald} \emph {et~al.}}]{McDonald:2004xn}%
  \BibitemOpen
  \bibfield  {author} {\bibinfo {author} {\bibfnamefont {P.}~\bibnamefont
  {McDonald}} \emph {et~al.} (\bibinfo {collaboration} {SDSS}),\ }\href
  {\doibase 10.1086/497563} {\bibfield  {journal} {\bibinfo  {journal}
  {Astrophys. J.}\ }\textbf {\bibinfo {volume} {635}},\ \bibinfo {pages} {761}
  (\bibinfo {year} {2005})},\ \Eprint {http://arxiv.org/abs/astro-ph/0407377}
  {arXiv:astro-ph/0407377 [astro-ph]} \BibitemShut {NoStop}%
\bibitem [{\citenamefont {Viel}\ \emph {et~al.}(2006)\citenamefont {Viel},
  \citenamefont {Haehnelt},\ and\ \citenamefont {Lewis}}]{Viel:2006yh}%
  \BibitemOpen
  \bibfield  {author} {\bibinfo {author} {\bibfnamefont {M.}~\bibnamefont
  {Viel}}, \bibinfo {author} {\bibfnamefont {M.~G.}\ \bibnamefont {Haehnelt}},
  \ and\ \bibinfo {author} {\bibfnamefont {A.}~\bibnamefont {Lewis}},\ }\href
  {\doibase 10.1111/j.1745-3933.2006.00187.x} {\bibfield  {journal} {\bibinfo
  {journal} {Mon. Not. Roy. Astron. Soc.}\ }\textbf {\bibinfo {volume} {370}},\
  \bibinfo {pages} {L51} (\bibinfo {year} {2006})},\ \Eprint
  {http://arxiv.org/abs/astro-ph/0604310} {arXiv:astro-ph/0604310 [astro-ph]}
  \BibitemShut {NoStop}%
\bibitem [{\citenamefont {Viel}\ \emph {et~al.}(2013)\citenamefont {Viel},
  \citenamefont {Becker}, \citenamefont {Bolton},\ and\ \citenamefont
  {Haehnelt}}]{Viel:2013apy}%
  \BibitemOpen
  \bibfield  {author} {\bibinfo {author} {\bibfnamefont {M.}~\bibnamefont
  {Viel}}, \bibinfo {author} {\bibfnamefont {G.~D.}\ \bibnamefont {Becker}},
  \bibinfo {author} {\bibfnamefont {J.~S.}\ \bibnamefont {Bolton}}, \ and\
  \bibinfo {author} {\bibfnamefont {M.~G.}\ \bibnamefont {Haehnelt}},\ }\href
  {\doibase 10.1103/PhysRevD.88.043502} {\bibfield  {journal} {\bibinfo
  {journal} {Phys. Rev.}\ }\textbf {\bibinfo {volume} {D88}},\ \bibinfo {pages}
  {043502} (\bibinfo {year} {2013})},\ \Eprint {http://arxiv.org/abs/1306.2314}
  {arXiv:1306.2314 [astro-ph.CO]} \BibitemShut {NoStop}%
\bibitem [{\citenamefont {Gnedin}\ and\ \citenamefont
  {Hui}(1998)}]{Gnedin:1997td}%
  \BibitemOpen
  \bibfield  {author} {\bibinfo {author} {\bibfnamefont {N.~Y.}\ \bibnamefont
  {Gnedin}}\ and\ \bibinfo {author} {\bibfnamefont {L.}~\bibnamefont {Hui}},\
  }\href {\doibase 10.1046/j.1365-8711.1998.01249.x} {\bibfield  {journal}
  {\bibinfo  {journal} {Mon. Not. Roy. Astron. Soc.}\ }\textbf {\bibinfo
  {volume} {296}},\ \bibinfo {pages} {44} (\bibinfo {year} {1998})},\ \Eprint
  {http://arxiv.org/abs/astro-ph/9706219} {arXiv:astro-ph/9706219 [astro-ph]}
  \BibitemShut {NoStop}%
\bibitem [{\citenamefont {Hui}\ and\ \citenamefont
  {Gnedin}(1997)}]{Hui:1997dp}%
  \BibitemOpen
  \bibfield  {author} {\bibinfo {author} {\bibfnamefont {L.}~\bibnamefont
  {Hui}}\ and\ \bibinfo {author} {\bibfnamefont {N.~Y.}\ \bibnamefont
  {Gnedin}},\ }\href {\doibase 10.1093/mnras/292.1.27} {\bibfield  {journal}
  {\bibinfo  {journal} {Mon. Not. Roy. Astron. Soc.}\ }\textbf {\bibinfo
  {volume} {292}},\ \bibinfo {pages} {27} (\bibinfo {year} {1997})},\ \Eprint
  {http://arxiv.org/abs/astro-ph/9612232} {arXiv:astro-ph/9612232 [astro-ph]}
  \BibitemShut {NoStop}%
\bibitem [{\citenamefont {Stone}\ and\ \citenamefont
  {Norman}(1992{\natexlab{a}})}]{Stone:1992wj}%
  \BibitemOpen
  \bibfield  {author} {\bibinfo {author} {\bibfnamefont {J.~M.}\ \bibnamefont
  {Stone}}\ and\ \bibinfo {author} {\bibfnamefont {M.~L.}\ \bibnamefont
  {Norman}},\ }\href {\doibase 10.1086/191680} {\bibfield  {journal} {\bibinfo
  {journal} {Astrophys. J. Suppl.}\ }\textbf {\bibinfo {volume} {80}},\
  \bibinfo {pages} {753} (\bibinfo {year} {1992}{\natexlab{a}})}\BibitemShut
  {NoStop}%
\bibitem [{\citenamefont {Stone}\ and\ \citenamefont
  {Norman}(1992{\natexlab{b}})}]{Stone:1992mq}%
  \BibitemOpen
  \bibfield  {author} {\bibinfo {author} {\bibfnamefont {J.~M.}\ \bibnamefont
  {Stone}}\ and\ \bibinfo {author} {\bibfnamefont {M.~L.}\ \bibnamefont
  {Norman}},\ }\href {\doibase 10.1086/191681} {\bibfield  {journal} {\bibinfo
  {journal} {Astrophys. J. Suppl.}\ }\textbf {\bibinfo {volume} {80}},\
  \bibinfo {pages} {791} (\bibinfo {year} {1992}{\natexlab{b}})}\BibitemShut
  {NoStop}%
\bibitem [{\citenamefont {Fry}(1984)}]{Fry:1983cj}%
  \BibitemOpen
  \bibfield  {author} {\bibinfo {author} {\bibfnamefont {J.~N.}\ \bibnamefont
  {Fry}},\ }\href {\doibase 10.1086/161913} {\bibfield  {journal} {\bibinfo
  {journal} {Astrophys. J.}\ }\textbf {\bibinfo {volume} {279}},\ \bibinfo
  {pages} {499} (\bibinfo {year} {1984})}\BibitemShut {NoStop}%
\bibitem [{\citenamefont {Goroff}\ \emph {et~al.}(1986)\citenamefont {Goroff},
  \citenamefont {Grinstein}, \citenamefont {Rey},\ and\ \citenamefont
  {Wise}}]{Goroff:1986ep}%
  \BibitemOpen
  \bibfield  {author} {\bibinfo {author} {\bibfnamefont {M.~H.}\ \bibnamefont
  {Goroff}}, \bibinfo {author} {\bibfnamefont {B.}~\bibnamefont {Grinstein}},
  \bibinfo {author} {\bibfnamefont {S.~J.}\ \bibnamefont {Rey}}, \ and\
  \bibinfo {author} {\bibfnamefont {M.~B.}\ \bibnamefont {Wise}},\ }\href
  {\doibase 10.1086/164749} {\bibfield  {journal} {\bibinfo  {journal}
  {Astrophys. J.}\ }\textbf {\bibinfo {volume} {311}},\ \bibinfo {pages} {6}
  (\bibinfo {year} {1986})}\BibitemShut {NoStop}%
\bibitem [{\citenamefont {Bernardeau}(1992)}]{Bernardeau:1992zw}%
  \BibitemOpen
  \bibfield  {author} {\bibinfo {author} {\bibfnamefont {F.}~\bibnamefont
  {Bernardeau}},\ }\href {\doibase 10.1086/171398} {\bibfield  {journal}
  {\bibinfo  {journal} {Astrophys. J.}\ }\textbf {\bibinfo {volume} {392}},\
  \bibinfo {pages} {1} (\bibinfo {year} {1992})}\BibitemShut {NoStop}%
\bibitem [{\citenamefont {Scoccimarro}(1997)}]{Scoccimarro:1996jy}%
  \BibitemOpen
  \bibfield  {author} {\bibinfo {author} {\bibfnamefont {R.}~\bibnamefont
  {Scoccimarro}},\ }\href {\doibase 10.1086/304578} {\bibfield  {journal}
  {\bibinfo  {journal} {Astrophys. J.}\ }\textbf {\bibinfo {volume} {487}},\
  \bibinfo {pages} {1} (\bibinfo {year} {1997})},\ \Eprint
  {http://arxiv.org/abs/astro-ph/9612207} {arXiv:astro-ph/9612207 [astro-ph]}
  \BibitemShut {NoStop}%
\bibitem [{\citenamefont {Bernardeau}\ \emph {et~al.}(2002)\citenamefont
  {Bernardeau}, \citenamefont {Colombi}, \citenamefont {Gaztanaga},\ and\
  \citenamefont {Scoccimarro}}]{Bernardeau:2001qr}%
  \BibitemOpen
  \bibfield  {author} {\bibinfo {author} {\bibfnamefont {F.}~\bibnamefont
  {Bernardeau}}, \bibinfo {author} {\bibfnamefont {S.}~\bibnamefont {Colombi}},
  \bibinfo {author} {\bibfnamefont {E.}~\bibnamefont {Gaztanaga}}, \ and\
  \bibinfo {author} {\bibfnamefont {R.}~\bibnamefont {Scoccimarro}},\ }\href
  {\doibase 10.1016/S0370-1573(02)00135-7} {\bibfield  {journal} {\bibinfo
  {journal} {Phys. Rept.}\ }\textbf {\bibinfo {volume} {367}},\ \bibinfo
  {pages} {1} (\bibinfo {year} {2002})},\ \Eprint
  {http://arxiv.org/abs/astro-ph/0112551} {arXiv:astro-ph/0112551 [astro-ph]}
  \BibitemShut {NoStop}%
\bibitem [{\citenamefont {Carrasco}\ \emph {et~al.}(2014)\citenamefont
  {Carrasco}, \citenamefont {Foreman}, \citenamefont {Green},\ and\
  \citenamefont {Senatore}}]{Carrasco:2013sva}%
  \BibitemOpen
  \bibfield  {author} {\bibinfo {author} {\bibfnamefont {J.~J.~M.}\
  \bibnamefont {Carrasco}}, \bibinfo {author} {\bibfnamefont {S.}~\bibnamefont
  {Foreman}}, \bibinfo {author} {\bibfnamefont {D.}~\bibnamefont {Green}}, \
  and\ \bibinfo {author} {\bibfnamefont {L.}~\bibnamefont {Senatore}},\ }\href
  {\doibase 10.1088/1475-7516/2014/07/056} {\bibfield  {journal} {\bibinfo
  {journal} {JCAP}\ }\textbf {\bibinfo {volume} {1407}},\ \bibinfo {pages}
  {056} (\bibinfo {year} {2014})},\ \Eprint {http://arxiv.org/abs/1304.4946}
  {arXiv:1304.4946 [astro-ph.CO]} \BibitemShut {NoStop}%
\end{thebibliography}%

\end{document}